\documentclass[aps, prd, twocolumn, superscriptaddress, nofootinbib, showpacs,preprintnumbers]{revtex4}
\usepackage{graphicx}
\usepackage{amsmath}
\usepackage{ulem} %\sout{hogehoge}
\usepackage{mathrsfs}
\def\fsl#1{\setbox0=\hbox{$#1$}                 % set a box for #1 
   \dimen0=\wd0                                 % and get its size
   \setbox1=\hbox{/} \dimen1=\wd1               % get size of /
   \ifdim\dimen0>\dimen1                        % #1 is bigger
      \rlap{\hbox to \dimen0{\hfil/\hfil}}      % so center / in box
      #1                                        % and print #1
   \else                                        % / is bigger
      \rlap{\hbox to \dimen1{\hfil$#1$\hfil}}   % so center #1
      /                                         % and print /
   \fi}                                         %
\newcommand{\tr}{\mbox{tr}}

\newcommand{\VEV}[1]{\langle #1 \rangle}
\newcommand{\gtrsim}{\mathop{>}\limits_{\displaystyle{\sim}}}
\newcommand{\lessim}{\mathop{<}\limits_{\displaystyle{\sim}}}

\newcommand{\lrvec}[1]{\overset{\leftrightarrow}{#1}}
\makeatletter
\@addtoreset{equation}{section}
\makeatother

\begin{document}
\preprint{KUNS-2512}
\title{
Does unitarity imply finiteness of electroweak oblique corrections at 
one-loop?\\
--- constraining extra neutral Higgs bosons ---
}  
\date{\today}
\pacs{ 
12.39.Fe, % Chiral Lagrangians
12.60.Fr,	% Extensions of electroweak Higgs sector
14.80.Bn, % Standard-model Higgs bosons
14.80.Ec % Other neutral Higgs bosons
}

\author{Ryo Nagai}
\affiliation{
  Department of Physics, Nagoya University, Nagoya 464-8602, Japan
}
\author{Masaharu Tanabashi}
\affiliation{
  Kobayashi-Maskawa Institute for the Origin of Particles and the Universe, 
  Nagoya University, Nagoya 464-8602, Japan
}
\affiliation{
  Department of Physics, Nagoya University, Nagoya 464-8602, Japan
}
\author{Koji Tsumura}
\affiliation{
  Department of Physics, Kyoto University, Kyoto 606-8502, Japan
}

\begin{abstract}
Introducing arbitrary number of neutral Higgs bosons
in the electroweak symmetry breaking sector, 
we derive a set of conditions 
among Higgs couplings which need
to be satisfied to maintain the unitarity of the high energy 
scattering amplitudes 
of weak gauge bosons
at the tree level (unitarity sum rules).
It is shown that the unitarity sum rules require the tree level 
$\rho$ parameter to be 1, without explicitly  
invoking the custodial symmetry arguments.
The one-loop finiteness of the electroweak oblique 
corrections is automatically guaranteed once these unitarity 
sum rules are imposed among Higgs couplings.
Severe constraints on the 
lightest Higgs coupling (125GeV Higgs coupling) and the mass of the 
second lightest Higgs boson are obtained from the unitarity and the 
results of the electroweak precision tests (oblique parameter measurements).
These results are compared with the effective theory of the light Higgs boson, 
and we find simple relationships between the mass of the second lightest 
Higgs boson in our framework and the ultraviolet cutoff in the effective 
theory framework.
\end{abstract}

\maketitle

\section{Introduction}

The year 2012 discovery of a Higgs boson at 125GeV at the Large 
Hadron Collider (LHC) experiments\cite{Aad:2012tfa,Chatrchyan:2012ufa}
completed the set of all 
particles predicted in the Standard Model (SM).
We now have a key particle to solve the
mystery of the origin of particle masses (electroweak symmetry
breaking). 
Due to the lack of mechanism to stabilize the electroweak
scale against the radiative corrections, however,
the SM electroweak symmetry breaking (EWSB) sector is 
believed to be incomplete.
Varieties of extended EWSB models have therefore been proposed. 
These extended models typically contain more particles 
other than the observed Higgs boson in their EWSB sector.

One of the key roles of the SM Higgs boson is to 
unitarize the high energy longitudinal weak gauge bosons' scattering 
amplitudes\cite{Cornwall:1973tb,Cornwall:1974km,Llewellyn Smith:1973ey,Lee:1977eg}.
The Higgs boson also makes the SM renormalizable, {\it i.e.,} it cancels 
non-renormalizable ultraviolet (UV) divergences appearing at the loop level.
The Higgs coupling strengths with the weak gauge bosons are
precisely adjusted in order to make the SM unitary and renormalizable.
Although experimental data accumulated so far on the 125GeV boson
are consistent with the SM Higgs
particle\cite{kn:ATLAS-CONF-2014-009,kn:CMS-PAS-HIG-14-009,Aad:2013wqa}
(See also Refs.\cite{Giardino:2012dp,Low:2012rj,Carmi:2012in,Falkowski:2013dza,Giardino:2013bma,Boudjema:2013qla,Artoisenet:2013puc,Cheung:2014noa}),
in the extended EWSB scenarios, the coupling strengths of the 125GeV
boson still have a chance to deviate largely from the predicted values in
the SM\@.
Actually, it has been pointed out that the 125GeV particle is,
within the accuracy of the present data, 
also consistent\cite{Matsuzaki:2012mk,Matsuzaki:2012xx}
 with a
techni-dilaton (light composite scalar particle) composed through
hypothetical walking technicolor dynamics.
This situation will change drastically in future.
Future LHC experiments with high luminosity will be able to measure
$hVV\, (V=W,Z)$ coupling more accurately\cite{Ref:HL-LHC,CMS:2013xfa}, 
where $h$ is the observed Higgs boson.
Various Higgs coupling strengths will be measured very precisely 
at the International Linear Collider (ILC) experiment\cite{Ref:ILC,Peskin:2012we}.

%%%%%%%%revised by RN 14.11.20%%%%%%%%%%%%%%%%%%%%%%%%
How can we utilize such high precision Higgs measurements?
If the measured value of Higgs coupling strengths turn out to deviate 
from the SM values, in order to make the theory unitary and
to keep consistency with the electroweak precision tests (EWPTs), 
new particles other than the 125GeV Higgs boson need to exist.
Can we make definite predictions for the properties of this required 
new particle in this case?
 In this paper, we try to answer this question from the viewpoint 
of the unitarity and the EWPTs.
We assume EWSB sector contains richer spectrum of particles, {\it i.e.,}
a zoo of ``Higgs'' bosons\footnote{
We call all scalar particles participating the unitarization of the longitudinal
gauge boson scattering as ``Higgs'' bosons.
% In the electroweak chiral Lagrangian approach, particles are classified
% by the unbroken $U(1)_{\rm em}$ symmetry.
}, in addition to the discovered 125GeV 
Higgs boson in order to make the deviation of Higgs couplings possible
without conflicting with the unitarity and the EWPTs.   
We do not assume particular Higgs potential models, however,
since we seek for clues of physics beyond the SM as model 
independent as possible.
Conditions to keep the scattering amplitudes (perturbatively) unitary
at high energy have been derived in Ref.\cite{Gunion:1990kf}
and are well-known as the ``unitarity sum rules''\footnote{
Unitarity sum rules in the Higgsless theories\cite{Csaki:2003dt}, in which 
a tower of spin-1 resonances exists instead of the spin-0 Higgs boson
in the Higgs sector, 
have been fully investigated in Ref.\cite{SekharChivukula:2008mj}.
Assuming simultaneous existence of both spin-0 and spin-1 particles,
Ref.\cite{Abe:2012fb} gave model independent sum rules.
See also Refs.\cite{Chivukula:2003kq,Chang:2003vs} for related topics.
}.
However, implications of such unitarity sum rules to the EWPTs at 
loop level have not been clarified until now.
In the former half of this paper, we study the implications of the 
unitarity sum rules to the finiteness of the electroweak precision
parameters (oblique correction parameters $S$, $T$, $U$) at the loop level.
For such a purpose, we reanalyze the unitarity sum rules by using the
electroweak chiral Lagrangian technique and the
equivalence theorem to keep manifest $SU(2)\times U(1)$ 
gauge invariance, which allow us to use the model not only for the 
tree level unitarity analysis but also for the loop level oblique correction
analysis.
We simplify our analysis assuming only neutral Higgs bosons in the EWSB sector.
We find the one-loop finiteness of the electroweak oblique correction 
parameters is automatically guaranteed by the unitarity sum rules within 
this setup. 
Extensions including charged Higgs bosons and fermions will be discussed 
elsewhere.

In the latter half of the paper, we study phenomenological implications
of the uniarity and the electroweak oblique parameter constraints.
We use these constraints to impose upper bounds on the second lightest 
Higgs boson mass as a function of the deviation of the 125GeV Higgs coupling
$\Delta\kappa_V^{} (\equiv \kappa_V-1)$. 
Here $\Delta\kappa_V^{}$ denotes the deviation of the 125GeV Higgs coupling 
with weak gauge bosons from its SM value.
Once the absence of the second lightest Higgs boson is confirmed below
1 TeV, the electroweak precision constraint will rule out $\Delta\kappa_V \lessim -0.02$ at 95\% CL\@.

We keep the tree-level $\rho$ parameter arbitrary in the 
unitarity analysis, which enables us to investigate theoretical 
structures which determine the value of $\rho$ parameter.
Especially, we are able to show,
without explicitly invoking the custodial 
symmetry arguments,
the unitarity of the scattering amplitudes requires the tree-level 
$\rho$ parameter 
to be unity in any EWSB model if it only possesses neutral Higgs bosons.
Custodial symmetry is not a required symmetry.
Instead, $\rho=1$ is considered as a result of the unitarity 
in this class of models.
This is consistent with the fact that $\rho=1$ is predicted in 
all the known renormalizable EWSB models which do not contain charged 
Higgs boson couplings with the electroweak gauge bosons. 
Our finding will be helpful to understand the reason of $\rho=1$ in the
septet Higgs extension model\cite{Hisano:2013sn,Kanemura:2013mc,Alvarado:2014jva}
which does not enjoy explicit custodial symmetry.
We will discuss the septet issue in our separate publication.

Our strategy described in this paper should not be confused
with the usual light Higgs effective field theory 
approaches\cite{Buchmuller:1985jz,De Rujula:1991se,Hagiwara:1992eh,Hagiwara:1993ck,Hagiwara:1993qt,Alam:1997nk,Barger:2003rs,Giudice:2007fh,Kanemura:2008ub,Boos:2013mqa,Grzadkowski:2010es,Grinstein:2007iv,Alonso:2012px,Buchalla:2012qq,Azatov:2012bz,Corbett:2012dm,Corbett:2012ja,Mebane:2013cra,Jenkins:2013fya,Grojean:2013kd,Elias-Miro:2013gya,Belanger:2013xza,Elias-Miro:2013mua,Buchalla:2013rka,Lopez-Val:2013yba,Corbett:2013pja,Contino:2013kra,Jenkins:2013zja,Jenkins:2013wua,Alonso:2013hga,Alonso:2014rga}.
In the effective field theory approach based on the linear
sigma model\cite{Buchmuller:1985jz,Grzadkowski:2010es,De Rujula:1991se,Hagiwara:1992eh,Hagiwara:1993ck,Hagiwara:1993qt,Alam:1997nk,Elias-Miro:2013mua,Barger:2003rs,Kanemura:2008ub,Corbett:2012dm,Corbett:2012ja,Corbett:2013pja,Mebane:2013cra,Belanger:2013xza,Lopez-Val:2013yba,Boos:2013mqa},
the discovered 125GeV Higgs boson field is assumed to be
a component of a doublet Higgs field just like in the SM\@.
The deviations of Higgs couplings are encoded in the higher
dimensional effective Lagrangian coefficients including
their renormalization group flow at the loop level\cite{De Rujula:1991se,Hagiwara:1992eh,Hagiwara:1993ck,Hagiwara:1993qt,Alam:1997nk,Grojean:2013kd,Elias-Miro:2013gya,Elias-Miro:2013mua,Jenkins:2013zja,Jenkins:2013wua,Alonso:2013hga,Alonso:2014rga}.

Due to the presence of such higher dimensional operators,
perturbative unitarity of the scattering amplitudes is violated 
at certain high energy scale (cutoff scale of the effective theory)
in the effective field theory\cite{Zhang:2003it,Chang:2013aya}. 
Yet unknown UV completion theory therefore needs to 
replace 
the
effective field theory above the cutoff scale.
In this sense, in addition to the studies of the effective field theory,
we need to study model dependently.
Actually, many model dependent studies have been 
performed\cite{Hisano:2013sn,Kanemura:2013mc,Cheon:2012rh,Craig:2012vn,Chang:2012ve,Bai:2012ex,Ferreira:2012nv,Chang:2012zf,Chen:2013kt,Celis:2013rcs,Grinstein:2013npa,Chen:2013rba,Craig:2013hca,Kanemura:2013eja,Ferreira:2014naa,Kanemura:2014bqa,Chang:2012gn,Chiang:2013rua,Earl:2013jsa,Killick:2013mya,Abe:2012fb,Pich:2012dv}.
In this paper, we try to establish a systematic classification of
possibilities of perturbative UV completions appearing at the cutoff
scale.\footnote{
Ref.\cite{Low:2009di}  found 
theoretical constraints on effective Lagrangian parameters 
assuming unitary UV completion behind the effective theory.
See also Ref.\cite{Falkowski:2012vh,Bellazzini:2014waa}. 
} 
Especially, we find simple relationships between bounds
on the  second lightest Higgs boson mass in our framework
and the UV cutoff in the effective field theory framework.

This paper is organized as follows:
In Sec.II, we describe the model we use in this paper.
For simplicity, we restrict ourselves only to the neutral
Higgs extension models.  
We next take the unitary gauge in Sec.III,
and compare our model with the gauge non-invariant 
model used in Ref.\cite{Gunion:1990kf}.
Sec.IV is devoted to the unitarity sum rules 
and their possible applications to physics.
We then evaluate the one-loop radiative corrections
to the $f\bar{f}\to f'\bar{f}'$ amplitudes
in Sec.V\@.
We explicitly show that the amplitudes automatically 
remain finite
at one-loop level if we impose the unitarity sum rules
among various Higgs couplings.
The explicit formulas of the electroweak oblique 
parameters\cite{Peskin:1990zt}
(Peskin-Takeuchi parameters) are presented in Sec.VI, 
and
we obtain bounds on the second lightest Higgs boson 
mass from the unitarity and the EWPTs in Sec.VII\@.
Sec.VIII discusses extra conditions other than the unitarity 
sum rules we need to impose to make the theory fully 
UV-complete.
Relationship between our approach and the effective field 
theory will be discussed in Sec.IX\@. 
Conclusions and outlook are given in Sec.X\@.

\section{The model}\label{sec:The model}

We use the electroweak chiral 
Lagrangian\cite{Appelquist:1980vg,Appelquist:1980ae}
technique to
describe the arbitrary interactions among weak gauge bosons and 
neutral ``Higgs'' bosons in an $SU(2)\times U(1)$ gauge invariant manner.
The Lagrangian $\mathcal{L}$ of this model can be decomposed as
\begin{equation}
\mathcal{L}
= \mathcal{L}_{\chi}
  +\mathcal{L}_{\rm gauge}
  +\mathcal{L}_{\rm Higgs},
\end{equation}
with ${\mathcal L}_{\chi}$, 
${\mathcal L}_{\rm gauge}$, and
$\mathcal{L}_{\rm Higgs}$ being 
the $SU(2)\times U(1)/U(1)$ non-linear sigma model Lagrangian, 
the $SU(2)\times U(1)$ gauge Lagrangian,
and the Higgs Lagrangian, respectively.
Hereafter, we restrict our model Lagrangian to contain only terms 
of mass dimension four or less and up to (at most) two derivatives 
(${\cal O}(\partial^2)$ terms) since we are interested in models in which 
scattering amplitudes remain unitary at high energy.

The spontaneous EWSB sector is 
described by using the electroweak chiral Lagrangian
\begin{eqnarray}
  {\mathcal L}_{\chi}
  &=& \dfrac{v^2}{4}\tr\!\left[(D_\mu U)^\dag(D^\mu U)\right]
  \nonumber\\
  & & 
  +\beta\dfrac{v^2}{4}\tr\!\left[U^\dag (D_\mu U)\tau_3\right]
                      \tr\!\left[U^\dag (D^\mu U)\tau_3\right] .
  \nonumber \\
\label{eq:EWchirallag}
\end{eqnarray}
We denote $v \simeq 246$GeV the decay constant of the charged
would-be Nambu-Goldstone boson (NGB).
The non-linear sigma model field $U$
\begin{equation}
  U =\exp{(i\tilde{w}^a\tau_a)},
\end{equation}
is introduced in Eq.(\ref{eq:EWchirallag}), so as to describe
the NGB field arising from the spontaneous EWSB.
Here $\tau_a$ ($a=1,2,3$) are the Pauli matrices, and 
$\tilde{w}^a$ are the NGB fields. 
Note that, under the $SU(2) \times U(1)$ gauge transformation, 
the NGB field $\tilde{w}^a\tau_a$ transforms non-linearly,
\begin{equation}
  U \to G_L U G_Y^\dagger, 
\end{equation}
with
\begin{equation}
  G_L \equiv \exp\left(i\dfrac{\tau_a}{2}\theta^a_L\right), \qquad
  G_Y {\equiv} \exp\left(i\dfrac{\tau_3}{2}\theta_Y\right).
\end{equation}
The covariant derivative $D_\mu U$ is defined as
\begin{equation}
 D_\mu U = \partial_\mu U+ig{\bf W}_\mu U-ig_Y^{}U{\bf B}_\mu , 
\label{eq:covariantU}
\end{equation}
with 
$SU(2)\times U(1)$ gauge fields ${\bf W}_\mu$ and ${\bf B}_\mu$ being defined
by
\begin{equation}
{\bf W}_\mu = W^a_\mu\dfrac{\tau_a}{2}, 
\qquad 
{\bf B}_\mu = B_\mu\dfrac{\tau_3}{2}.
\end{equation}
The gauge transformation of Eq.(\ref{eq:covariantU}) is 
\begin{equation}
  D_\mu U \rightarrow G_L (D_\mu U) G_Y^\dagger ,
\end{equation}
where the gauge fields transform as
\begin{eqnarray}
  {\bf W}_\mu &\to& G_L {\bf W}_\mu G_L^\dagger
                +\dfrac{i}{g} (\partial_\mu G_L) G_L^\dagger, 
\label{eq:gaugetransf1}  \\
  {\bf B}_\mu &\to& G_Y {\bf B}_\mu G_Y^\dagger
                +\dfrac{i}{g_Y} (\partial_\mu G_Y) G_Y^\dagger .
\label{eq:gaugetransf2}
\end{eqnarray}
The gauge invariance of the electroweak chiral Lagrangian
Eq.(\ref{eq:EWchirallag}) is manifest.

The vacuum expectation value (VEV) of $U$,
\begin{equation}
  \langle U \rangle = 1,
\end{equation}
breaks the electroweak symmetry spontaneously
\begin{equation}
  \langle U \rangle \to
  \langle G_L U G_Y^\dagger \rangle
  = G_L G_Y^\dagger \ne 1
  = \langle U \rangle .
\end{equation}
The spectrum of physical particles can be obtained by taking 
the unitary gauge $U=1$, with which the electroweak 
chiral Lagrangian Eq.(\ref{eq:EWchirallag}) leads to the
mass terms of $W$ and $Z$ bosons,
\begin{equation}
  M_W^2 = \dfrac{g^2}{4}v^2, \qquad
  M_Z^2 = \dfrac{g_Z^2}{4} v_Z^2, \qquad
\end{equation}
with
\begin{equation}
  v^2_Z \equiv v^2(1-2\beta),
\end{equation}
and
\begin{equation}
  g_Z^2 \equiv g^2 + g_Y^2.
\end{equation}
Here the charged $W$ boson field ($W_\mu$), the neutral $Z$ boson field
($Z_\mu$) and the photon field $A_\mu$ are given by
\begin{equation}
  W^{\pm}=\dfrac{1}{\sqrt{2}}\left(W^1_\mu \mp iW^2_\mu \right),
\end{equation}
and
\begin{equation}
\left(
\begin{array}{ccc}
Z_\mu\\
A_\mu\\
\end{array}
\right) 
= 
\left(
\begin{array}{ccc}
c&-s\\
s&c\\
\end{array}
\right)\left(
\begin{array}{ccc}
W^3_\mu\\
B_\mu\\
\end{array}
\right),
\end{equation}
with
\begin{equation}
s\equiv\frac{g_Y^{}}{\sqrt{g^2+g^2_Y}}, \qquad 
c\equiv\frac{g}{\sqrt{g^2+g^2_Y}}.
\end{equation}
The QED coupling strength $e$ is given by
\begin{equation}
  e \equiv gs .
\end{equation}

The coefficient $\beta$ in the electroweak chiral Lagrangian
Eq.(\ref{eq:EWchirallag}) can be related with the tree-level
$\rho$ parameter, which is defined as
\begin{equation}
  \rho_0 \equiv \dfrac{g_Z^2/M_Z^2}{g^2/M_W^2}
       = \dfrac{v^2}{v_Z^2}
       = \dfrac{1}{1-2\beta}.
\end{equation}
We keep $\rho_0$ arbitrary in our analysis of longitudinal gauge boson
scattering amplitudes, which makes it possible to investigate the effects
of $\rho_0 \ne 1$ in the longitudinal gauge boson scattering amplitudes.
This is in contrast to the analysis of Ref.\cite{Gunion:1990kf} in which
$\rho_0 = 1$ is assumed in their practical applications of the unitarity 
sum rules to the EWSB models.

We investigate the longitudinal gauge boson scattering amplitudes
using their equivalence with the NGB scattering 
amplitudes\cite{Cornwall:1974km,Lee:1977eg,Chanowitz:1985hj}.
We define the NGB fields $w^\pm$ (charged NGB)
and $z$ (neutral NGB)
\begin{equation}
  w^{\pm} = \dfrac{v}{\sqrt{2}}\left(\tilde{w}_1\mp i\tilde{w}_2\right), \qquad
  z = v_Z^{} \tilde{w}_3, 
\end{equation}
to make the kinetic terms of $w^\pm$ and $z$ normalized canonically.
We then obtain
\begin{equation}
  \tilde{w}^a\tau_a 
 = \dfrac{\sqrt{2}}{v} \left( w^+ \tau_+ + w^- \tau_- \right)
  +\dfrac{1}{v_Z^{}} z\, \tau_3,
\end{equation}
with
\begin{equation}
  \tau_\pm \equiv \dfrac{1}{2} \left(\tau_1 \pm i \tau_2 \right).
\end{equation}

The $SU(2)\times U(1)$ gauge Lagrangian ${\mathcal L}_{\rm gauge}$ is given by
\begin{equation}
  {\mathcal L}_{\rm gauge}
  = -\frac{1}{2}\tr[{\bf W}_{\mu\nu}{\bf W}^{\mu\nu}]
    -\frac{1}{2}\tr[{\bf B}_{\mu\nu}{\bf B}^{\mu\nu}] .
  \label{eq:gauge}
\end{equation}
Here $SU(2)\times U(1)$ field strengths ${\bf W}_{\mu\nu}$,
${\bf B}_{\mu\nu}$ are
\begin{equation}
{\bf W}_{\mu\nu} \equiv 
\partial_\mu {\bf W}_\nu-\partial_\nu {\bf W}_\mu+ig[{\bf W}_\mu,{\bf W}_\nu],
\end{equation}
\begin{equation}
{\bf B}_{\mu\nu}\equiv\partial_\mu {\bf B}_\nu-\partial_\nu {\bf B}_\mu .
\end{equation}
Note the gauge field strengths behave 
\begin{equation}
  {\bf W}_{\mu\nu}\to G_L{\bf W}_{\mu\nu}G^\dag_L, \qquad
  {\bf B}_{\mu\nu}\to {\bf B}_{\mu\nu},
\end{equation}
under the gauge transformation given 
in Eq.(\ref{eq:gaugetransf1}) and Eq.(\ref{eq:gaugetransf2}).
The Lagrangian Eq.(\ref{eq:gauge}) is
therefore invariant under the gauge transformation.
%  given 
% in Eq.(\ref{eq:gaugetransf1}) and Eq.(\ref{eq:gaugetransf2}).

We next incorporate neutral spin-0 ``Higgs'' bosons
($\phi^0_n$, $n=1,2,\cdots N_0$) as ``matter'' particles in 
the chiral Lagrangian, which keep the model unitary at high energy,
\begin{eqnarray}
 {\mathcal L}_{\rm Higgs} 
  &=& -V + 
       \frac{1}{2} \sum_{n_1=1}^{N_0} \sum_{n_2=1}^{N_0} K_{n_1^{} n_2^{}}
       (\partial_\mu \phi_{n_1^{}}^0)(\partial^\mu \phi_{n_2^{}}^0)
  \nonumber\\
  & & 
      + {\cal L}_{\rm int} , 
\end{eqnarray}
with $V$, $K$ being functions of $\phi^0_n$.

The masses of these ``Higgs'' particles and their self-interactions 
are described by $V(\phi^0)$.\footnote{
Note that $\phi_n^0$ in Eq.(\ref{eq:VEVphi0}) is classified by the 
unbroken $U(1)_{\rm em}$ symmetry.
Eq.(\ref{eq:VEVphi0}) does not imply the absence of the vacuum expectation 
values of the linearly realized Higgs multiplet field
to which $\phi_n^0$ is considered to belong.
}
We assume 
\begin{equation}
  \langle \phi^0_n \rangle = 0, 
\label{eq:VEVphi0}
\end{equation}
for $n=1,2,\cdots N_0$.
$V(\phi^0)$ is therefore
\begin{equation}
  V(\phi^0)
  = \frac{1}{2} \sum_{n=1}^{N_0} M_{\phi_n^0}^2 \phi_n^0 \phi_n^0 
   +\cdots,
\label{eq:higgs_potential}
\end{equation}
with ``$\cdots$'' being terms of self-interactions among
these ``Higgs'' particles.
We take $K_{n_1^{} n_2^{}}$ so as to make the Higgs kinetic 
term canonically normalized\footnote{In general, $K_{n1,n2}$ induces higher dimensional operators which are function of Higgs fields. We ignore these operators, however, since they violate the perturbative unitarity of ${\phi+\phi \to \phi+\phi}$ scattering amplitudes explicitly at high energy.}
\begin{equation}
  K_{n_1^{} n_2^{}}(\phi^0) = \delta_{n_1^{} n_2^{}}^{}.
\end{equation}

Interactions of these ``Higgs'' particles with the electroweak 
gauge bosons are described by ${\mathcal L}_{\rm int}$,
\begin{equation}
  {\mathcal L}_{\rm int} 
  = {\mathcal L}_\phi + {\mathcal L}_{{\phi \lrvec{\partial} \phi}} + {\mathcal L}_{\phi\phi}~,
\label{eq:higgsint}
\end{equation}
where
\begin{eqnarray}
\lefteqn{
\mathcal{L}_{{\phi}} = 
} \nonumber\\
 & & 
   -v\sum_{n=1}^{N_0} \kappa_{WW}^{\phi^0_n}\phi^0_n
    \tr\!\left[U^\dag (D_\mu U)\tau_+]\tr[U^\dag (D^\mu U)\tau_-\right]
\nonumber\\
&  & -\dfrac{v}{4}\sum_{n=1}^{N_0} \kappa_{ZZ}^{\phi^0_n}\phi^0_n
     \tr\!\left[U^\dag (D_\mu U)\tau_3\right]\tr\!\left[U^\dag (D^\mu U)\tau_3\right], 
\nonumber\\
& &
\label{eq:lagphi}
\\
\lefteqn{
\mathcal{L}_{{\phi \lrvec{\partial} \phi}} =
} \nonumber\\
  & & 
   -\dfrac{i}{4} \sum_{n=1}^{N_0} \sum_{m=1}^{N_0}
   {{\kappa}_{Z}^{\phi^0_n\phi^{0}_m}} 
   (\phi^0_n\lrvec{\partial}_\mu \phi^{0}_{m} )
   \tr\!\left[U^\dag (D^\mu U)\tau_3\right],
\nonumber\\
& &
\label{eq:lagphidelphi}
\\
\lefteqn{
\mathcal{L}_{{\phi\phi}} =
} \nonumber\\
& &
  -\frac{1}{2}\sum_{n=1}^{N_0} \sum_{m=1}^{N_0}
%\sum_{n,m}
   \kappa_{WW}^{\phi^0_n\phi^{0}_m}\phi^0_n\phi^{0}_{m}
   \times
\nonumber\\
& & \qquad \times
   \tr\!\left[U^\dag (D_\mu U)\tau_+\right]
   \tr\!\left[U^\dag (D^\mu U)\tau_-\right]
\nonumber\\
& & -\frac{1}{8}\sum_{n=1}^{N_0} \sum_{m=1}^{N_0}
%\sum_{n,m}
    \kappa_{ZZ}^{\phi^{0}_n\phi^{0}_m}
    \phi^0_n\phi^{0}_{m}
   \times
\nonumber\\
& & \qquad \times
    \tr\!\left[U^\dag (D_\mu U)\tau_3\right]
    \tr\!\left[U^\dag (D^\mu U)\tau_3\right] ,
\nonumber\\
& &
\label{eq:lagphiphi}
\end{eqnarray}
with 
\begin{equation}
  \phi_n^0 \lrvec{\partial}_\mu \phi_m^0
  \equiv
  \phi_n^0 (\partial_\mu \phi_m^0)
 -(\partial_\mu\phi_n^0)  \phi_m^0 .
\end{equation}

Note that our ``Higgs'' $\phi_n^0$ are all {\it real} scalar fields.
The Higgs coupling parameters 
$\kappa_{WW}^{\phi_n^0}$, 
$\kappa_{ZZ}^{\phi_n^0}$, 
$\kappa_{Z}^{\phi_n^0 \phi_m^0}$, 
$\kappa_{WW}^{\phi_n^0 \phi_m^0}$
and
$\kappa_{ZZ}^{\phi_n^0 \phi_m^0}$
are therefore required to be real.  We also
note the $n\leftrightarrow m$ antisymmetry of $\kappa_{Z}^{\phi_n^0
\phi_m^0}$, {\it i.e.,}  
\begin{equation}
  \kappa_{Z}^{\phi_n^0 \phi_m^0} = -\kappa_{Z}^{\phi_m^0 \phi_n^0} ,
\label{eq:kappa-antisymmetry}
\end{equation}
and the $n\leftrightarrow m$ symmetry of $\kappa_{VV}^{\phi_n^0 \phi_m^0}$,
{\it i.e.,}
\begin{equation}
  \kappa_{WW}^{\phi_n^0 \phi_m^0} = \kappa_{WW}^{\phi_m^0 \phi_n^0} ,
  \qquad
  \kappa_{ZZ}^{\phi_n^0 \phi_m^0} = \kappa_{ZZ}^{\phi_m^0 \phi_n^0} .
\label{eq:kappa-symmetry}
\end{equation}
Although the interaction Lagrangian Eq.(\ref{eq:higgsint}) has some
similarity with the light Higgs effective theory realized in 
the non-linear sigma model\cite{Grinstein:2007iv,Azatov:2012bz,Alonso:2012px,Buchalla:2012qq,Buchalla:2013rka}, our approach differs from the low energy 
effective theory, since we
explicitly introduce heavy Higgs bosons in order to keep the 
model unitary at high energy as we stressed before.

We here make a couple of comments on the $CP$ transformation
properties of the model.  
We know
\begin{eqnarray*}
  & & (CP) w^+(x^\mu)  (CP)^{-1} = - w^-(x_\mu), \\
  & &  (CP) w^-(x^\mu)  (CP)^{-1} = - w^+(x_\mu), \\
  & & (CP) z(x^\mu) (CP)^{-1} = - z(x_\mu),
\end{eqnarray*}
and thus\footnote{Precisely speaking, we choose the convention for charged NGBs 
under the $CP$ transformation by Eq.(\ref{Eq:CP}).}
\begin{equation}
  (CP) \tilde{w}^a(x^\mu) \tau_a (CP)^{-1} 
= \tau_2 (\tilde{w}^a(x_\mu) \tau_a ) \tau_2. \label{Eq:CP}
\end{equation}
The $CP$ transformation of the non-linear sigma model field is 
therefore given by
\begin{equation}
  (CP) U(x^\mu) (CP)^{-1}
                  = \tau_2 U(x_\mu) \tau_2.
\end{equation}
In order to keep the electroweak chiral Lagrangian Eq.(\ref{eq:EWchirallag})
invariant under the $CP$ transformation, ${\bf W}^\mu$ and ${\bf B}_\mu$
need to transform as
\begin{equation}
  (CP) {\bf W}^\mu(x^\mu) (CP)^{-1} = 
  \tau_2 {\bf W}_\mu(x_\mu) \tau_2,
\label{eq:CPW}
\end{equation}
and
\begin{equation}
  (CP) {\bf B}^\mu(x^\mu) (CP)^{-1} = 
  \tau_2 {\bf B}_\mu(x_\mu) \tau_2 .
\label{eq:CPB}
\end{equation}
It is easy to check that Eq.(\ref{eq:CPW}) and Eq.(\ref{eq:CPB})
are consistent with conventional $CP$ quantum number assignments
of the electroweak gauge bosons.
We also find
\begin{equation}
  (CP)\, \tr\left[ U^\dagger (D_\mu U) \tau_\pm \right] (CP)^{-1}
  = - \tr\left[ U^\dagger (D^\mu U) \tau_\mp \right], 
\end{equation}
\begin{equation}
  (CP)\, \tr\left[ U^\dagger (D_\mu U) \tau_3 \right] (CP)^{-1}
  = - \tr\left[ U^\dagger (D^\mu U) \tau_3 \right] . 
\end{equation}

We are now ready to discuss the $CP$ transformation properties
of neutral ``Higgs'' bosons in our model.  
We assign
\begin{equation}
  (CP) \phi_n^0(x^\mu) (CP)^{-1}
  = \eta_n \phi_n^0(x_\mu),
\end{equation}
with 
\begin{equation}
  \eta_n =
  \begin{cases}
    +1 & \text{ for $CP$ even}\\
    -1 & \text{ for $CP$ odd}
  \end{cases}.
\end{equation}
Requiring the Lagrangians Eqs.(\ref{eq:lagphi}),
(\ref{eq:lagphidelphi}) and (\ref{eq:lagphiphi}) invariant
under the $CP$ transformation, we obtain
\begin{equation}
  \kappa_{WW}^{\phi_n^0} \eta_n = \kappa_{WW}^{\phi_n^0}, \qquad
  \kappa_{ZZ}^{\phi_n^0} \eta_n = \kappa_{ZZ}^{\phi_n^0}, 
\label{eq:CPkappaWW1}
\end{equation}
\begin{equation}
  - \kappa_{Z}^{\phi_n^0 \phi_m^0} \eta_n \eta_m  = 
    \kappa_{Z}^{\phi_n^0 \phi_m^0}, 
\label{eq:CPkappaZ2}
\end{equation}
and
\begin{equation}
  \kappa_{WW}^{\phi_n^0 \phi_m^0} \eta_n \eta_m =
  \kappa_{WW}^{\phi_n^0 \phi_m^0},
  \qquad
  \kappa_{ZZ}^{\phi_n^0 \phi_m^0} \eta_n \eta_m =
  \kappa_{ZZ}^{\phi_n^0 \phi_m^0}. 
\label{eq:CPkappaWW2}
\end{equation}
From Eq.(\ref{eq:CPkappaWW1}), it is easy to see
\begin{equation}
  \kappa_{WW}^{\phi_n^0} = \kappa_{ZZ}^{\phi_n^0} = 0, \quad
  \mbox{for $\eta_n = -1$}.
\end{equation}
Also, combining Eq.(\ref{eq:kappa-symmetry}) and Eq.(\ref{eq:CPkappaWW2}),
we obtain
\begin{equation}
  \kappa_{WW}^{\phi_n^0 \phi_m^0}=\kappa_{ZZ}^{\phi_n^0 \phi_m^0}=0,
  \quad
  \mbox{for $\eta_n \eta_m = -1$},
\end{equation}
if the Higgs sector preserves the $CP$ invariance.

\section{Lagrangian in the unitary gauge}
Unitarity sum rules of longitudinal weak boson scattering 
amplitudes\cite{Cornwall:1973tb,Cornwall:1974km,Llewellyn Smith:1973ey}
were thoroughly investigated by Ref.\cite{Gunion:1990kf} 
in the context of the $SU(2)\times U(1)$ gauge theory with arbitrary
Higgs multiplets.
Ref.\cite{Gunion:1990kf} performed their analysis without introducing
unphysical would-be NGBs, however, in contrast
to our chiral Lagrangian analysis in which $SU(2)\times U(1)$ gauge 
invariance is kept manifest.
In order to make direct comparisons between the results of 
Ref.\cite{Gunion:1990kf} and the results presented in this paper,
it is convenient to rewrite our model in the unitary gauge
\begin{equation}
  U = 1,
\end{equation}
in which unphysical would-be  NGBs are absent.
We then find
\begin{eqnarray}
  {\cal L}_\chi &=& M_W^2 W^+_\mu W^{-\mu} + \dfrac{1}{2} M_Z^2 Z_\mu Z^\mu,
  \\
{\cal L}_{\phi}
  &=& g M_W \sum_{n=1}^{N_0}\kappa_{WW}^{\phi^0_n}\phi^0_nW^+_\mu W^{-\mu}
\nonumber\\
  & &+ \dfrac{g_Z^{}}{2} \dfrac{v}{v_Z^{}} M_Z \sum_{n=1}^{N_0}
       \kappa_{ZZ}^{\phi^0_n}\phi^0_nZ_\mu Z^\mu,
  \\
{\cal L}_{\phi \lrvec{\partial} \phi}
  &=& \dfrac{g_Z^{}}{4}\sum_{n=1}^{N_0}\sum_{m=1}^{N_0}{\kappa}_{Z}^{\phi^0_n\phi^{0}_m}
      (\phi^0_n\lrvec{\partial}_\mu\phi^{0}_m)Z^\mu,
  \\
{\cal L}_{\phi\phi}
  &=&\dfrac{g^2}{4}\sum_{n=1}^{N_0}\sum_{m=1}^{N_0}
     \kappa_{WW}^{\phi^0_n\phi^{0}_m}\phi^0_n\phi^{0}_mW^{+}_{\mu} W^{-\mu}
  \nonumber
  \\
  & &+\dfrac{g^2_Z}{8}\sum_{n=1}^{N_0}\sum_{m=1}^{N_0}
      \kappa_{ZZ}^{\phi^0_n\phi^{0}_m}\phi^0_n\phi^{0}_mZ_{\mu} Z^\mu,
\end{eqnarray}
which correspond to the masses of vector bosons ($V$),
the Higgs-$V$-$V$ vertices, the Higgs-Higgs-$V$ vertex,
and the Higgs-Higgs-$V$-$V$ vertices of Ref.\cite{Gunion:1990kf},
respectively.
It is easy to see that the $CP$ properties 
Eqs.(\ref{eq:CPkappaWW1}--\ref{eq:CPkappaWW2}) are identical to the
$CP$ properties of $WW\phi$, $ZZ\phi$, $Z\phi\phi$, $WW\phi\phi$, 
$ZZ\phi\phi$ couplings obtained in Ref.\cite{Gunion:1990kf}.

%-------------------------------------------------------------------------------------------
%\subsection{${VV\phi\phi}$ coupling}

\section{Unitarity sum rules}
The cancellation of the unitarity violating high energy
scattering amplitudes
of longitudinally polarized gauge bosons requires a set of conditions 
among Higgs couplings (``unitarity sum
rules'')\cite{Cornwall:1973tb,Cornwall:1974km,Llewellyn Smith:1973ey}.
The unitarity sum rules in the $SU(2)\times U(1)$ gauge theory were
studied a couple of decades ago by Ref.\cite{Gunion:1990kf} and 
recently by Ref.\cite{Grinstein:2013fia}.
In this section, using the equivalence theorem of the amplitudes
of longitudinal gauge bosons and the would-be NGBs,
we rederive the sum rules\cite{Gunion:1990kf} in our gauge invariant 
Lagrangian through the NGB scattering amplitudes.
We will then check explicitly the equivalence of our results with 
the sum rules derived in Ref.\cite{Gunion:1990kf}, which supports
the consistency of our method using the gauge invariant Lagrangian.

\subsection{${NGB+NGB\to NGB+NGB}$}

The NGB scattering amplitudes are calculated in 
Appendix~\ref{sec:amp_NGB} in the case of $g=g_Y=0$ (gaugeless limit).
Mandelstam variables $s$, $t$, and $u$ are also defined 
in the Appendix~\ref{sec:amp_NGB}. 
Requiring the cancellation of the ${\cal O}(u)$ divergence
in the high energy $w^+w^-\to w^+ w^-$ scattering amplitude 
Eq.(\ref{eq:amp_ww2ww}), we obtain
\begin{equation}
\label{eq:wwww}
-4+3\dfrac{v_Z^2}{v^2}
   +\sum_{n=1}^{N_0} \kappa_{WW}^{\phi^0_n} \kappa_{WW}^{\phi^0_n} =0 ,
\end{equation}
which agrees with Eq.(4.1) of Ref.\cite{Gunion:1990kf} in the
absence of doubly-charged Higgs bosons. 
Although we here impose the cancellation of scattering amplitude up to 
the ultimately high energy scale, the energy (cutoff) dependent modifications  
of ${\cal O}(M_V^2/s)$ to the sum rules may be allowed. On the other hand, 
as we will see later, exact sum rules are required to maintain the finiteness 
of the oblique corrections.
We see, from Eq.(\ref{eq:wwww}), an inequality
\begin{equation}
  v_Z^2 \le \dfrac{4}{3} v^2, 
\label{eq:misc1}
\end{equation}
which is satisfied in the SM $v_Z^2 = v^2$.
However, Eq.(\ref{eq:misc1}) is not satisfied in the triplet Higgs model 
($I=1$, $Y=1$), in which $v_Z^2 = 2v^2$ is predicted\footnote{
The (pure) triplet Higgs model does not accommodate mass generation 
mechanisms for SM fermions and cannot be accepted as a phenomenologically 
viable EWSB model. }. 
Actually, the triplet Higgs model contains (doubly) charged Higgs bosons
coupled with electroweak gauge bosons in its spectrum, and thus
cannot be covered by the analysis presented in this manuscript.

In a similar manner, using the $w^+ w^- \to zz$ amplitude 
Eq.(\ref{eq:amp_ww2zz}), we find a sum rule,
\begin{equation}\label{eq:wwzz}
  \dfrac{v_Z^2}{v^2}
  -\dfrac{v^2}{v_Z^2}\sum_{n=1}^{N_0}
   \kappa_{ZZ}^{\phi^0_n}\kappa_{WW}^{\phi^0_n}=0 .
\end{equation} 
Again, it is straightforward to see the equivalence of Eq.(\ref{eq:wwzz})
with Eq.(4.2) of Ref.\cite{Gunion:1990kf} in the absence of charged Higgs 
bosons.

We note that the $zz\to zz$ amplitude Eq.(\ref{eq:amp_zz2zz}) 
does not produce extra conditions because of $s+t+u=0$.
Note NGBs are massless in the gaugeless limit. 

\subsection{${NGB+NGB\to \phi+\phi}$}

We next consider the $w^+ w^- \to \phi^0_{n_1^{}} \phi^0_{n_2^{}}$ amplitude  
Eq.(\ref{eq:amp_ww2phiphi}).
The amplitude can be decomposed into two pieces, depending on the
relative angular momentum between two scalar bosons in the final state.
Requiring the cancellation of the ${\cal O}(s)$ enhanced term in the
$S$-wave amplitude, we obtain a relation between
$WW\phi\phi$ and $WW\phi$ interaction terms,
\begin{equation}
\kappa^{\phi^0_{n_1^{}}\phi^{0}_{n_2^{}}}_{WW} 
     -\kappa^{\phi^0_{n_1^{}}}_{WW} \kappa^{\phi^0_{n_2^{}}}_{WW}=0 .
\label{eq:wwphiphi2}
\end{equation}
On the other hand, requiring the cancellation of the 
${\cal O}(t-u)$ term in the the $P$-wave amplitude, 
we obtain
\begin{equation}
\kappa_Z^{\phi_{n_1^{}}^0\phi_{n_2^{}}^0} = 0.   
\label{eq:wwphiphi1}
\end{equation}
Presence of $\kappa_Z^{\phi_{n_1^{}}^0\phi_{n_2^{}}^0}$ 
without introducing extra particles other than the neutral Higgs bosons
would therefore cause a violation of unitarity in the $WW \to \phi\phi$ 
scattering amplitude.

These relations Eq.(\ref{eq:wwphiphi2}) and Eq.(\ref{eq:wwphiphi1}) 
correspond to a single equation Eq.(A3) of Ref.\cite{Gunion:1990kf}, 
which reads
\begin{equation}
\kappa^{\phi^0_{n_1^{}}\phi^{0}_{n_2^{}}}_{WW} 
     -\kappa^{\phi^0_{n_1^{}}}_{WW} \kappa^{\phi^0_{n_2^{}}}_{WW}
     + i \kappa_Z^{\phi^0_{n_1^{}}\phi^0_{n_2^{}}} =0 , 
\label{eq:wwphiphi}
\end{equation}
in the notation of the present manuscript.
Using Eq.(\ref{eq:kappa-antisymmetry}) and Eq.(\ref{eq:kappa-symmetry}), 
however,  Eq.(\ref{eq:wwphiphi}) can be decomposed into 
$n_1 \leftrightarrow n_2$ symmetric and anti-symmetric parts, which can 
be shown to be identical to our Eq.(\ref{eq:wwphiphi2}) and 
Eq.(\ref{eq:wwphiphi1}), respectively. 

We next move to the $zz \to \phi^0_{n_1^{}} \phi^0_{n_2^{}}$ amplitude
Eq.(\ref{eq:amp_zz2phiphi}).
We find a sum rule
\begin{equation}\label{eq:zzphiphi}
\kappa_{ZZ}^{\phi^{0}_{n_1^{}}\phi^{0}_{n_2^{}}}
-\sum_{m=1}^{N_0}{\kappa}_{Z}^{\phi^{0}_{n_1^{}}\phi^{0}_{m}}
                 {\kappa}_{Z}^{\phi^{0}_{n_2^{}}\phi^{0}_{m}}
-\dfrac{v^2}{v_Z^2}\kappa_{ZZ}^{\phi^0_{n_1^{}}}\kappa_{ZZ}^{\phi^0_{n_2^{}}}=0, 
\end{equation}
which is required to cancel the ${\cal O}(s)$ divergence of the
amplitude.
Eq.(\ref{eq:zzphiphi}) is identical to Eq.(A18) of 
Ref.\cite{Gunion:1990kf}.

\subsection{${NGB+NGB\to \phi+NGB}$}

The $w^+w^- \to \phi^0_n z$ amplitude also possesses $S$-wave and
$P$-wave contributions in Eq.(\ref{eq:amp_ww2phiz}).
The cancellation of the high energy $P$-wave amplitude requires
\begin{equation}
\kappa^{\phi^0_n}_{WW}-\dfrac{v^2}{v_Z^2}\kappa^{\phi^0_n}_{ZZ}=0,
\label{eq:wwphiz}
\end{equation}
while the $S$-wave amplitude requires
\begin{equation}
\sum_{m=1}^{N_0} \kappa^{\phi^0_n\phi^0_m}_{Z}\kappa^{\phi^0_m}_{WW}=0 .
\label{eq:wwphiz-s}
\end{equation}

Again, we note that the $zz\to \phi_n^0 z$ amplitude 
Eq.(\ref{eq:amp_zz2phiz}) does not produce extra conditions.
It is also easy to check the equivalence of Eqs.(\ref{eq:wwphiz}) 
and Eq.(\ref{eq:wwphiz-s}) with Eq.(4.5) of Ref.\cite{Gunion:1990kf}.

%======================================
\subsection{Applications}
\label{sec-application}

As emphasized in Ref.\cite{Gunion:1990kf}, the unitarity sum rules
can be applied to constrain various extended Higgs models.
For an example, as Ref.\cite{Gunion:1990kf} argued, assuming $v=v_Z^{}$, 
that the future observation of the Higgs-$W$-$W$ coupling larger than 
the SM value would suggest the existence of charged Higgs 
particles.
This fact can be seen from Eq.(\ref{eq:wwww}), which leads to an 
upper bound of Higgs-$W$-$W$ coupling $\kappa_{WW}^{\phi_n^0}\le 1$ 
for $v=v_Z^{}$ in any model only having extra neutral Higgs 
particles.

In this subsection, we list a couple of observations in 
the unitarity sum rules which have not been 
stressed in earlier 
literature.

Let us start with an implication of the unitarity sum rules
to the $\rho$ parameter $\rho_0 = v^2/v_Z^2$.
Combining Eq.(\ref{eq:wwww}) and Eq.(\ref{eq:wwzz}), we find
\begin{equation}
\sum_{n=1}^{N_0}
\kappa^{\phi^0_n}_{WW}(\kappa^{\phi^0_n}_{WW}-\rho_0\kappa^{\phi^0_n}_{ZZ})
=\dfrac{4}{\rho_0}(\rho_0-1).
\label{eq:rho0constraint1}
\end{equation}
On the other hand, the unitarity sum rules for $w^+w^-\to \phi z$ 
Eq.(\ref{eq:wwphiz}) reads
\begin{equation}
  \kappa^{\phi^0_n}_{WW}=\rho_0\kappa^{\phi^0_n}_{ZZ}.
\label{eq:rho0constraint2}
\end{equation}
Plugging Eq.(\ref{eq:rho0constraint2}) into 
Eq.(\ref{eq:rho0constraint1}), we obtain a 
condition on the $\rho_0$ parameter,
\begin{equation}\label{rho0}
\dfrac{1}{\rho_0}(\rho_0-1)=0, 
\end{equation}
solely from the unitarity requirements.
The $\rho_0$ parameter needs to be $1$  in order to
unitarize the $w^+w^-\to w^+w^-$, $w^+w^-\to zz$ and $w^+w^-\to \phi z$ 
scattering amplitudes in any EWSB model 
with $v\ne 0$, $v_Z^{}\ne 0$ that only has neutral Higgs 
particles.
Note that this argument cannot be applied to the triplet Higgs mixing 
model (a doublet and a triplet Higgs fields)\cite{Konetschny:1977bn,Magg:1980ut,Cheng:1980qt,Lazarides:1980nt,Mohapatra:1979ia}, 
since we restrict ourselves within the neutral Higgs extension cases only.
However, the unitarity argument will be useful when we understand 
%\sout{$\rho=1$}  
$\rho_0=1$ in the septet Higgs case\cite{Hisano:2013sn,Kanemura:2013mc,Alvarado:2014jva}, in which we do not have 
manifest custodial symmetry.
We will discuss the issue in our subsequent paper, in which 
we extend our analysis including the charged Higgs bosons.

It is also intriguing that the unitarity sum rule for the 
$w^+ w^- \to zz$ amplitude Eq.(\ref{eq:wwzz}) is sensitive to the 
sign of $\kappa^{\phi^0_n}_{ZZ}\kappa_{WW}^{\phi^0_n}$.
Note that the current experimental results on the 125GeV Higgs 
boson ($h$) are sensitive only to the absolute values of $hZZ$ and
$hWW$ couplings ($|\kappa^h_{ZZ}|$ and $|\kappa^h_{WW}|$), not
to their relative sign.\footnote{%\sout{In the $h\to \gamma\gamma$ channel, 
%the relative sign of $\kappa_W^{}$ to $\kappa_t$ can be determined 
%by fitting the data in the SM.}
This fact is in contrast to the case of 
the relative sign between $\kappa_W$ 
and $\kappa_t$ (top-Higgs coupling), which can be determined using
the $h\to \gamma\gamma$ channel in the SM\@.
}
As shown in Eq.(\ref{eq:wwzz}), a wrong sign 
$\kappa^h_{ZZ} \kappa^h_{WW}$ would cause a violation of unitarity
in the $WW\to ZZ$ amplitude.
Future measurements on the $WW\to ZZ$ (or $ZZ\to WW$
or $WZ \to WZ$) cross section
can thus be used to check whether the 
$\kappa^{\phi^0_n}_{ZZ}\kappa_{WW}^{\phi^0_n}$ sign is like the SM 
or not.

The condition Eq.(\ref{eq:wwphiphi1}) gives us an insight
on the hypothetical $CP$-odd neutral Higgs boson properties in 
a model independent manner.
Existence of such a $CP$-odd Higgs boson $a$, 
having non-vanishing $haZ$ 
coupling without introducing extra charged Higgs boson, 
would contradict with the unitarity relation Eq.(\ref{eq:wwphiphi1})
and would therefore cause an enhancement of the $WW \to ha$ cross 
section.

We finally make an important comment on the implications of the 
unitarity sum rules to the electroweak radiative corrections.
As we will see in the sections below, a violation of the unitarity 
sum rules often causes a UV divergence in the electroweak 
radiative corrections. It is therefore severely constrained by the
existing precision measurements on the electroweak interactions.
The issue is studied extensively in this manuscript 
in sections~\ref{sec:finiteness}
and \ref{sec:oblique}.

\section{Finiteness of $f\bar{f}\to f'\bar{f}'$ amplitudes
including oblique corrections at one loop}
\label{sec:finiteness}

Thanks to the gauge invariance of the 
non-linear sigma model Lagrangian we use, 
in the present framework, 
effects of radiative corrections can be studied
without causing unphysical negative metric particle
problems even in the $R_\xi$ gauge fixing method.
Lack of the renormalizability of the non-linear sigma model,
however, causes UV divergences in the amplitudes, which cannot
be renormalized by the redefinitions of the Lagrangian 
parameters.
As we show in this section, one-loop UV divergences
in the massless fermion scattering amplitudes disappear
after appropriate redefinitions of gauge coupling strengths
and the VEVs, only when a set of 
sum rules is satisfied among the Higgs coupling strengths.
In this section, we write down such a set of sum rules
explicitly. 
We find these sum rules are automatically satisfied once
the Higgs coupling strengths satisfy the unitarity sum rules we found 
in the previous section.

Before going details in the loop analysis, we briefly summarize 
the relationships between the vacuum polarization functions $\Pi_{33}$,
$\Pi_{3Q}$, $\Pi_{QQ}$ and $\Pi_{11}$
and 
the $f\bar{f} \to f' \bar{f'}$ scattering amplitudes.
We assume here the vacuum polarization functions evaluated in the
background gauge fixing method, with which the cancellation of the 
divergences between the one-loop vertex corrections and the fermion 
wave function renormalizations is guaranteed, thanks to 
the na\"{\i}ve Ward-Takahashi identities.

We first discuss the relationship between the
vacuum polarization functions 
$\Pi_{33}$, $\Pi_{3Q}$, $\Pi_{QQ}$ and $\Pi_{11}$,
\begin{eqnarray}
  \Pi_{33}(p^2) &=& \Pi_{33}(0) + p^2 \Pi'_{33}(p^2), \\
  \Pi_{11}(p^2) &=& \Pi_{11}(0) + p^2 \Pi'_{11}(p^2), \\
  \Pi_{3Q}(p^2) &=& p^2 \Pi'_{3Q}(p^2), \\
  \Pi_{QQ}(p^2) &=& p^2 \Pi'_{QQ}(p^2),
\end{eqnarray}
and the $f\bar{f} \to f' \bar{f'}$ scattering amplitudes.
Here $\Pi_{33}(p^2)$, $\Pi_{11}$, and 
$\Pi_{QQ}$ are neutral and charged
weak $SU(2)$ current correlators, 
and  the electromagnetic current correlator, respectively.
$\Pi_{3Q}$ is the correlator between the 
neutral weak $SU(2)$ current and the electromagnetic current.
These current correlators can be related with the 
vacuum polarization  functions of the electroweak gauge bosons,
\begin{eqnarray}
  \Pi_{11} &=& \dfrac{1}{g^2} \Pi_{WW}, 
\label{eq:def_11}
\\
  \Pi_{33} &=& \dfrac{1}{g_Z^2} \left[
    \Pi_{ZZ} + \dfrac{g_Y^2}{g^2} \Pi_{AA} + 2 \dfrac{g_Y^{}}{g} \Pi_{ZA}
  \right], 
\label{eq:def_33}
  \\
  \Pi_{3Q} &=&
    \dfrac{1}{g^2} \Pi_{AA} + \dfrac{1}{g g_Y^{}} \Pi_{ZA}, 
\label{eq:def_3Q}
  \\
  \Pi_{QQ} &=&
  \dfrac{g_Z^2}{g^2 g_Y^2} \Pi_{AA}. %,
\label{eq:def_QQ}
\end{eqnarray}
The na\"{\i}ve Ward-Takahashi identities arising from the conservation 
of the electromagnetic current gives
\begin{equation}
  \Pi_{3Q}(p^2=0) = 
  \Pi_{QQ}(p^2=0) = 0.
\end{equation}
By using these vacuum polarization functions, 
the neutral and charged current $f\bar{f} \to f' \bar{f}'$ scattering 
amplitudes ($f\ne f'$) including these oblique corrections can be expressed as
\begin{eqnarray}
\lefteqn{
  -{\cal M}_{\rm NC}  =
} \nonumber\\
  & & 
  e_*^2 \dfrac{{\cal Q}{\cal Q}'}{-p^2}
  % \nonumber\\
  % & &
   +\dfrac{(I_3-s_*^2 {\cal Q})(I'_3-s_*^2 {\cal Q}')}
          {-\left(\dfrac{s_*^2 c_*^2}{e_*^2}-\Pi'_{33}+\Pi'_{3Q}\right)p^2
           +\dfrac{v_{Zr}^2}{4}},
  \nonumber\\
  & &
  \\
\lefteqn{
  -{\cal M}_{\rm CC} =
} \nonumber\\
  & &
  \dfrac{(I_+ I'_-+I_- I'_+)/2}
          {-\left(\dfrac{s_*^2}{e_*^2}-\Pi'_{11}+\Pi'_{3Q}\right)p^2
           +\dfrac{v_r^2}{4}},
  % \nonumber\\
  % & &
\end{eqnarray}
with renormalized parameters  
$v_{Zr}^2$, $v_r^2$, $e_*^2$, $s_*^2$ and $c_*^2$ are defined by
\begin{eqnarray}
  v_{Zr}^2 &=& v_Z^2 + \delta v_Z^2 + 4\,\Pi_{33}(0), 
  \\
  v_{r}^2 &=& v^2 + \delta v^2 + 4\,\Pi_{11}(0), 
  \\
  \dfrac{1}{e_*^2} &=& \dfrac{1}{g^2} + \dfrac{1}{g_Y^2} 
  +\delta\left(\dfrac{1}{g^2}\right)
  +\delta\left(\dfrac{1}{g_Y^2}\right)- \Pi'_{QQ}, \nonumber \\ \\
  \dfrac{s_*^2}{e_*^2} &=& \dfrac{1}{g^2} 
  +\delta\left(\dfrac{1}{g^2}\right)- \Pi'_{3Q}, \\
  c_*^2 &=& 1- s_*^2,
\end{eqnarray}
with $\delta v_Z^2$, $\delta v^2$, 
$\delta(1/g^2)$, $\delta(1/g_Y^2)$ being counter terms 
to renormalize the divergences in $\Pi_{33}(0)$, $\Pi_{11}(0)$, 
$\Pi'_{QQ}$ and $\Pi'_{3Q}$.
Here the amplitudes are described by using a simplified version
of notations of Ref.\cite{Kennedy:1988sn}.
The definitions of $I_3$, $I_\pm$, and ${\cal Q}$ are given in
Ref.\cite{Chivukula:2004pk}.
Finiteness of the scattering amplitudes thus requires
\begin{equation}
  \dfrac{\delta v_Z^2}{4} + \Pi_{33}(0), 
\label{eq:vzrenormalize}
\end{equation}
\begin{equation}
  \dfrac{\delta v^2}{4} + \Pi_{11}(0), 
\label{eq:vwrenormalize}
\end{equation}
\begin{equation}
  \Pi'_{33}- \Pi'_{3Q},
\label{eq:S_renormalize}
\end{equation}
\begin{equation}
  \Pi'_{11}- \Pi'_{3Q},
\end{equation}
are all finite.
We study these conditions in the subsections below.

\subsection{$\Pi_{33}(0)$ and $\Pi_{11}(0)$}

We investigate the conditions of finiteness of 
Eq.(\ref{eq:vzrenormalize}) and Eq.(\ref{eq:vwrenormalize}).
The UV divergences in $\Pi_{11}(0)$ and 
$\Pi_{33}(0)$ can be absorbed into the renormalizations
of $v_Z^{}$ and $v$ if these two parameters are independently
adjustable parameters.
Triplet Higgs mixing models\cite{Konetschny:1977bn,Magg:1980ut,Cheng:1980qt,Lazarides:1980nt,Mohapatra:1979ia}
 including Georgi-Machacek scenario\cite{Georgi:1985nv,Chanowitz:1985ug,Gunion:1989ci,Gunion:1990dt}
fall into this category.
In multi-Higgs doublet models\cite{Branco:2011iw} including 
the SM, and 
the doublet-septet mixing model\cite{Hisano:2013sn,Kanemura:2013mc,Alvarado:2014jva},
on the other hand, $v_Z^{}$ and $v$ are linearly related parameters,
\begin{equation}
  v_Z^2 = \dfrac{1}{\rho_0^{}} v^2,
\label{eq:rho0def}
\end{equation}
with $\rho_0^{}$ being a positive constant.
Although the parameter $\rho_0^{}$ is phenomenologically required to 
be\footnote{
Strictly speaking, what we need to require is $\rho\simeq 1$ 
after taking account of the quantum corrections allowing experimental
uncertainty of $10^{-3}$ level.
}
\begin{equation}
  \rho_0^{} = 1,
\end{equation}
in this manuscript, we keep this parameter arbitrary for a while in order 
to clarify the theoretical structure of Eq.(\ref{eq:vzrenormalize})
and Eq.(\ref{eq:vwrenormalize}).

In models satisfying the requirement Eq.(\ref{eq:rho0def}),
the counter terms we can introduce should satisfy
\begin{equation}
  \delta v_Z^2 = \dfrac{1}{\rho_0} \delta v^2.
\end{equation}
In this class of models, we therefore find
\begin{equation}
  v_Z^2 \Pi_{11}(0) - v^2 \Pi_{33}(0)
\label{eq:rho-correction}
\end{equation}
needs to be finite in order to keep the $f\bar{f}\to f' \bar{f}'$ 
amplitude finite at the loop level.
In this subsection, we focus on the conditions guarantee
the 
finiteness of Eq.(\ref{eq:rho-correction}) at the one-loop level.

We evaluate the vacuum polarization functions 
$\Pi_{11}(p^2)$ and $\Pi_{33}(p^2)$ at one-loop level.
It is convenient to decompose these functions into two pieces,
\begin{eqnarray}
  \Pi_{11}(p^2) &=& \tilde{\Pi}_{11}(p^2) 
   + \Pi_{11}^{\rm Higgs}(p^2; \kappa),
  \\
  \Pi_{33}(p^2) &=& \tilde{\Pi}_{33}(p^2) 
  + \Pi_{33}^{\rm Higgs}(p^2; \kappa),
\end{eqnarray}
where 
$\tilde{\Pi}_{11}(p^2)$ and $\tilde{\Pi}_{33}(p^2)$ 
are contributions arising from loops containing solely 
the gauge bosons and NGBs, and are independent of the
Higgs coupling strengths $\kappa$.
These contributions are evaluated by using the background gauge
fixing method with 't Hooft-Feynman gauge $\xi=1$.
See Appendix~\ref{sec:eval3311} for details. 
Using the dimensional regularization, we obtain
\begin{eqnarray}
\lefteqn{
 \tilde{\Pi}_{11}(p^2=0) = 
}
  \nonumber\\
  & & (D-2)\left[ A(M_W) + \dfrac{g^2}{g_Z^2} A(M_Z) 
                         + \dfrac{g_Y^2}{g_Z^2} A(0)
      \right.
  \nonumber\\
  & & \quad \left. 
        +\dfrac{g^2}{g_Z^2} B_0(M_W, M_Z; 0)
        +\dfrac{g_Y^2}{g_Z^2} B_0(M_W, 0; 0) \right]
  \nonumber\\
  & & +\dfrac{v_Z^2}{4v^2} \left[
         A(M_W) + A(M_Z) + B_0(M_W, M_Z; 0)
      \right]
  \nonumber\\
  & & -\dfrac{1}{4}\left(4-3\dfrac{v_Z^2}{v^2}\right) A(M_W)
      -\dfrac{1}{4}\dfrac{v_Z^2}{v^2} A(M_Z)
  \nonumber\\
  & & -\dfrac{1}{g_Z^2}\left[
        g^2 \dfrac{v}{2}\left(2-\dfrac{v_Z^2}{v^2}\right)
        - g_Y^2 \dfrac{v_Z^2}{2v}
       \right]^2 B(M_W, M_Z; 0)
  \nonumber\\
  & &  -\dfrac{g^2 g_Y^2}{g_Z^2} \left[
        \dfrac{v}{2} \left(2-\dfrac{v_Z^2}{v^2} \right)
       +\dfrac{v_Z^2}{2v} 
       \right]^2 B(M_W, 0; 0)
  \nonumber\\
  & &  -\dfrac{g^2 v_Z^2}{4} B(M_W, M_Z; 0),
\label{eq:tildePi_11}
  \\
\lefteqn{
 \tilde{\Pi}_{33}(p^2=0) =
} 
 \nonumber\\
  & & -\dfrac{1}{2} \dfrac{v_Z^4}{v^4} A(M_W)
      -\dfrac{1}{2} g^2 \dfrac{v_Z^4}{v^2} B(M_W, M_W; 0),
\label{eq:tildePi_33}
\end{eqnarray}
with $D$ being the number of space-time dimensions.
Here UV divergent loop functions $A$, $B$ and $B_0$ are 
defined by Eq.(\ref{eq:def_Abare}), Eq.(\ref{eq:def_Bbare}) 
and Eq.(\ref{eq:def_B0bare}).

\begin{figure}[t]
 \begin{center}
   \begin{minipage}{0.5\textwidth}
     \begin{center}
       \includegraphics[width=55mm, bb=0 0 559 264]{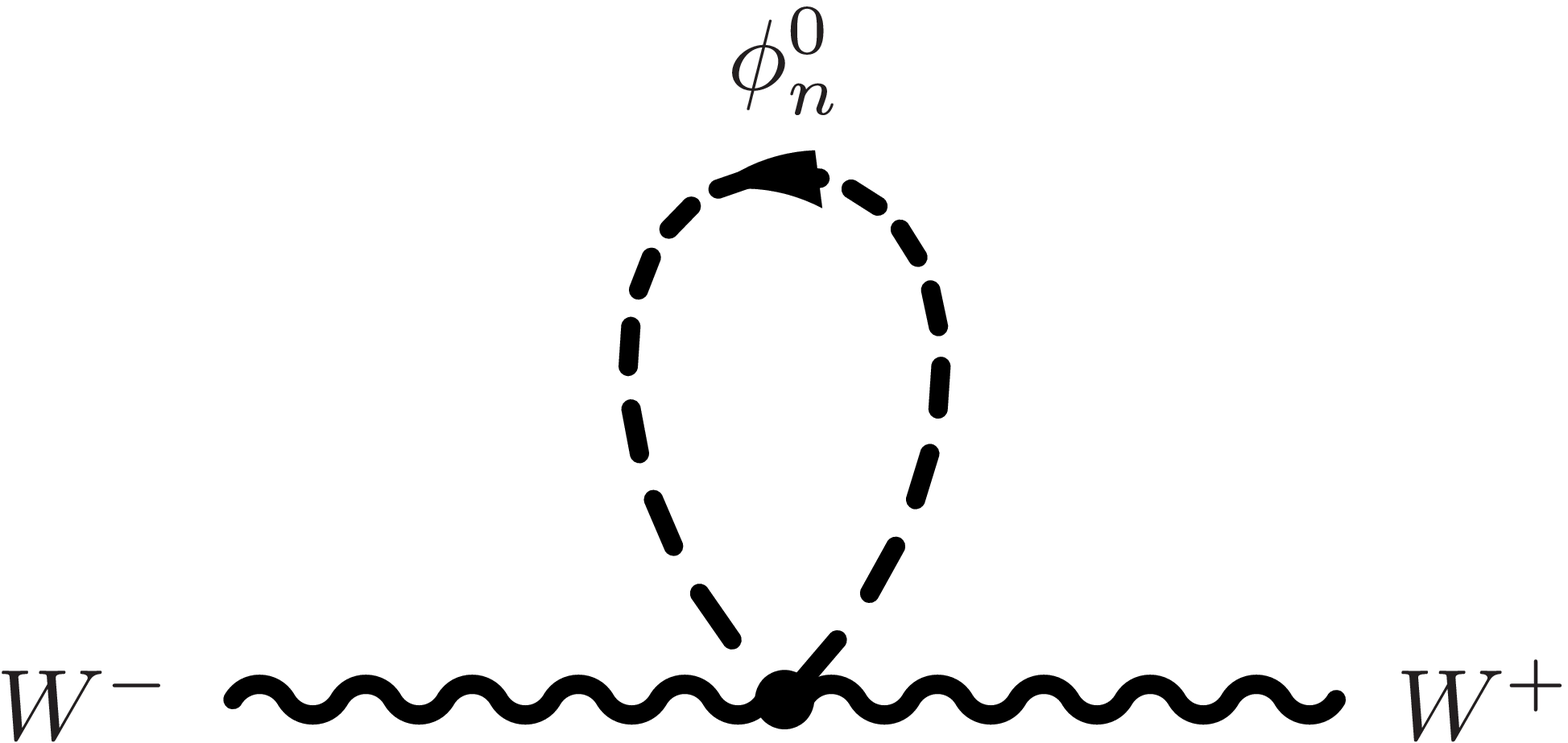}\\[2ex]
       (a)
     \end{center}
   \end{minipage}
   \vspace*{5ex}

   \begin{minipage}{0.5\textwidth}
     \begin{center}
       \includegraphics[width=50mm, bb=0 0 451 271]{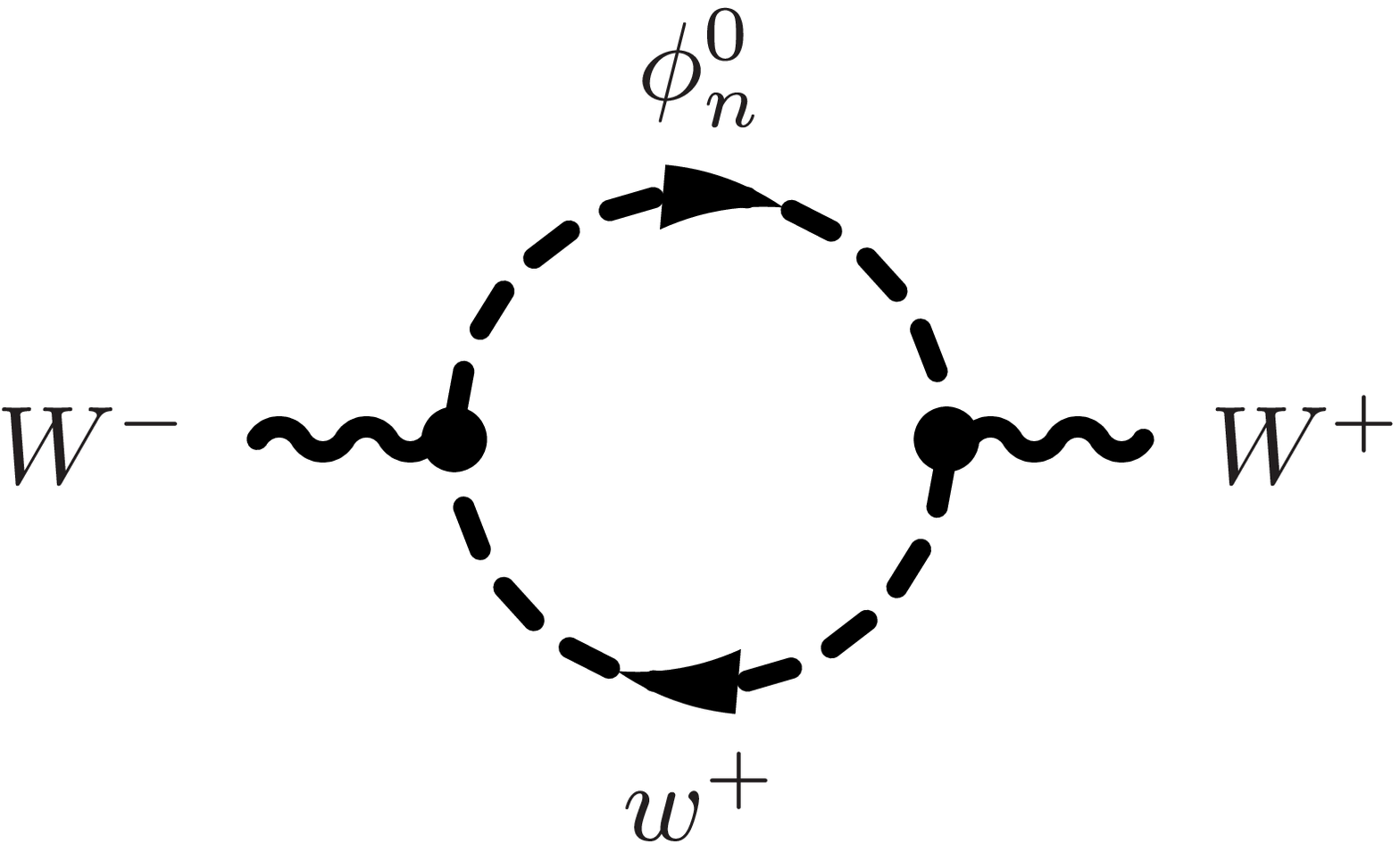}\\[2ex]
       (b)
     \end{center}
   \end{minipage}
   \vspace*{5ex}

   \begin{minipage}{0.5\textwidth}
     \begin{center}
       \includegraphics[width=50mm, bb=0 0 451 272]{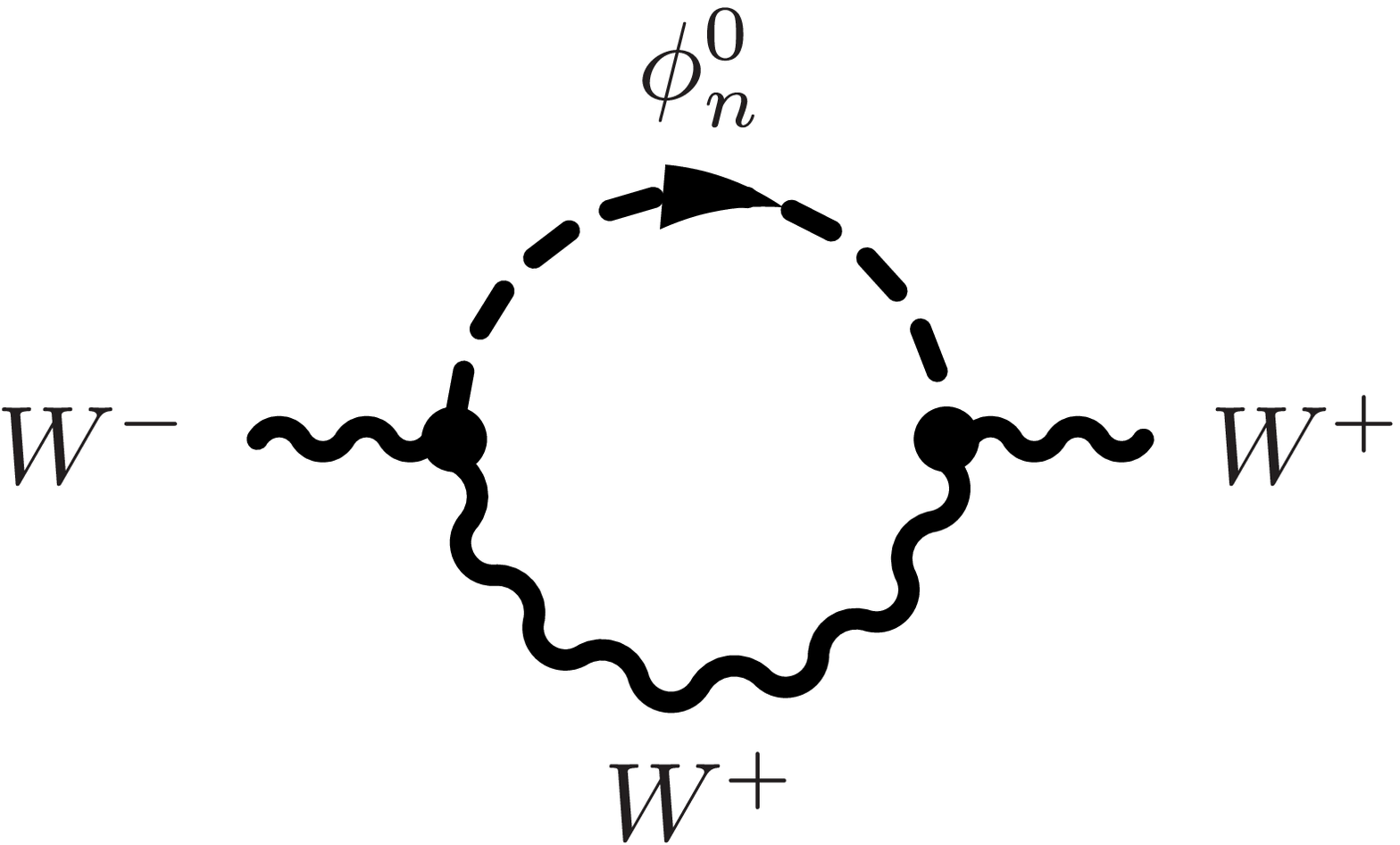}\\[2ex]
       (c)
     \end{center}
   \end{minipage}
 \end{center}
 \caption{One-loop diagrams for the $W$ boson self-energies 
$\Pi^{\mbox{Higgs}}_{WW}$ in our model.}
\label{fig-vac11}
\end{figure}

We next evaluate the Higgs loop contributions to $\Pi_{11}^{\rm Higgs}$.
The corresponding Feynman diagrams are given in Fig.\ref{fig-vac11}.
In the 't Hooft-Feynman gauge, we find
\begin{eqnarray}
\lefteqn{
  \Pi_{11}^{\rm Higgs}(p^2=0; \kappa)
  = 
  \dfrac{1}{g^2} \Pi_{WW}^{\rm Higgs}(0) =
}  \nonumber\\
& & \dfrac{1}{4} \sum_{n=1}^{N_0} \kappa_{WW}^{\phi_n^0 \phi_n^0} A(M_{\phi_n^0})
  \nonumber\\
& &
      +\dfrac{1}{4} \sum_{n=1}^{N_0} \kappa_{WW}^{\phi_n^0} \kappa_{WW}^{\phi_n^0} \left\{
       B_0(M_{\phi_n^0}, M_W; 0) 
       \right.
  \nonumber\\
& & \qquad\qquad \left.
- 4M_W^2 B(M_{\phi_n^0}, M_W; 0)
      \right\},
  \nonumber\\
  & & 
\label{eq:Pi_11_0}
\end{eqnarray}
where the first, the second and the third terms are from
Fig.\ref{fig-vac11}(a), Fig.\ref{fig-vac11}(b), and Fig.\ref{fig-vac11}(c), 
respectively.

\begin{figure}[t]
 \begin{center}
   \begin{minipage}{0.5\textwidth}
     \begin{center}
       \includegraphics[width=50mm, bb=0 0 490 262]{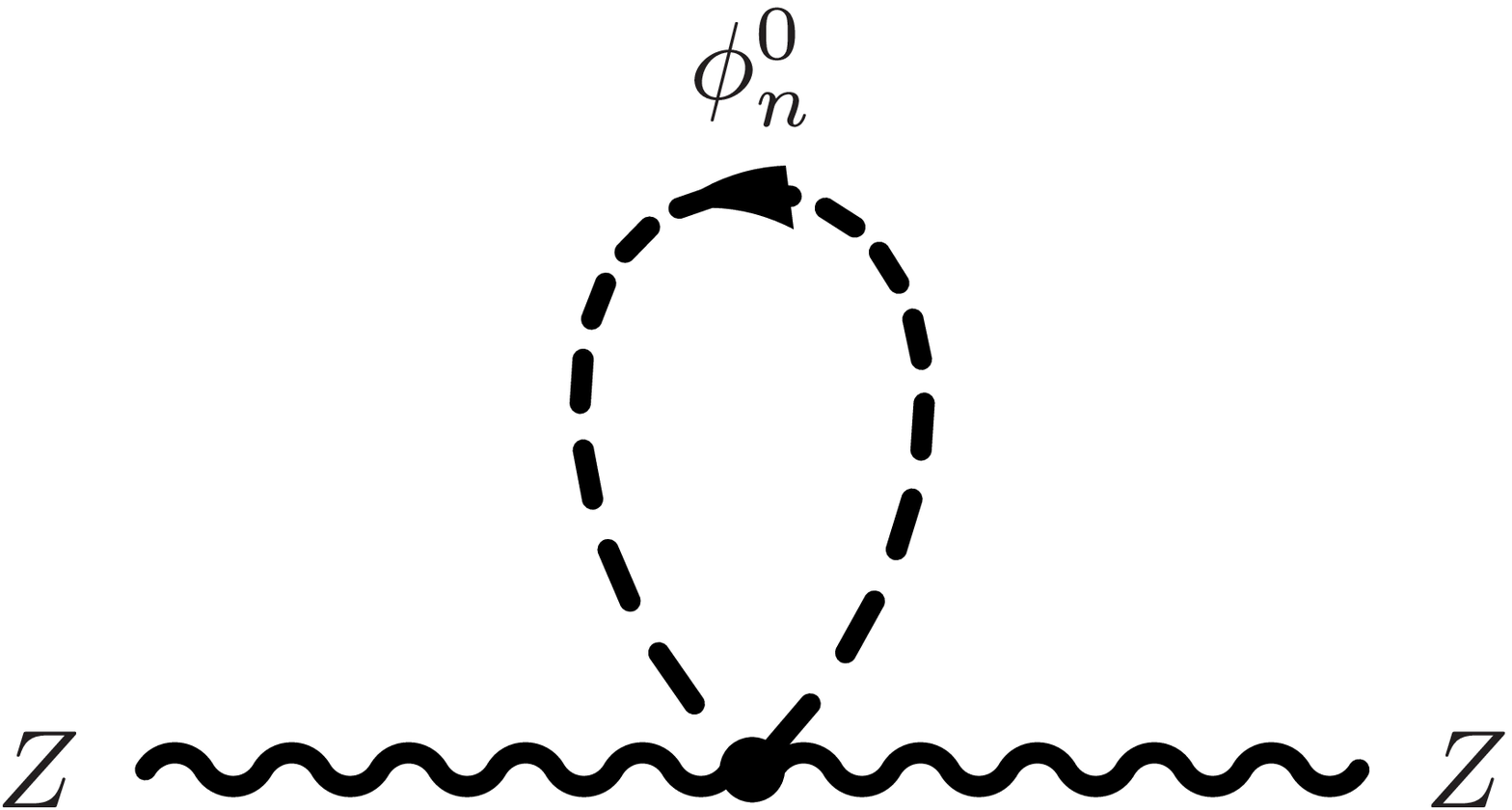}\\[2ex]
       (a)
     \end{center}
   \end{minipage}
   \vspace*{5ex}

   \begin{minipage}{0.5\textwidth}
     \begin{center}
       \includegraphics[width=45mm, bb=0 0 382 257]{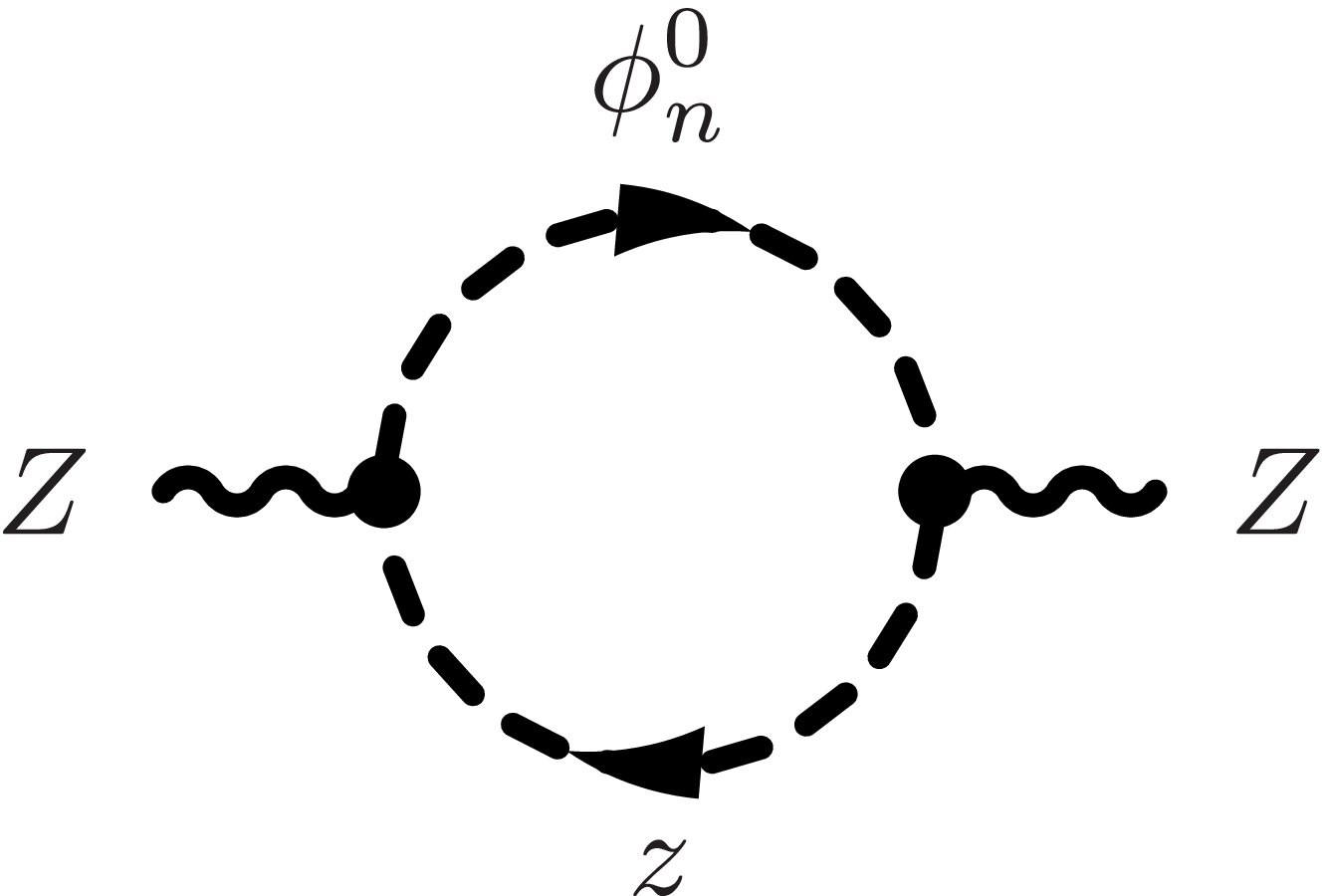}\\[2ex]
       (b)
     \end{center}
   \end{minipage}
   \vspace*{5ex}

   \begin{minipage}{0.5\textwidth}
     \begin{center}
       \includegraphics[width=45mm, bb=0 0 382 266]{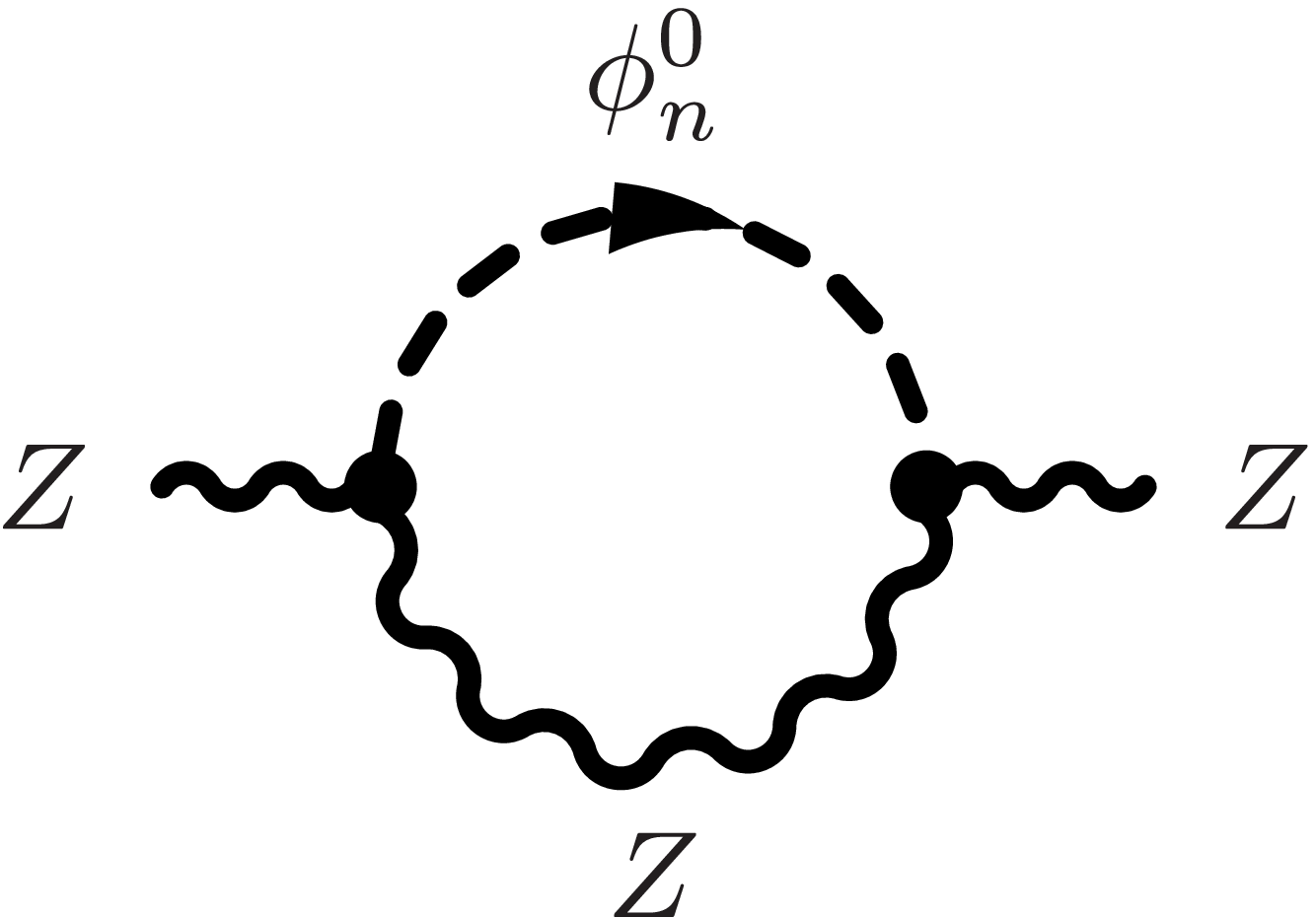}\\[2ex]
       (c)
     \end{center}
   \end{minipage}
   \vspace*{5ex}

   \begin{minipage}{0.5\textwidth}
     \begin{center}
       \includegraphics[width=45mm, bb=0 0 382 280]{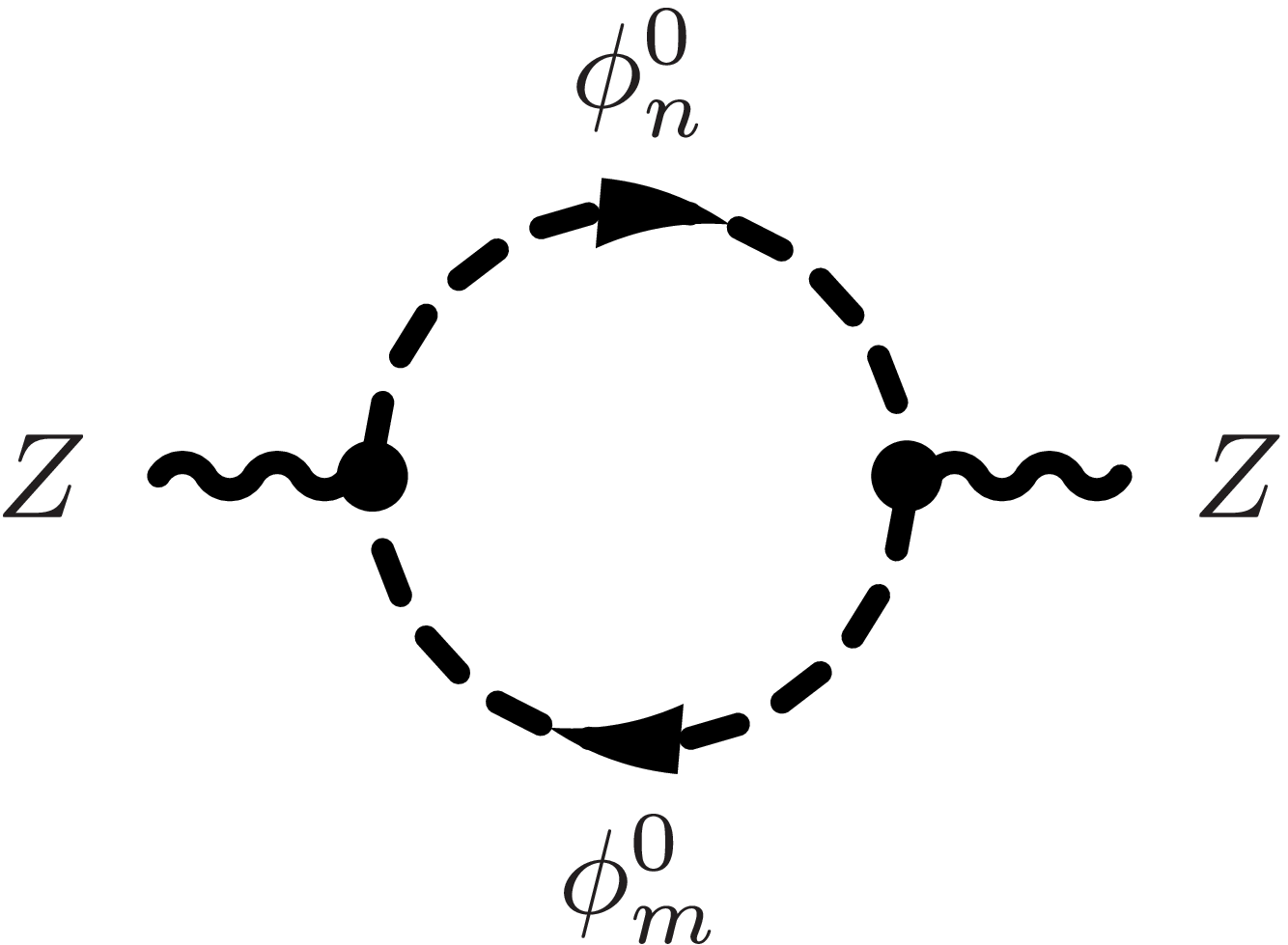}\\[2ex]
       (d)
     \end{center}
   \end{minipage}

 \end{center}
 \caption{One-loop diagrams for the $Z$ boson self-energies $\Pi^{\mbox{Higgs}}_
{ZZ}$ in our model. }
 \label{fig-vac33}
\end{figure}

In a similar manner, evaluating the Feynman diagrams Fig.\ref{fig-vac33}, 
we obtain
\begin{eqnarray}
\lefteqn{
  \Pi_{33}^{\rm Higgs}(p^2=0; \kappa)
  =
  \dfrac{1}{g_Z^2} \Pi_{ZZ}^{\rm Higgs}(0) =
} \nonumber\\
& & \dfrac{1}{8} \sum_{n=1}^{N_0} \sum_{m=1}^{N_0}
%\sum_{n,m}
\kappa_Z^{\phi_n^0 \phi_m^0}\kappa_Z^{\phi_n^0 \phi_m^0}
      B_0(M_{\phi_n^0}, M_{\phi_m^0}; 0)
  \nonumber\\
& &
      +\dfrac{1}{4} \sum_{n=1}^{N_0} \kappa_{ZZ}^{\phi_n^0 \phi_n^0} A(M_{\phi_n^0})
  \nonumber\\
  & &
     \quad +\dfrac{v^2}{4v_Z^2} \sum_{n=1}^{N_0} \kappa_{ZZ}^{\phi_n^0} \kappa_{ZZ}^{\phi_n^0} 
     \left\{
       B_0(M_{\phi_n^0}, M_Z; 0) \right.
  \nonumber\\
  & & \qquad \qquad \left. - 4M_Z^2 B(M_{\phi_n^0}, M_Z; 0)
      \right\}.
\label{eq:Pi_33_0}
\end{eqnarray}

There may also exist tadpole graphs if $\phi_n^0$ fields acquire
their VEVs at one-loop.
We assume these one-loop VEVs of $\phi_n^0$ are eliminated by introducing
appropriate linear potential counter terms in the Higgs potential
Eq.(\ref{eq:higgs_potential}).

Our results of the vacuum polarization functions, 
Eq.(\ref{eq:tildePi_11}), Eq.(\ref{eq:tildePi_33}), 
Eq.(\ref{eq:Pi_11_0}), and Eq.(\ref{eq:Pi_33_0}), 
can be compared with the SM,
taking $N_0=1$, $v=v_Z^{}$, $\kappa_{WW}^{\phi_1^0}=
\kappa_{ZZ}^{\phi_1^0}= \kappa_{WW}^{\phi_1^0\phi_1^0}=
\kappa_{ZZ}^{\phi_1^0\phi_1^0}=1$.\footnote{
$\kappa_{Z}^{\phi_1^0\phi_1^0}=0$ is automatic
because of the antisymmetry $n_1\leftrightarrow n_2$ in
$\kappa_{Z}^{\phi_{n_1^{}}^0\phi_{n_2^{}}^0}=0$.}
Comparing them with the SM results of 
Hagiwara-Matsumoto-Haidt-Kim (HMHK)\cite{Hagiwara:1994pw}
which employs
the pinch technique in their evaluation of 
the vacuum polarization functions, we find
\begin{eqnarray}
  \Pi_{11}^{\rm NTT}(0) - \Pi_{11}^{\rm HMHK}(0) 
  &=& -\dfrac{1}{2} A(M_W) -\dfrac{1}{4} A(M_Z) ,
  \nonumber\\
  & &
  \\
  \Pi_{33}^{\rm NTT}(0) - \Pi_{33}^{\rm HMHK}(0) 
  &=& -\dfrac{1}{2} A(M_W) -\dfrac{1}{4} A(M_Z) ,
  \nonumber\\
  & &
\end{eqnarray}
where $\Pi_{11}^{\rm NTT}(0)$ and $\Pi_{33}^{\rm NTT}(0)$
denote the results presented in this section with the
assumptions above, while 
$\Pi_{11}^{\rm HMHK}(0)$ and $\Pi_{33}^{\rm HMHK}(0)$ are
the SM pinch technique results of 
Ref.\cite{Hagiwara:1994pw}.
These difference do not affect physical consequences, however.
They actually can be considered to arise from 
the difference of conventions for the choice of normal ordering 
in the $WW$-NGB-NGB and the $ZZ$-NGB-NGB vertices in the linear 
sigma model Lagrangian (HMHK)
and in the non-linear sigma model Lagrangian (NTT). 

Let's go back to our non-linear sigma model Lagrangian
with arbitrary Higgs coupling strengths $\kappa$.
Note that the loop functions $A$, $B$, and $B_0$ diverge in the
ultraviolet.  Introducing the UV cutoff momentum $\Lambda$, they can
be expressed by using Eq.(\ref{eq:defA}), Eq.(\ref{eq:defB}) and
Eq.(\ref{eq:defB0}).
It is now straightforward to obtain the UV divergences
in $\Pi_{11}$ and $\Pi_{33}$.

We find
\begin{eqnarray}
\lefteqn{
  \left. \Pi_{11}(0) \right|_{\rm div}  =
} \nonumber\\
  & &
    \dfrac{1}{4} \left( 
      \sum_{n=1}^{N_0} \kappa_{WW}^{\phi_n^0 \phi_n^0} 
    - 2\sum_{n=1}^{N_0} \kappa_{WW}^{\phi_n^0}\kappa_{WW}^{\phi_n^0} -4 + 2\dfrac{v_Z^2}{v^2}
    \right) \dfrac{\Lambda^2}{(4\pi)^2}
    \nonumber\\
    & &
    + \left\{ 
       \dfrac{1}{4} \sum_{n=1}^{N_0} \left(
        -\kappa_{WW}^{\phi_n^0 \phi_n^0}
         +\kappa_{WW}^{\phi_n^0}\kappa_{WW}^{\phi_n^0}
        \right) M_{\phi_n^0}^2 
      \right.
    \nonumber\\
    & & \quad
       -\dfrac{3}{16} \sum_{n=1}^{N_0} \kappa_{WW}^{\phi_n^0} \kappa_{WW}^{\phi_n^0}  g^2 v^2
%       \right.
    \nonumber\\
    & &
      \quad \left.  
        -\dfrac{3}{16} g_Z^2 \dfrac{v_Z^4}{v^2} - \dfrac{3}{4} g^2 v^2
        +\dfrac{9}{16} g^2 v_Z^2 \right\} \dfrac{1}{(4\pi)^2} 
         \ln \dfrac{\Lambda^2}{\mu^2},
\nonumber\\
& &
\end{eqnarray}
and
\begin{eqnarray}
\lefteqn{
  \left. \Pi_{33}(0) \right|_{\rm div}  =
} \nonumber\\
& & \dfrac{1}{4} \left(
   -
%\sum_{n,m} 
\sum_{n=1}^{N_0} \sum_{m=1}^{N_0}
\kappa_{Z}^{\phi_n^0 \phi_m^0}\kappa_{Z}^{\phi_n^0 \phi_m^0}
   +\sum_{n=1}^{N_0} \kappa_{ZZ}^{\phi_n^0 \phi_n^0} 
   \right.
\nonumber\\
& & \quad
   \left.
   -2\dfrac{v^2}{v_Z^2} 
    \sum_{n=1}^{N_0} \kappa_{ZZ}^{\phi_n^0}\kappa_{ZZ}^{\phi_n^0} 
   -2\dfrac{v_Z^4}{v^4}
   \right) \dfrac{\Lambda^2}{(4\pi)^2}
  \nonumber\\
  & &
  +\left\{
    \dfrac{1}{4} \sum_{n=1}^{N_0} \left(
       \sum_{m=1}^{N_0} \kappa_{Z}^{\phi_n^0 \phi_m^0}\kappa_{Z}^{\phi_n^0 \phi_m^0}
      -\kappa_{ZZ}^{\phi_n^0 \phi_n^0} 
    \right.\right.
  \nonumber\\
  & &
    \qquad \left.\left.
      + \dfrac{v^2}{v_Z^2} \kappa_{ZZ}^{\phi_n^0}\kappa_{ZZ}^{\phi_n^0}
    \right) M_{\phi_n^0}^2 \right.
  \nonumber\\
  & & \quad \left. 
     -\dfrac{3}{16} g_Z^2 v^2 \sum_{n=1}^{N_0} 
      \kappa_{ZZ}^{\phi_n^0}\kappa_{ZZ}^{\phi_n^0}
     -\dfrac{3}{8} g^2 \dfrac{v_Z^4}{v^2}
     \right\} \dfrac{1}{(4\pi)^2} \ln \dfrac{\Lambda^2}{\mu^2}. %,
\nonumber\\
& &
\end{eqnarray}

We are now ready to derive conditions to guarantee the finiteness
of Eq.(\ref{eq:rho-correction}).
We obtain a condition,
\begin{eqnarray}
  0 &=&   \dfrac{v_Z^4}{v^2} - v_Z^2 
  \nonumber\\
    & & +
  \dfrac{v_Z^2}{4} \sum_{n=1}^{N_0}
    (\kappa_{WW}^{\phi_n^0\phi_n^0} 
  - 2\kappa_{WW}^{\phi_n^0}\kappa_{WW}^{\phi_n^0} )
  \nonumber\\
  & &
 -\dfrac{v^2}{4} \sum_{n=1}^{N_0}
    \biggl(\kappa_{ZZ}^{\phi_n^0\phi_n^0} 
  - 2\dfrac{v^2}{v_Z^2} \kappa_{ZZ}^{\phi_n^0}\kappa_{ZZ}^{\phi_n^0} 
  \nonumber\\
  & & 
  \qquad
  - \sum_{m=1}^{N_0} \kappa_Z^{\phi_n^0\phi_m^0}\kappa_Z^{\phi_n^0\phi_m^0}\biggr),
\nonumber\\
& & 
\label{eq:T_finiteness1}
\end{eqnarray}
which guarantees the cancellation of the 
$\Lambda^2$ divergence, and
\begin{eqnarray}
  0&=& 
   -\dfrac{3}{16} g_Z^2 \dfrac{v_Z^6}{v^2}
   +\dfrac{3}{16} g^2 v_Z^2(5v_Z^2-4v^2) 
  \nonumber\\
  & &
 -\dfrac{3}{16}v^2 \sum_{n=1}^{N_0} \left(
    g^2 v_Z^2 \kappa_{WW}^{\phi_n^0}\kappa_{WW}^{\phi_n^0}
   -g_Z^2 v^2 \kappa_{ZZ}^{\phi_n^0}\kappa_{ZZ}^{\phi_n^0} 
    \right)
  \nonumber\\
  & &+
  \left[ -\dfrac{v_Z^2}{4}\sum_{n=1}^{N_0} 
   (\kappa_{WW}^{\phi_n^0 \phi_n^0}
   -\kappa_{WW}^{\phi_n^0}\kappa_{WW}^{\phi_n^0} ) M_{\phi_n^0}^2
   \right.
  \nonumber\\
  & & \quad 
  +\dfrac{v^2}{4}\sum_{n=1}^{N_0}
   \biggl(\kappa_{ZZ}^{\phi_n^0 \phi_n^0}
   -\dfrac{v^2}{v_Z^2}\kappa_{ZZ}^{\phi_n^0}\kappa_{ZZ}^{\phi_n^0} 
  \nonumber\\
  & & \quad \left.
   -\sum_{m=1}^{N_0} \kappa_{Z}^{\phi_n^0 \phi_m^0}\kappa_{Z}^{\phi_n^0 \phi_m^0}
   \biggr) M_{\phi_n^0}^2 \right]  ,
\label{eq:T_finiteness2}
\end{eqnarray}
for the cancellation of the $\ln \Lambda^2$ divergence.
If we impose conditions that terms proportional to $g_Z^2$,
$g^2$ and $M_{\phi_n^0}^2$ should vanish separately in 
Eq.(\ref{eq:T_finiteness2}), we obtain
\begin{eqnarray}
  0 &=& -\dfrac{v_Z^6}{v^6} 
        + \sum_{n=1}^{N_0} \kappa_{ZZ}^{\phi_n^0} \kappa_{ZZ}^{\phi_n^0},
\label{eq:T_finiteness3}
    \\
  0 &=& 5\dfrac{v_Z^2}{v^2} -4 - \sum_{n=1}^{N_0} \kappa_{WW}^{\phi_n^0}\kappa_{WW}^{\phi_n^0},
\label{eq:T_finiteness4}
    \\
  0 &=& -\left(
    \kappa_{WW}^{\phi_n^0 \phi_n^0}-\kappa_{WW}^{\phi_n^0}\kappa_{WW}^{\phi_n^0}
    \right)
    \nonumber\\
    & & 
        +\dfrac{v^2}{v_Z^2} \left(
          \kappa_{ZZ}^{\phi_n^0 \phi_n^0}-\dfrac{v^2}{v_Z^2}
          \kappa_{ZZ}^{\phi_n^0}\kappa_{ZZ}^{\phi_n^0} \right)
   \nonumber\\
   & & 
        -\dfrac{v^2}{v_Z^2} \sum_{m=1}^{N_0} 
         \kappa_Z^{\phi_n^0 \phi_m^0}\kappa_Z^{\phi_n^0 \phi_m^0} .
   \nonumber\\
   & &
\label{eq:T_finiteness5}
\end{eqnarray}

\subsection{$\Pi'_{33}(0)-\Pi'_{3Q}(0)$}

We next turn to the finiteness of Eq.(\ref{eq:S_renormalize}).
In a similar manner to the previous subsection, we decompose
\begin{equation}
  \Pi_{3Q}(p^2) = \tilde{\Pi}_{3Q}(p^2) + \Pi_{3Q}^{\rm Higgs}(p^2; \kappa).
\end{equation}
It is evident
\begin{equation}
  \Pi_{AA}^{\rm Higgs} = \Pi_{ZA}^{\rm Higgs} = 0,
\end{equation}
since the neutral Higgs bosons do not couple with the photon.
Using Eq.(\ref{eq:def_33}), Eq.(\ref{eq:def_3Q}) and Eq.(\ref{eq:def_QQ}),
we therefore obtain
\begin{equation}
  \Pi^{\rm Higgs}_{33} = \dfrac{1}{g_Z^2} \Pi_{ZZ}^{\rm Higgs}, \qquad
  \Pi^{\rm Higgs}_{3Q} = \Pi^{\rm Higgs}_{QQ} = 0.
\end{equation}
Analysis similar to Eq.(\ref{eq:Pi_33_0}) then gives 
the divergent part of $\Pi'^{\rm Higgs}_{33}(0; \kappa)$ as
\begin{eqnarray}
\lefteqn{
  \left. \Pi'^{\rm Higgs}_{33}(0; \kappa) \right|_{\rm div}
  = 
} \nonumber\\
  & &
-\left[
      \dfrac{1}{24} 
        %\sum_{n,m}
\sum_{n=1}^{N_0} \sum_{m=1}^{N_0}
\kappa_Z^{\phi_n^0 \phi_m^0} \kappa_Z^{\phi_n^0 \phi_m^0} 
  % \right. 
  % \nonumber\\
  % & & \qquad \left.
      +
      \dfrac{1}{12}\dfrac{v^2}{v_Z^2}
        \sum_{n=1}^{N_0} \kappa_{ZZ}^{\phi_n^0}\kappa_{ZZ}^{\phi_n^0}
    \right]  \times \nonumber\\
  & & \quad \times 
  \dfrac{1}{(4\pi)^2} \ln \dfrac{\Lambda^2}{\mu^2} .
\label{eq:PiDifDivHiggs33}
\end{eqnarray}
Note that the vacuum polarization function $\Pi'^{\rm Higgs}_{3Q}$ is also trivial
\begin{equation}
  \left. \Pi'^{\rm Higgs}_{3Q}(0; \kappa) \right|_{\rm div} = 0 .
\label{eq:PiDifDivHiggs3Q}
\end{equation}

The $\kappa$ independent contributions to
the divergent coefficients to $\Pi'_{33}$ and $\Pi'_{3Q}$ have 
been evaluated in appendix of Ref.\cite{Sekhar Chivukula:2007ic}.
They are
\begin{eqnarray}
\lefteqn{
  \left. \tilde{\Pi}'_{33}(0) \right|_{\rm div}
  = 
} \nonumber\\
& &
\left[ 
       \left(\dfrac{22}{3}-\dfrac{1}{12}\dfrac{v_Z^2}{v^2}
       \right)
      -\dfrac{1}{12}\left(1-\dfrac{v_Z^2}{v^2}\right) 
                    \left(4-\dfrac{v_Z^2}{v^2}\right) 
    \right] \times 
\nonumber\\
 & & \qquad \times
    \dfrac{1}{(4\pi)^2} \ln \dfrac{\Lambda^2}{\mu^2},
\label{eq:PiDifDivNLSM33}
\end{eqnarray}
and
\begin{eqnarray}
\lefteqn{
  \left. \tilde{\Pi}'_{3Q}(0) \right|_{\rm div}
  = 
} \nonumber\\
 & &
\left[ 
       \left(\dfrac{22}{3}-\dfrac{1}{12}\dfrac{v_Z^2}{v^2}
       \right)
      -\dfrac{1}{3} + \dfrac{1}{4}\dfrac{v_Z^2}{v^2}
    \right]  
  % \times
  % \nonumber\\
  % & & \qquad \times
    \dfrac{1}{(4\pi)^2} \ln \dfrac{\Lambda^2}{\mu^2}.
  \nonumber\\
  & & 
\label{eq:PiDifDivNLSM3Q}
\end{eqnarray}

It is now straightforward to obtain a condition guaranteeing the 
cancellation of the $\ln \Lambda^2$ divergence in 
Eq.(\ref{eq:S_renormalize}),
\begin{eqnarray}
 0&=& \dfrac{1}{12}\dfrac{v_Z^2}{v^2}
   \left(2-\dfrac{v_Z^2}{v^2}\right)
  % \nonumber\\
  % & & 
  -\dfrac{1}{24}
%\sum_{n,m} 
\sum_{n=1}^{N_0} \sum_{m=1}^{N_0}
   \kappa_Z^{\phi_n^0 \phi_m^0} \kappa_Z^{\phi_n^0 \phi_m^0} 
  \nonumber\\
  & & 
  -\dfrac{1}{12}\dfrac{v^2}{v_Z^2} \sum_{n=1}^{N_0} 
   \kappa_{ZZ}^{\phi_n^0} \kappa_{ZZ}^{\phi_n^0} .
\label{eq:S_finiteness}
\end{eqnarray}

\subsection{$\Pi'_{11}(0)-\Pi'_{3Q}(0)$}

The finiteness condition of $\Pi'_{11}(0)-\Pi'_{3Q}(0)$
can be studied in a similar manner.
We find
\begin{equation}
  \left. \Pi'^{\rm Higgs}_{11}(0; \kappa) \right|_{\rm div}
  = - \dfrac{1}{12} 
      \sum_{n=1}^{N_0} \kappa_{WW}^{\phi_n^0}\kappa_{WW}^{\phi_n^0}
      \dfrac{1}{(4\pi)^2} \ln \dfrac{\Lambda^2}{\mu^2},
\label{eq:PiDifDivHiggs11}
\end{equation}
and 
\begin{equation}
  \left. \tilde{\Pi}'_{11}(0) \right|_{\rm div}
  =    \left(\dfrac{22}{3}-\dfrac{1}{12}\dfrac{v_Z^2}{v^2}
       \right)
       \dfrac{1}{(4\pi)^2} \ln \dfrac{\Lambda^2}{\mu^2} .
\label{eq:PiDifDivNLSM11}
\end{equation}
Using Eq.(\ref{eq:PiDifDivHiggs3Q}), Eq.(\ref{eq:PiDifDivNLSM3Q}),
Eq.(\ref{eq:PiDifDivHiggs11}) and Eq.(\ref{eq:PiDifDivNLSM11}),
we find a condition guaranteeing the finiteness 
of $\Pi'_{11}(0)-\Pi'_{3Q}(0)$:
\begin{equation}
  0 = 
   \dfrac{1}{3} - \dfrac{1}{4} \dfrac{v_Z^2}{v^2} 
  -\dfrac{1}{12} \sum_{n=1}^{N_0} \kappa_{WW}^{\phi_n^0}\kappa_{WW}^{\phi_n^0} .
\label{eq:S+U_finiteness}
\end{equation}

\subsection{Unitarity Sum Rules vs. finiteness of $f\bar{f}\to f'\bar{f'}$}

It is easy to show that the conditions of the finiteness of the 
$f\bar{f} \to f' \bar{f'}$ amplitudes, {\it i.e.,}
Eqs.(\ref{eq:T_finiteness1}), (\ref{eq:T_finiteness3}), 
(\ref{eq:T_finiteness4}),
(\ref{eq:T_finiteness5}),
(\ref{eq:S_finiteness}), and
(\ref{eq:S+U_finiteness}), are automatically satisfied if the
Higgs coupling parameters satisfy the unitarity sum rules
Eqs.(\ref{eq:wwww}), (\ref{eq:wwzz}), (\ref{eq:wwphiphi2}), 
(\ref{eq:wwphiphi1}), (\ref{eq:zzphiphi}) and
(\ref{eq:wwphiz}) in the present framework.

Even though we do not require the renormalizability of the model
in its construction, any unitary EWSB model with 
neutral Higgs extension only thus leads to finite 
$f\bar{f} \to f'\bar{f'}$ amplitude at the one-loop level.
This fact enables us to perform the EWPTs 
for any unitary model using $f\bar{f} \to f'\bar{f'}$ 
amplitudes at one-loop level.

Let us next consider the converse of the problem:  Does
a model satisfying the finiteness constraints 
Eqs.(\ref{eq:T_finiteness1}),
(\ref{eq:T_finiteness3}),
(\ref{eq:T_finiteness4}),
(\ref{eq:T_finiteness5}), (\ref{eq:S_finiteness}), and
(\ref{eq:S+U_finiteness}), automatically satisfy the unitarity sum rules?
Evidently, the answer is negative.
There is a large class of models which satisfy the finiteness
constraints Eqs.(\ref{eq:T_finiteness1}),
(\ref{eq:T_finiteness3}), (\ref{eq:T_finiteness4}), 
(\ref{eq:T_finiteness5}), (\ref{eq:S_finiteness}), and
(\ref{eq:S+U_finiteness}), but do not satisfy 
the unitarity sum rules
Eqs.(\ref{eq:wwww}), (\ref{eq:wwzz}), (\ref{eq:wwphiphi2}), 
(\ref{eq:wwphiphi1}), (\ref{eq:zzphiphi}) and
(\ref{eq:wwphiz}).
To give an example, 
the $\kappa_{WW}^{\phi_{n_1}^0 \phi_{n_2}^0}$ coupling 
cannot be constrained by the
finiteness conditions 
Eqs.(\ref{eq:T_finiteness1}),
(\ref{eq:T_finiteness3}), (\ref{eq:T_finiteness4}), 
(\ref{eq:T_finiteness5}), (\ref{eq:S_finiteness}), and
(\ref{eq:S+U_finiteness})
for $n_1 \ne n_2$.
On the other hand, the 
$\kappa_{WW}^{\phi_{n_1^{}}^0 \phi_{n_2^{}}^0}$ coupling not satisfying
Eq.(\ref{eq:wwphiphi2}) violates the perturbative unitarity 
in the $WW \to \phi_{n_1^{}}^0 \phi_{n_2^{}}^0$ amplitude.
Although the great success of the EWPTs,
which use the $f\bar{f} \to f'\bar{f'}$ processes, suggests
the validity of the finiteness conditions,
Eqs.(\ref{eq:T_finiteness1}),
(\ref{eq:T_finiteness3}), (\ref{eq:T_finiteness4}), 
(\ref{eq:T_finiteness5}), (\ref{eq:S_finiteness}), and
(\ref{eq:S+U_finiteness}), with very high accuracy, it does 
not imply the perturbative unitarity in the 
$WW \to \phi_{n_1^{}}^0 \phi_{n_2^{}}^0$ process.

It should also be noted that the finiteness conditions
are only sensitive to the absolute values of the Higgs-$V$-$V$ 
couplings ($\kappa_{ZZ}^{\phi_n^0}$ and $\kappa_{WW}^{\phi_n^0}$)
and insensitive to their relative sign 
$\kappa_{ZZ}^{\phi_n^0}\kappa_{WW}^{\phi_n^0}$.
If we adequately choose the other parameters, the finiteness
conditions can be satisfied even with a wrong signed
$\kappa_{ZZ}^{\phi_n^0}\kappa_{WW}^{\phi_n^0} < 0$.
On the other hand, the wrong signed 
$\kappa_{ZZ}^{\phi_n^0}\kappa_{WW}^{\phi_n^0} < 0$
clearly contradicts with the unitarity sum rule in the $WW\to ZZ$ 
process Eq.(\ref{eq:wwzz}) as we stressed in 
{Sec}.~\ref{sec-application}.

The numerical comparison between the unitarity sum rules
and the finiteness conditions will be performed in 
{Sec}.~\ref{sec:heavy higgs} and 
{Sec}.~\ref{sec:effect-field-theory} in this manuscript.

\section{Oblique correction parameters}
\label{sec:oblique}
In order to compare our models with the electroweak precision 
measurements of the $f\bar{f} \to f' \bar{f'}$ processes, it 
is most convenient to introduce the electroweak precision 
parameters such as the oblique correction parameters of 
Ref.\cite{Peskin:1990zt}
($S$, $T$ and $U$).
Hereafter we assume
\begin{equation}
  v_Z^{} = v,
\label{eq:rho=1}
\end{equation}
and
the bare parameters $v$ and $v_Z$ cannot be adjusted independently 
to renormalize the UV divergences of $\Pi_{33}(0)$ and $\Pi_{11}(0)$.
The electroweak oblique correction parameters 
are defined by
\begin{eqnarray}
  \dfrac{1}{16\pi}S
     &=& (\Pi'_{33}(0) - \Pi'_{3Q}(0)) 
     \nonumber\\
     & & \quad
      - \left. (\Pi'_{33}(0) - \Pi'_{3Q}(0)) \right|_{\rm SM},
\label{eq:Spara1}
\\
  \alpha T 
     &=& \dfrac{4}{v^2} \left(\Pi_{11}(0) - \Pi_{33}(0)\right)
     \nonumber\\
     & & \quad
      - \dfrac{4}{v^2} \left. \left(\Pi_{11}(0) - \Pi_{33}(0)\right)
        \right|_{\rm SM},
\\
  \dfrac{1}{16\pi}U
     &=& (\Pi'_{11}(0) - \Pi'_{33}(0)) 
     \nonumber\\
     & & \quad
      - \left. (\Pi'_{11}(0) - \Pi'_{33}(0)) \right|_{\rm SM},
\end{eqnarray}
where $\left. \Pi \right|_{\rm SM}$ denotes the vacuum polarization
function in the SM. 

As we did in the previous section, we decompose
\begin{equation}
  \Pi(p^2) = \tilde{\Pi}(p^2) + \Pi^{\rm Higgs}(p^2; M_{\phi^0}, \kappa)  . 
\end{equation}
Under the assumption of Eq.(\ref{eq:rho=1}), $\tilde{\Pi}$
in our generalized model is identical to that of the SM.
Also, since the neutral Higgs bosons have no coupling with the 
photon, we can easily show
\begin{equation}
  \Pi^{\rm Higgs}_{3Q} = \Pi^{\rm Higgs}_{QQ} = 0.
\end{equation}
Eq.(\ref{eq:Spara1}) can therefore be rewritten as
\begin{equation}
  \dfrac{1}{16\pi}S
   = \Pi'^{\rm Higgs}_{33}(0; M_{\phi^0}, \kappa) 
  - \Pi'^{\rm Higgs}_{33}(0; M_h, \kappa_{\rm SM}^{}) ,
\end{equation}
%\sout{with $\kappa_{\rm SM}^{}=1$.}
with $\kappa_{\rm SM}^{}$ denoting the SM values of the Higgs coupling
strengths. 
In a similar manner, we obtain
\begin{eqnarray}
  \alpha T
  &=& \dfrac{4}{v^2} \left(
      \Pi_{11}^{\rm Higgs}(0; M_{\phi^0}, \kappa) 
    - \Pi_{11}^{\rm Higgs}(0; M_h, \kappa_{\rm SM}^{})
    \right)
  \nonumber\\
  & &
  -\dfrac{4}{v^2} \left(
      \Pi_{33}^{\rm Higgs}(0; M_{\phi^0}, \kappa) 
    - \Pi_{33}^{\rm Higgs}(0; M_h, \kappa_{\rm SM}^{})
    \right) , 
  \nonumber\\
  & &
\end{eqnarray}
and
\begin{eqnarray}
  \dfrac{1}{16\pi} U
  &=& \left(
      \Pi'^{\rm Higgs}_{11}(0; M_{\phi^0}, \kappa) 
    - \Pi'^{\rm Higgs}_{11}(0; M_h, \kappa_{\rm SM}^{})
    \right)
 \nonumber\\
  & &
  - \left(
      \Pi'^{\rm Higgs}_{33}(0; M_{\phi^0}, \kappa) 
    - \Pi'^{\rm Higgs}_{33}(0; M_h, \kappa_{\rm SM}^{})
    \right) .
  \nonumber\\
  & &
\end{eqnarray}

We are now ready to write down the one-loop formulas
for the oblique correction parameters,
\begin{eqnarray}
  S &=& S_{\rm log} + S_f, \\
  T &=& T_{\rm quad} + T_{\rm log} + T_f, \\
  U &=& U_{\rm log} + U_f .
\end{eqnarray}
Here $T_{\rm quad}$ denotes the $\Lambda^2$ divergent term.
$S_{\rm log}$, $T_{\rm log}$ and $U_{\rm log}$ are the $\ln \Lambda^2$ terms.
$S_f$, $T_f$ and $U_f$ are the finite terms.
\begin{widetext}
We find
\begin{eqnarray}
  S_{\rm log} 
  &=& \dfrac{1}{12\pi} 
    \left[1 - \sum_{n=1}^{N_0} \kappa_{ZZ}^{\phi_n^0}\kappa_{ZZ}^{\phi_n^0} 
            - \dfrac{1}{2} 
%\sum_{n,m} 
\sum_{n=1}^{N_0} \sum_{m=1}^{N_0}
              \kappa_Z^{\phi_n^0 \phi_m^0}\kappa_Z^{\phi_n^0 \phi_m^0}
    \right] 
  % \times
  % \nonumber\\
  % & & \qquad \times
  \ln \dfrac{\Lambda^2}{\mu^2},
\\
  S_f
  &=& \dfrac{1}{4\pi} \sum_{n=1}^{N_0} \kappa_{ZZ}^{\phi_n^0}\kappa_{ZZ}^{\phi_n^0}
    {G^{Z\phi_n^0}}'
   -\dfrac{1}{4\pi} {G^{Zh}}'
  % \nonumber\\
  % & &
   +\dfrac{1}{8\pi} \sum_{n=1}^{N_0} \sum_{m=1}^{N_0} 
    \kappa_{Z}^{\phi_n^0\phi_m^0}\kappa_{Z}^{\phi_n^0 \phi_m^0}
    {F^{\phi_n^0\phi_m^0}}', 
\end{eqnarray}
\begin{eqnarray}
  \alpha T_{\rm quad}
  &=& \sum_{n=1}^{N_0} \Biggl[ 
      \kappa_{WW}^{\phi_n^0\phi_n^0}
    -2\kappa_{WW}^{\phi_n^0} \kappa_{WW}^{\phi_n^0}
    - \kappa_{ZZ}^{\phi_n^0\phi_n^0}
    +2\kappa_{ZZ}^{\phi_n^0} \kappa_{ZZ}^{\phi_n^0}
  % \nonumber\\
  % & & \qquad 
    + \sum_{m=1}^{N_0} \kappa_{Z}^{\phi_n^0\phi_m^0}\kappa_{Z}^{\phi_n^0\phi_m^0}
   \Biggr] \dfrac{\Lambda^2}{(4\pi)^2 v^2},
  \\
  \alpha T_{\rm log}
  &=& \Bigg\{ 
  \sum_{n=1}^{N_0} \Biggl[
    \left(
      -\kappa_{WW}^{\phi_n^0 \phi_n^0} 
      +\kappa_{WW}^{\phi_n^0} \kappa_{WW}^{\phi_n^0}
      +\kappa_{ZZ}^{\phi_n^0 \phi_n^0} 
      -\kappa_{ZZ}^{\phi_n^0} \kappa_{ZZ}^{\phi_n^0} 
  % \right.
  % \nonumber\\
  % & & \qquad \left.
      -\sum_{m=1}^{N_0} \kappa_{Z}^{\phi_n^0\phi_m^0} \kappa_{Z}^{\phi_n^0\phi_m^0}
    \right) \dfrac{M_{\phi_n^0}^2}{v^2}
  \nonumber\\
  & & \qquad 
      -\dfrac{3}{4} \left(
       g^2 \kappa_{WW}^{\phi_n^0}\kappa_{WW}^{\phi_n^0} 
      -g_Z^2\kappa_{ZZ}^{\phi_n^0}\kappa_{ZZ}^{\phi_n^0}
      \right) \Biggr]  -\dfrac{3}{4}g_Y^2
      \Biggr\} \dfrac{1}{(4\pi)^2} \ln \dfrac{\Lambda^2}{\mu^2},
  \nonumber\\
  & &
\end{eqnarray}
\begin{eqnarray}
  \alpha T_f
  &=& \dfrac{1}{(4\pi)^2 v^2} 
      %\sum_{n,m}
      \sum_{n=1}^{N_0} \sum_{m=1}^{N_0}
\kappa_Z^{\phi_n^0 \phi_m^0}\kappa_Z^{\phi_n^0 \phi_m^0} 
  % \times
  % \nonumber\\
  % & & \times
      \left(
        -\dfrac{1}{2} F^{\phi_n^0 \phi_m^0} 
        +M_{\phi_n^0}^2 \left(\ln \dfrac{M_{\phi_n^0}^2}{\mu^2}-1\right)
      \right)
  \nonumber\\
  & & 
      + \dfrac{1}{(4\pi)^2 v^2} 
        \sum_{n=1}^{N_0}
        \left(
          \kappa_{WW}^{\phi_n^0\phi_n^0} 
         -\kappa_{WW}^{\phi_n^0} \kappa_{WW}^{\phi_n^0} 
  %        \right.
  % \nonumber\\
  % & & \left.
         -\kappa_{ZZ}^{\phi_n^0\phi_n^0} 
         +\kappa_{ZZ}^{\phi_n^0} \kappa_{ZZ}^{\phi_n^0} 
        \right) M_{\phi_n^0}^2 \left(\ln \dfrac{M_{\phi_n^0}^2}{\mu^2}-1\right)
  \nonumber\\
  & & +\dfrac{1}{2(4\pi)^2v^2} \sum_{n=1}^{N_0}\left(
        \kappa_{WW}^{\phi_n^0} \kappa_{WW}^{\phi_n^0} 
       -\kappa_{ZZ}^{\phi_n^0} \kappa_{ZZ}^{\phi_n^0} 
       \right) M_{\phi_n^0}^2
  \nonumber\\
  & & +\dfrac{1}{(4\pi)^2v^2} \sum_{n=1}^{N_0} \kappa_{WW}^{\phi_n^0} \kappa_{WW}^{\phi_n^0} 
  %      \times
  % \nonumber\\
  % & & \quad \times
       \left[
         G^{W\phi_n^0} - \dfrac{1}{2} M_{\phi_n^0}^2 
       - M_W^2 \left(\ln \dfrac{M_W^2}{\mu^2}-1\right)
       \right]
  \nonumber\\
  & & -\dfrac{1}{(4\pi)^2v^2} \sum_{n=1}^{N_0} \kappa_{ZZ}^{\phi_n^0} \kappa_{ZZ}^{\phi_n^0} 
  %      \times
  % \nonumber\\
  % & & \quad \times
       \left[
         G^{Z\phi_n^0} - \dfrac{1}{2} M_{\phi_n^0}^2 
       - M_Z^2 \left(\ln \dfrac{M_Z^2}{\mu^2}-1\right)
       \right]
  \nonumber\\
  & & +\dfrac{1}{(4\pi)^2 v^2} \left[
    -G^{Wh} 
    + M_W^2 \left(\ln \dfrac{M_W^2}{\mu^2}-1\right)
  %   \right.
  % \nonumber\\
  % & & \left.
    + G^{Zh} 
    - M_Z^2 \left(\ln \dfrac{M_Z^2}{\mu^2}-1\right)
    \right],
\end{eqnarray}
and
\begin{eqnarray}
  U_{\rm log}
  &=& \dfrac{1}{12\pi} \left[
      \sum_{n=1}^{N_0} \left( -\kappa_{WW}^{\phi_n^0}\kappa_{WW}^{\phi_n^0}
                    +\kappa_{ZZ}^{\phi_n^0}\kappa_{ZZ}^{\phi_n^0}
             \right)
  % \right.
  % \nonumber\\
  % & &
  % \left.
     +\frac{1}{2} %\sum_{n,m} 
\sum_{n=1}^{N_0} \sum_{m=1}^{N_0}
\kappa_Z^{\phi_n^0\phi_m^0}\kappa_Z^{\phi_n^0\phi_m^0}
    \right] \ln \dfrac{\Lambda^2}{\mu^2},
  \\
  U_f
  &=& \dfrac{1}{4\pi} \sum_{n=1}^{N_0} \kappa_{WW}^{\phi_n^0}\kappa_{WW}^{\phi_n^0}
      {G^{W\phi_n^0}}'
      -\dfrac{1}{4\pi} {G^{Wh}}'
  % \nonumber\\
  % & & 
    -\dfrac{1}{4\pi} \sum_{n=1}^{N_0} \kappa_{ZZ}^{\phi_n^0}\kappa_{ZZ}^{\phi_n^0}
      {G^{Z\phi_n^0}}'
      +\dfrac{1}{4\pi} {G^{Zh}}'
 % \nonumber\\
 %  & &
     -\dfrac{1}{8\pi} \sum_{n=1}^{N_0}\sum_{m=1}^{N_0}
      \kappa_{Z}^{\phi_n^0\phi_m^0}\kappa_{Z}^{\phi_n^0 \phi_m^0}
      {F^{\phi_n^0\phi_m^0}}' .
 \nonumber\\
  & &
\end{eqnarray}
It is obvious $T_{\rm quad}= S_{\rm log} = T_{\rm log} = U_{\rm log} = 0$
in models satisfying the conditions Eq.(\ref{eq:T_finiteness1}), 
Eq.(\ref{eq:T_finiteness2}), Eq.(\ref{eq:S_finiteness})
and Eq.(\ref{eq:S+U_finiteness}).

\end{widetext}

\section{Constraints on a Heavy Higgs Boson}
\label{sec:heavy higgs}

If the masses of the extra Higgs bosons become extremely heavy 
keeping their non-vanishing $\kappa$s,
the %\sout{perturbative unitarity of} 
longitudinal electroweak gauge boson 
scattering amplitude is enhanced and the perturbative unitarity 
can be violated even in the models 
which satisfy the unitarity sum rules.
In a similar manner, the heavy extra Higgs boson mass induces
large finite correction to the electroweak precision parameters 
($S$ and $T$) even in the model which satisfy the finiteness 
conditions.
The mass of the extra Higgs boson can therefore be constrained 
by the perturbative unitarity and the EWPTs.

In this section, we assume models in which the unitarity sum rules 
Eqs.(\ref{eq:wwww}), (\ref{eq:wwzz}), (\ref{eq:wwphiphi2}), 
(\ref{eq:wwphiphi1}), (\ref{eq:zzphiphi}) and
(\ref{eq:wwphiz}) are satisfied. 
We also identify the 125 GeV Higgs boson ($h$) discovered 
by the LHC experiments as the lightest Higgs boson in our 
framework ($\phi_1^0$), {\it i.e.,} $M_{\phi_1^0}=M_h=125$GeV\@.
The second lightest Higgs boson $\phi_2^0$ is denoted by $H$.
In the following subsections, constraints on the second lightest 
Higgs boson mass 
$M_H=M_{\phi_2^0}$ are investigated by using the perturbative 
unitarity argument and the results of EWPTs.

\subsection{Unitarity constraints}
\label{sec:unit-constr}

Thanks to the equivalence theorem between the high energy longitudinal
gauge boson scattering amplitudes and the NGB scattering amplitudes, 
$S$-wave amplitude of the $W_L W_L \to W_L W_L$ processes 
is evaluated as an integral over the scattering angle $\theta$
of the corresponding NGB amplitude,
\begin{equation}
  t_0^{W_L^+ W_L^- \to W_L^+ W_L^-}
  = \dfrac{1}{32\pi} \int_{-1}^{1} d\cos\theta {\cal A}_{w^+ w^- \to w^+ w^-}, %.
\label{eq:pwa_ww2ww}
\end{equation}
where the validity of the equivalence is of ${\cal O}(M_V^2/s)$. 
The scattering angle $\theta$ is related with the
Mandelstam variable $s$, $t$ as
\begin{equation}
  t = -\frac{s}{2}(1-\cos\theta).
\end{equation}
Similarly, the $S$-wave 
$Z_L Z_L \to W_L W_L$ and $Z_L Z_L \to Z_L Z_L$ amplitudes
are 
\begin{eqnarray}
  t_0^{Z_L Z_L \to W_L^+ W_L^-}
  &=& \dfrac{1}{32\pi} \dfrac{1}{\sqrt{2}}\int_{-1}^{1} d\cos\theta {\cal A}_{w^+ w^- \to zz},
  \nonumber\\
  & & 
\label{eq:pwa_zz2ww}
  \\
  t_0^{Z_L Z_L \to Z_L Z_L}
  &=& \dfrac{1}{32 \pi} \dfrac{1}{2} \int_{-1}^{1} d\cos\theta {\cal A}_{zz \to zz}.
  \nonumber\\
  & & 
\label{eq:pwa_zz2zz}
\end{eqnarray}
Factors $1/\sqrt{2}$ in Eq.(\ref{eq:pwa_zz2ww}) and 
$1/2$ in Eq.(\ref{eq:pwa_zz2zz}) arise from the Bose statistics of identical 
particles in the initial and final states.

We assume the unitarity sum rules
Eqs.(\ref{eq:wwww}), (\ref{eq:wwzz}), (\ref{eq:wwphiphi2}), 
(\ref{eq:wwphiphi1}), (\ref{eq:zzphiphi}) and
(\ref{eq:wwphiz}).  
The Higgs coupling constants therefore satisfy
\begin{equation}
  \kappa_{WW}^{\phi_n^0} = \kappa_{ZZ}^{\phi_n^0},  \quad
  \kappa_{Z}^{\phi_{n_1}^0 \phi_{n_2}^0} = 0,
\end{equation}
\begin{equation}
  \kappa_{WW}^{\phi_{n_1}^0 \phi_{n_2}^0}
  = \kappa_{WW}^{\phi_{n_1}^0}\kappa_{WW}^{\phi_{n_2}^0}, \quad
  \kappa_{ZZ}^{\phi_{n_1}^0 \phi_{n_2}^0}
  = \kappa_{ZZ}^{\phi_{n_1}^0}\kappa_{ZZ}^{\phi_{n_2}^0} ,
\end{equation}
and the tree-level $\rho$ parameter restricted to be unity.
Plugging these relations into the NGB scattering amplitudes
Eqs.(\ref{eq:amp_ww2ww}), (\ref{eq:amp_ww2zz}), 
and (\ref{eq:amp_zz2zz}) and computing the integrals of
Eqs.(\ref{eq:pwa_ww2ww}), 
(\ref{eq:pwa_zz2ww}), 
and (\ref{eq:pwa_zz2zz}), for sufficiently high energy scale
$s \gg M_{\phi_n^0}^2$, we obtain
\begin{eqnarray}
  t_0^{W_L^+ W_L^- \to W_L^+ W_L^-}
   &=&  {\mathscr T}, 
  \\
  t_0^{Z_L Z_L \to W_L^+W_L^-}
   &=& \dfrac{1}{2\sqrt{2}} {\mathscr T},
  \\
  t_0^{Z_LZ_L \to Z_L Z_L}
   &=& \dfrac{3}{4} {\mathscr T},
\end{eqnarray}
with
\begin{equation}
  {\mathscr T} \equiv -\dfrac{G_F}{4\sqrt{2} \pi} 
    \sum_{n=1}^{N_0} (\kappa_{V}^{\phi_n^0})^2 M_{\phi_n^0}^2 , \qquad
  G_F = \dfrac{1}{\sqrt{2} v^2} .
\end{equation}
Here the Higgs-$V$-$V$ coupling is denoted by $\kappa_V^{\phi_n^0}$,
\begin{equation}
  \kappa_V^{\phi_n^0}
  \equiv \kappa_{WW}^{\phi_n^0} = \kappa_{ZZ}^{\phi_n^0} .
\label{eq:def_kappaV1}
\end{equation}
Using the unitarity sum rule
\begin{equation}
   \sum_{n=1}^{N_0} (\kappa_V^{\phi_n^0})^2 = 1, 
\label{eq:unitarity-sr}
\end{equation}
and our ordering of neutral Higgs bosons
\begin{equation}
  M_h=M_{\phi_1^0} < M_H=M_{\phi_2^0} \le M_{\phi_3^0} \le \cdots ,  
\label{eq:higgs-ordering}
\end{equation}
we see
\begin{equation}
  |{\mathscr T}| \ge \dfrac{G_F}{4\sqrt{2}\pi} \left[
    \kappa_V^2 M_h^2 + (1-\kappa_V^2) M_H^2
  \right],
\end{equation}
with $\kappa_V^{}$ being defined as
\begin{equation}
  \kappa_V^{} \equiv \kappa_V^h = \kappa_V^{\phi_1^0} = 
  \kappa_{WW}^{\phi_1^0} = \kappa_{ZZ}^{\phi_1^0}.
\label{eq:def_kappaV2}
\end{equation}

We next deduce the bound on $M_H$ from the perturbative unitarity
in the $S$-wave transition matrix among $W_L^+ W_L^-$
and $Z_L Z_L$ states,
\begin{eqnarray}
  {\cal T} 
  &=& \left(
    \begin{array}{cc}
      t_0^{W_L^+ W_L^- \to W_L^+ W_L^-} & t_0^{W_L^+ W_L^- \to Z_L Z_L}  \\
      t_0^{Z_L Z_L \to W_L^+ W_L^-}  & t_0^{Z_L Z_L \to Z_L Z_L}
    \end{array}
  \right)
  \nonumber\\
  &=& 
      \left(
        \begin{array}{cc}
          {\mathscr T} & \dfrac{1}{2\sqrt{2}}{\mathscr T} \\
          \dfrac{1}{2\sqrt{2}}{\mathscr T} & \dfrac{3}{4} {\mathscr T}
        \end{array}
      \right) .
\end{eqnarray}
It is easy to calculate the maximum eigenvalue of the transition 
matrix ${\cal T}$,
\begin{equation}
  t_0^{\rm max} = \dfrac{5}{4} {\mathscr T}.
\end{equation}
Perturbative unitarity requires $|t_0^{\rm max}|$ should 
satisfy
\begin{equation}
  |t_0^{\rm max}| < \dfrac{1}{2},
\end{equation}
in the off-resonant energy region, which immediately
leads to a mass constraint on the second lightest Higgs boson,
\begin{equation}
  M_H^2 (1-\kappa_V^2) + M_h^2 \kappa_V^2 < \dfrac{16\pi}{5} v^2. 
\label{eq:perturbative_unitarity}
\end{equation}
Once the deviation of the 125GeV Higgs boson coupling $\kappa_V^{}$
from its SM value $\kappa_V^{}=1$ is experimentally confirmed in 
future experiment, Eq.(\ref{eq:perturbative_unitarity}) provides
a mass upper bound on the extra Higgs boson.

We here make a comment comparing 
Eq.(\ref{eq:perturbative_unitarity}) with the famous 
Lee-Quigg-Thacker bound\cite{Lee:1977eg}
on the Higgs boson mass in the SM 
\begin{equation}
  M_h^2 < \dfrac{8\pi}{3} v^2 .
\label{eq:LQT}
\end{equation}
The difference of a factor $5/6$ between the RHS of 
Eq.(\ref{eq:perturbative_unitarity}) and Eq.(\ref{eq:LQT})
arises from our neglect of the
$hh$, $hH$ and $HH$ channels in the ${\cal T}$-matrix.  The 
amplitudes including these channels depend on the triple-Higgs
and quartic-Higgs coupling strengths, which we did not incorporated
in our theory, however.  We will discuss the issue in our forthcoming
publications.

\subsection{Electroweak Precision Tests}

We next study the constraints on the heavier Higgs boson mass $M_H$ 
given by the EWPTs.
In a model with $v=v_Z^{}$ and satisfying the unitarity sum rules, 
as we found in  Sec.\ref{sec:oblique}, 
the cancellation of UV divergences in the oblique correction 
parameters,
\begin{equation}
  T_{\rm quad}=
  S_{\rm log}=
  T_{\rm log}=
  U_{\rm log}= 0,
\end{equation}
takes place at the one-loop level.
Moreover, the expressions of 
finite corrections to the oblique parameters
are greatly simplified thanks to the unitarity sum rules.
We find
\begin{eqnarray}
  S&=& 
      -\dfrac{1}{4\pi} (1-\kappa_V^2) {G^{Zh}}'
      +\dfrac{1}{4\pi} \sum_{n=2}^{N_0} (\kappa_V^{\phi_n^0})^2 {G^{Z\phi_n^0}}',
   \nonumber\\
  & &
\label{eq:S1}
  \\
 T&=& \dfrac{1-\kappa_V^2}{16\pi^2 v^2 \alpha} [G^{Zh}-G^{Wh}]
  \nonumber\\
  & & -\dfrac{1}{16\pi^2 v^2 \alpha} 
      \sum_{n=2}^{N_0} (\kappa_V^{\phi_n^0})^2 [G^{Z\phi_n^0}-G^{W\phi_n^0}],
  \nonumber\\
  & &
\label{eq:T1}
  \\
 U&=& \dfrac{1-\kappa_V^2}{4\pi}[{G^{Zh}}'-{G^{Wh}}']
  \nonumber\\
  & & -\dfrac{1}{4\pi} \sum_{n=2}^{N_0} (\kappa_V^{\phi_n^0})^2 
       [{G^{Z\phi_n^0}}'-{G^{W\phi_n^0}}'].
\label{eq:U1}
\end{eqnarray}
Here we used the notations Eq.(\ref{eq:def_kappaV1}) and 
Eq.(\ref{eq:def_kappaV2}).  
The loop functions $G^{V\phi}$ and 
${G^{V\phi}}'$ are defined in Appendix~\ref{sect-loop_integrals}.

For sufficiently heavy $\phi_n^0$ ($n \ge 2$), 
Eqs.(\ref{eq:S1}), (\ref{eq:T1}) and (\ref{eq:U1}) can be 
approximated by
\begin{eqnarray}
  S &\simeq& 
    \dfrac{1}{12\pi} \sum_{n=2}^{N_0} (\kappa_V^{\phi_n^0})^2 \left[
    \ln \dfrac{M_{\phi_n^0}^2}{M_h^2} +0.86\right],
\label{eq:S-para2}
    \\
  T &\simeq&
    -\dfrac{3}{16\pi^2 v^2 \alpha} (M_Z^2-M_W^2) \times
    \nonumber\\
    & & \quad \times \sum_{n=2}^{N_0}
     (\kappa_V^{\phi_n^0})^2 \left[
     \ln \dfrac{M_{\phi_n^0}^2}{M_h^2} -1.05 \right],
    % \nonumber\\
    % & &
\label{eq:T-para2}
    \\
  U &\simeq&
     \dfrac{1-\kappa_V^2}{3\pi} \times (-0.028)
    +\dfrac{1}{3\pi}\sum_{n=2}^{N_0} (\kappa_V^{\phi_n^0})^2 
   \dfrac{M_Z^2-M_W^2}{M_{\phi_n^0}^2} ,
   \nonumber\\
   & &
\label{eq:U-para2}
\end{eqnarray}
where we used $M_Z=91.2$ GeV, $M_W=80.4$ GeV in the estimates
of the numerical coefficients. 
As we see from Eq.(\ref{eq:U-para2}), typical value of
$U$ parameter prediction is $|U| \lesssim 3\times 10^{-3}$, which 
is well below the present value of the measured value of $U$ parameter
uncertainty $10^{-2}$.
We are thus allowed to perform a two dimensional fit in the 
$S$-$T$ plane neglecting the $U$ parameter constraint.
 
Using the unitarity sum rule Eq.(\ref{eq:unitarity-sr}) and
the ordering of the Higgs mass Eq.(\ref{eq:higgs-ordering}), 
$S$ and $T$ parameters given in Eq.(\ref{eq:S-para2}) and 
Eq.(\ref{eq:T-para2}) can be shown to satisfy
\begin{eqnarray}
  S &\ge& S_H \simeq 
    \dfrac{1-\kappa_V^2}{12\pi}  \left[ \ln \dfrac{M_H^2}{M_h^2} 
     +0.86 \right] 
   >0,
\nonumber\\
& &
\label{eq:S_H}
    \\
  T &\le& T_H \simeq 
    -\dfrac{3(1-\kappa_V^2)}{16\pi^2 v^2 \alpha} (M_Z^2-M_W^2) \times
    \nonumber\\
    & & \qquad\qquad\quad
     \times
     \left[ \ln \dfrac{M_H^2}{M_h^2} -1.05\right] <0,
\nonumber\\
& &
\label{eq:T_H}
\end{eqnarray}
with $H$ being the second lightest neutral Higgs boson in the 
model.
Here $S_H$ and $T_H$ denote $S$ and $T$ parameters, respectively, 
in a model with two neutral Higgs bosons ($N_0=2$ model).
The inequalities in Eqs.(\ref{eq:S_H}) and (\ref{eq:T_H}) guarantee
that the limits on $M_H$ deduced from the EWPTs
can be regarded as conservative bounds.

\begin{figure}[t]
  \centering
  \includegraphics[width=85mm, bb=0 0 578 540]{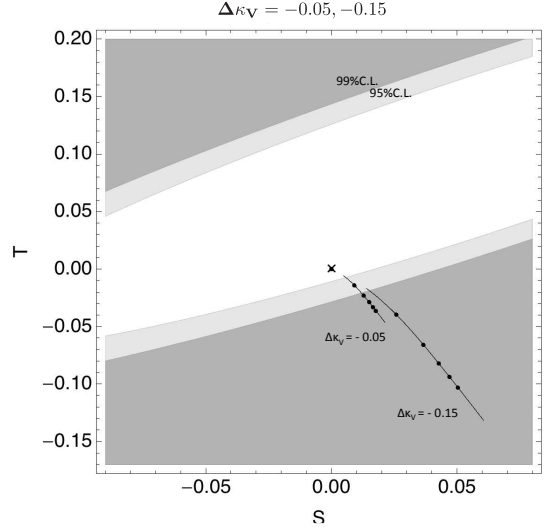}
  \caption{
The behaviors of $(S_H, T_H)$.
Contours of likelihood in $S$-$T$ plane, corresponding to 
95\% (gray) and 99\% (dark-gray) CL\@, assuming 
$M_h=125$GeV and $m_{\rm top}=173$GeV\@,
are also shown.
}
  \label{fig:MHonST}
\end{figure}

Figure~\ref{fig:MHonST} shows contours of the likelihood function of
$S$ and $T$ corresponding to 95\% and 99\% confidence level (CL) probability,  
derived from the present limit\cite{Baak:2014ora}
\begin{equation}
  S = 0.06 \pm 0.09, \qquad
  T = 0.10 \pm 0.07,
\label{eq:STlimits}
\end{equation}
with
\begin{equation}
  \rho_{ST}^{} = 0.91 .
\end{equation}
Two lines in Figure~\ref{fig:MHonST} show behaviors of $(S_H,T_H)$. 
The shorter line is for $\Delta\kappa_V^{} %\equiv \kappa_V - 1 
= -0.05$, and the 
longer one is for $\Delta\kappa_V = -0.15$, varying the second lightest 
Higgs boson mass $M_H$ from 250GeV to 5TeV\@.
Five dots on each line starting from the origin of this figure toward
the right-bottom direction correspond to the points
$M_H=500$GeV, $1.0$TeV, $1.5$TeV, $2.0$TeV and $2.5$TeV,
respectively. 
Note that these lines are not straight, since we do not use
the large $M_H$ approximation 
Eq.(\ref{eq:S_H}) and Eq.(\ref{eq:T_H}) in this figure.
Also, we obtain $(S_H, T_H)=(0,0)$ as we expect when we take
$M_H = M_h$.
If the 125GeV Higgs boson coupling $\kappa_V^{}$ turns out to deviate
sizably from the SM prediction $\kappa_V^{}=1$, then we will obtain 
an upper bound on the extra Higgs boson mass from the EWPTs.
Actually, as we see from Figure~\ref{fig:MHonST}, $M_H=283$ GeV (836 GeV)
with $\Delta \kappa_V^{} = -0.05$, and
$M_H=171$ GeV (265 GeV) with $\Delta \kappa_V^{} = -0.15$
are ruled out in the present model at 95\%CL (99\%CL).
\subsection{Unitarity vs. EWPTs}

\begin{figure}[t]
  \centering
  \includegraphics[width=85mm, bb=0 0 578 540]{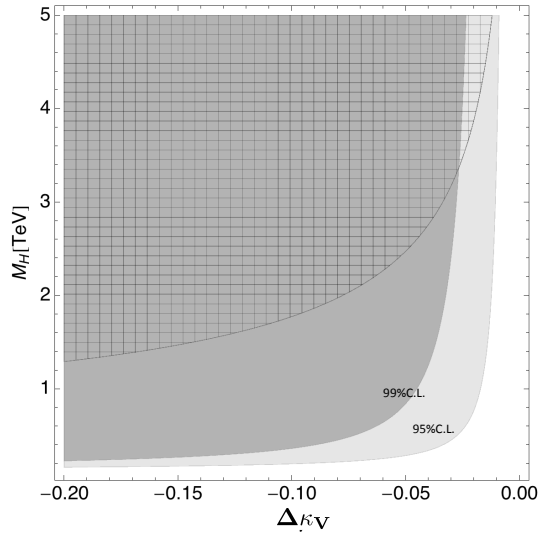}
  \caption{
Limits on the second lightest Higgs boson mass as function 
of $\Delta\kappa_V^{}\equiv \kappa_V^{} -1$.
The hatched area is disfavored from the perturbative unitarity.
The 95\%  and 99\% CL excluded areas from EWPTs are 
shown by gray and dark-gray, respectively.
}
  \label{fig:MHlimit}
\end{figure}

We are now ready to compare the unitarity limit on $M_H$ 
Eq.(\ref{eq:perturbative_unitarity}) and the EWPT limit 
shown in Figure~\ref{fig:MHonST}.
These limits on $M_H$ are depicted in Figure~\ref{fig:MHlimit} as
functions of $\Delta\kappa_V^{}% = \kappa_V - 1$
$. We note, for $-0.008 \lesssim\Delta\kappa_V<0$
($-0.03\lesssim\Delta\kappa_V<0$), the unitarity limit gives a constraint 
stronger than that of EWPTs at 95\% CL (99\% CL)\@. 
Note here that, for $M_H$ heavier than the unitarity bound, the 
theory becomes highly non-perturbative.
We cannot make reliable perturbative calculations of $S$ and $T$ 
parameters in this case. 

On the other hand, if the deviation of the Higgs-$V$-$V$ coupling
from its SM value is relatively large, {\it e.g.,} 
$\Delta\kappa_V^{} \lesssim -0.03$, then Figure~\ref{fig:MHlimit} shows 
EWPTs give a limit, $M_H \lesssim 450$GeV at 95\%CL 
($M_H \lesssim 2.4$TeV at 99\%CL), which is
stronger than the unitarity limit.
In this case, the theory remains perturbative and the bounds 
from EWPTs are considered to be trustable.

It is also interesting to compare Figure~\ref{fig:MHlimit} 
with the present
experimental value of $\kappa_V^{}$ measured for the 125GeV Higgs boson.
The ATLAS Collaboration reported
\begin{equation}
  \kappa_V^{} = 1.15 \pm 0.08,  
\label{eq:ATLAS-kappa_V-bound}
\end{equation}
in Ref.\cite{kn:ATLAS-CONF-2014-009},
while the CMS Collaboration\cite{kn:CMS-PAS-HIG-14-009} gave a bound
\begin{equation}
  \kappa_V^{} = 1.01 \pm 0.07 .
  \label{eq:CMS-kappa_V-bound}
\end{equation}
Results of ATLAS and CMS are both consistent 
with the SM value $\Delta\kappa_V=0$, though
positive $\Delta\kappa_V^{}=\kappa_V^{}-1$ is slightly favored by ATLAS, 
while CMS experiment prefers %\sout{slightly negative $\Delta\kappa_V^{}$} 
the SM prediction.  

If the positive $\Delta\kappa_V^{}$ (as favored by the present 
ATLAS result) would be established by the upgraded LHC in future,
since our model is constrained to be $\Delta\kappa_V^{} < 0$, 
then we could claim we need a framework of models to include
new particles other than the neutral Higgs bosons.
On the other hand, in the case of negative 
$\Delta\kappa_V^{}$, if the observed discrepancy were of 
order $|\Delta\kappa_V^{}| \simeq 0.02$ or below, it would be difficult to 
identify the origin of the difference.  
In this case, as shown in Figure~\ref{fig:MHlimit}, 
even a very heavy extra Higgs boson 
($M_H \gtrsim 1$TeV) can explain the EWPT result if we allow 
95\%CL uncertainty.
We are able to predict new neutral Higgs particle below 1TeV 
or less only in the case of negative $\Delta\kappa_V$ with
\begin{math}
  |\Delta\kappa_V| \gtrsim 0.02 
\end{math}.

\begin{figure}[t]
  \centering
  \includegraphics[width=85mm, bb=0 0 578 540]{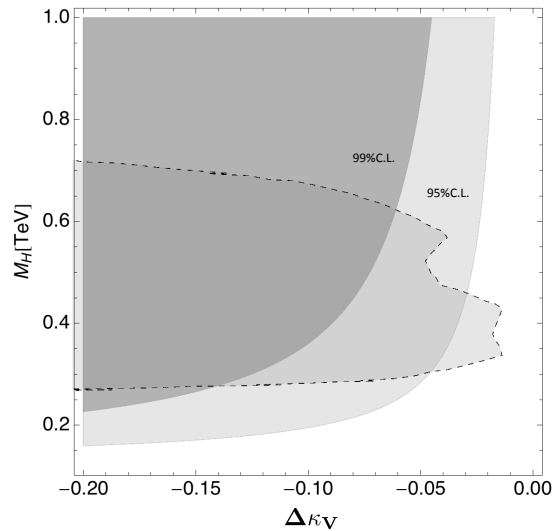}
  \caption{
Limits on the extra Higgs boson of the singlet extension 
of the SM, as function of $\Delta\kappa_V^{}\equiv \kappa_V^{} -1$.
%The hatched area is disfavored from the perturbative unitarity.
The 95\%  and 99\% CL excluded areas from EWPTs are 
shown by gray and dark-gray, respectively.
The region surrounded by the dashed contour is excluded by 
the CMS direct search\cite{kn:CMS-PAS-HIG-13-014} at the 
95\% CL\@.
}
  \label{fig:MHlimit2}
\end{figure}

\subsection{Comparison with the CMS Direct Search}

The LHC experiments continue to search for an extra heavy 
Higgs boson in various 
channels\cite{kn:ATLAS-CONF-2013-027,kn:ATLAS-CONF-2013-067,kn:ATLAS-CONF-2014-031,Chatrchyan:2013yoa,kn:CMS-PAS-HIG-13-008,kn:CMS-PAS-HIG-13-014,CMS:2013eua,kn:CMS-PAS-HIG-14-006},
after the discovery of the 125GeV Higgs 
particle.
Among them, Ref.\cite{kn:CMS-PAS-HIG-13-014} searched for the 
hypothetical heavy extra Higgs boson which arises in a singlet
extension of the SM in the 
$H\to ZZ \to 2\ell 2\nu$ channel, and gave non-trivial constraints 
in its mass-coupling plane, especially in its high mass region.
Note that the heavy Higgs coupling is related with the couplings of 
the 125GeV Higgs boson through the unitarity argument,
\begin{equation}
  (\kappa_V^h)^2 + (\kappa_V^H)^2 = 1.
\end{equation}
The constraint of Ref.\cite{kn:CMS-PAS-HIG-13-014} can therefore 
be superimposed on our Figure~\ref{fig:MHlimit},
as shown in Figure~\ref{fig:MHlimit2}.
Here we assumed that,
in addition to the bosonic amplitudes we discussed in this paper,
$Z h \to t\bar{t}$ and $Z H \to t\bar{t}$ amplitudes
are unitarized solely by two Higgs bosons (125GeV Higgs boson $h$ 
and an additional heavy Higgs boson $H$).
This assumption makes it possible to relate the $Ht\bar{t}$ coupling,
which affects the $gg \to H$ production cross section, with the
value of $\Delta\kappa_V$.
See the fermionic unitarity sum rules of Ref.\cite{Gunion:1990kf}.
It is quite interesting that, assuming the 
extra Higgs boson mass $M_H \simeq 400$ GeV, 
Figure~\ref{fig:MHlimit2} excludes 
$|\Delta \kappa_V| \gtrsim 0.016$ at the 95\% CL, 
which is stronger
than the present signal strength constraints on the 125GeV Higgs boson 
coupling $\Delta \kappa_V$ Eqs.(\ref{eq:ATLAS-kappa_V-bound}) and
(\ref{eq:CMS-kappa_V-bound}).
On the other hand, for $M_H \lesssim 300$ GeV and $M_H \gtrsim 460$ GeV, 
the strongest constraint comes from EWPTs. 
EWPTs have good sensitivity for constraining the Higgs coupling 
deviations for wider range of the extra Higgs boson mass.

\section{A UV completion and Self-interactions among Higgs bosons}
\label{sec:uv-completion}

Although the model we analyze in this paper is based on the 
non-linear sigma model, once the unitarity sum rules
Eqs.(\ref{eq:wwww}), (\ref{eq:wwzz}), (\ref{eq:wwphiphi2}), 
(\ref{eq:wwphiphi1}), (\ref{eq:zzphiphi}) and (\ref{eq:wwphiz}) 
are imposed among its Higgs coupling strengths $\kappa$s, 
the longitudinal gauge boson scattering amplitudes can be 
perturbative enough to satisfy the unitarity constraints.
Moreover, the electroweak oblique correction parameters
$S$, $T$ and $U$ are shown to be finite at one-loop level 
thanks to these unitarity sum rules.

Can the model we analyze in this paper be regarded as a renormalizable 
model, which does not need further UV completion, then? 
The answer depends on the assumptions on the Higgs self-interactions.
In this section, we take an example of $N_0=2$ to study 
what kind of constraints we need to impose among the self-interactions
of the Higgs particles, so as to make the model completely 
renormalizable.

In the case of $N_0=2$, the unitarity sum rules severely 
constrain the Higgs-gauge boson interaction Lagrangian,
\begin{eqnarray}
  {\cal L}_{\rm int} &=& 
  \dfrac{v}{2} \sum_{n=1,2} \kappa_V^{\phi_n^{0}} \phi_n^{0}
   \mbox{tr} \left[ (D_\mu U)^\dagger (D^\mu U) \right]
  \nonumber\\
  & &
  +\dfrac{1}{4}\sum_{n=1,2}\sum_{m=1,2} \kappa_V^{\phi_n^{0}}\kappa_V^{\phi_m^{0}}
   \phi_n^{0}\phi_m^{0}
   \mbox{tr} \left[ (D_\mu U)^\dagger (D^\mu U) \right],
  \nonumber\\
  & &
\label{eq:unitary_model}
\end{eqnarray}
with
\begin{equation}
  \sum_{n=1,2} (\kappa_V^{\phi_n^{0}})^2 = 1.
\end{equation}
On the other hand, Higgs self-interaction Lagrangian is left arbitrary
from the unitarity arguments:
\begin{eqnarray}
  V &=& \frac{1}{2} \sum_{n=1,2} M_{\phi_n^0}^2 \phi_n^0\phi_n^0
    % \nonumber\\
    % & &
    +\dfrac{1}{3!} \sum_{n_1^{}, n_2^{}, n_3^{}}
    \lambda_{n_1^{} n_2^{} n_3^{}} \phi_{n_1^{}}^0\phi_{n_2^{}}^0\phi_{n_3^{}}^0
   \nonumber\\
   & &
    +\dfrac{1}{4!} \sum_{n_1^{}, n_2^{}, n_3^{}, n_4^{}}
    \lambda_{n_1^{} n_2^{} n_3^{} n_4^{}} 
    \phi_{n_1^{}}^0\phi_{n_2^{}}^0\phi_{n_3^{}}^0\phi_{n_4^{}}^0,
\label{eq:12parameter_model}
\end{eqnarray}
in which we have 12 free parameters in total
(one free parameter in $\kappa_V$; two free parameters in $M_{\phi_n^0}^2$;
four in triple Higgs couplings $\lambda_{n_1^{} n_2^{} n_3^{}}$; and five in quartic
couplings $\lambda_{n_1^{} n_2^{} n_3^{} n_4^{}}$.)

In the absence of heavier particles other than these two neutral 
Higgs bosons, the model above should be described by the
doublet-singlet mixing scenario\footnote{
Ref.\cite{Henning:2014gca} studied the oblique electroweak corrections
in the doublet-singlet mixing scenario by using the effective theory 
framework.
Unitarity constraints of this model is discussed in Ref.\cite{Kang:2013zba}.
See also Ref.\cite{Gupta:2013zza} for the studies of radiative corrections 
in the doublet-singlet mixing model with an extra $U(1)$.
}, which possesses 
an $SU(2)$ doublet Higgs field $(\phi)$ and 
a real singlet Higgs field $(\sigma_2)$ with $Y=0$.
The Lagrangian of the doublet-singlet mixing scenario is given by
\begin{equation}
  {\cal L} = (D_\mu \phi)^\dagger (D^\mu \phi)
            +\dfrac{1}{2} (\partial_\mu \sigma_2)^2 - V .
\label{eq:Lsinglet1}
\end{equation}
Requiring the renormalizability, the Higgs potential $V$ should be
given by
\begin{eqnarray}
  V&=& \dfrac{\lambda}{2} 
       \left(\phi^\dagger \phi - \dfrac{1}{2}v^2\right)^2
      +\dfrac{M_{\sigma_2}^2}{2} \sigma_2^2
  % \nonumber\\
  % & & 
      +\dfrac{\lambda_{\sigma\sigma\sigma}}{3!}  \sigma_2^3
      +\dfrac{\lambda_{\sigma\sigma\sigma\sigma}}{4!}  \sigma_2^4
  \nonumber\\
  & & +\lambda_{\phi^\dagger\phi \sigma} 
       \left(\phi^\dagger \phi - \dfrac{1}{2}v^2\right) \sigma_2
  \nonumber\\
  & &
      +\frac{1}{2}\lambda_{\phi^\dagger\phi \sigma \sigma}
       \left(\phi^\dagger \phi - \dfrac{1}{2}v^2\right) \sigma_2^2.
  % \nonumber\\
  % & &
\end{eqnarray}
Minimizing the Higgs potential $V$, the doublet Higgs field 
%$\phi$ 
acquires its VEV 
\begin{equation}
  \VEV{\phi} = \left(
    \begin{array}{c}
      0 \\
      \dfrac{1}{\sqrt{2}} v 
    \end{array}
  \right) .
\end{equation}
Note that this model is described only by 6 free parameters.
In order for Eq.(\ref{eq:12parameter_model}) to be regarded as a
renormalizable theory, the free parameters in 
Eq.(\ref{eq:12parameter_model}) should satisfy $12-6=6$ constraints.

Hereafter we investigate such constraints.
For such a purpose,
we introduce the $SU(2)$ matrix field $U$,
\begin{equation}
  \phi = \dfrac{1}{\sqrt{2}}(v+\sigma_1) U \left(
    \begin{array}{c}
      0 \\
      1
    \end{array}
  \right),
\end{equation}
with $v$ being the VEV of the doublet Higgs field.
Using the chiral field $U$, 
the Lagrangian Eq.(\ref{eq:Lsinglet1}) can be rewritten as
\begin{eqnarray}
  {\cal L} &=& 
  \frac{1}{2}(\partial_\mu \sigma_1)^2
 +\frac{1}{2}(\partial_\mu \sigma_2)^2
  \nonumber\\
  & & 
  +\dfrac{v}{2} \sigma_1
   \mbox{tr} \left[ (D_\mu U)^\dagger (D^\mu U) \right]
  \nonumber\\
  & &
  +\dfrac{1}{4} \sigma_1 \sigma_1 
   \mbox{tr} \left[ (D_\mu U)^\dagger (D^\mu U) \right]
  -V,
\label{eq:uvcomplete_model}
\end{eqnarray}
with
\begin{eqnarray}
  V &=& \dfrac{1}{2} (\sigma_1, \sigma_2) M^2 \left(
        \begin{array}{c}
          \sigma_1 \\
          \sigma_2
        \end{array}
        \right) 
   \nonumber\\
   & &
       +\dfrac{\lambda}{2} v \sigma_1^3 
       +\dfrac{1}{2}\lambda_{\phi^\dagger\phi\sigma} \sigma_1^2 \sigma_2
       +\dfrac{1}{2}\lambda_{\phi^\dagger\phi\sigma\sigma} v \sigma_1 \sigma_2^2
   \nonumber\\
   & &
       +\dfrac{1}{3!} \lambda_{\sigma\sigma\sigma}\sigma_2^3
       +\dfrac{\lambda}{8} \sigma_1^4 
       +\dfrac{1}{4} \lambda_{\phi^\dagger\phi\sigma\sigma} \sigma_1^2 \sigma_2^2
   \nonumber\\
   & & 
       +\dfrac{1}{4!} \lambda_{\sigma\sigma\sigma\sigma}\sigma_2^4 .
\label{eq:6parameter_model}
\end{eqnarray}
Here the $2\times 2$ mass matrix $M^2$ is given by
\begin{equation}
  M^2 \equiv \left(
    \begin{array}{cc}
      \lambda v^2 & \lambda_{\phi^\dagger \phi \sigma} v \\
      \lambda_{\phi^\dagger \phi \sigma} v & M_{\sigma_2}^2  
    \end{array}
  \right) .
\label{eq:mass_matrix}
\end{equation}
We diagonalize the mass matrix 
Eq.(\ref{eq:mass_matrix}):
\begin{equation}
  {\cal V}^\dagger M^2 {\cal V}
  = \left(
    \begin{array}{cc}
      M_{\phi_1^0}^2 & 0 \\
      0 & M_{\phi_2^0}^2
    \end{array}
    \right),
\end{equation}
and identify
\begin{equation}
  \left(
    \begin{array}{c}
      \sigma_1 \\
      \sigma_2 
    \end{array}
  \right)
  = 
  {\cal V}
  \left(
    \begin{array}{c}
      \phi_1^0 \\
      \phi_2^0
    \end{array}
  \right),
\end{equation}
with ${\cal V}$ being an orthogonal matrix to make the mass matrix
diagonal.
Comparing the Higgs couplings in 
Eq.(\ref{eq:unitary_model}) and those 
in Eq.(\ref{eq:uvcomplete_model}), we see ${\cal V}$ should be 
expressed by $\kappa_V^{}$,
\begin{equation}
  {\cal V}
   = \left(
    \begin{array}{cc}
      \kappa_V^{\phi_1^0} & \kappa_V^{\phi_2^0} \\
     -\kappa_V^{\phi_2^0} & \kappa_V^{\phi_1^0}
    \end{array}
  \right).
\end{equation}
We next rewrite
\begin{equation}
  \left(
    \begin{array}{c}
      \phi_1^0 \\
      \phi_2^0
    \end{array}
  \right)
  = 
  {\cal V}^\dagger
  \left(
    \begin{array}{c}
      \sigma_1 \\
      \sigma_2 
    \end{array}
  \right), \quad
  {\cal V}^\dagger
   = \left(
    \begin{array}{cc}
      \kappa_V^{\phi_1^0} & -\kappa_V^{\phi_2^0} \\
      \kappa_V^{\phi_2^0} & \kappa_V^{\phi_1^0}
    \end{array}
  \right),
\label{eq:phi2sigma}
\end{equation}
and put Eq.(\ref{eq:phi2sigma}) into Eq.(\ref{eq:12parameter_model}).
We obtain
\begin{eqnarray}
  V &=& \dfrac{1}{2} (\sigma_1, \sigma_2) \tilde{M}^2 \left(
        \begin{array}{c}
          \sigma_1\\
          \sigma_2
        \end{array}
        \right)
   \nonumber\\
   & &
       +\dfrac{\tilde{\lambda}_{111}}{3!} \sigma_1^3
       +\dfrac{\tilde{\lambda}_{112}}{2} \sigma_1^2 \sigma_2
       +\dfrac{\tilde{\lambda}_{122}}{2} \sigma_1 \sigma_2^2
   \nonumber\\
   & &
       +\dfrac{\tilde{\lambda}_{222}}{3!} \sigma_2^3
       +\dfrac{\tilde{\lambda}_{1111}}{4!} \sigma_1^4
       +\dfrac{\tilde{\lambda}_{1112}}{3!} \sigma_1^3\sigma_2
    \nonumber\\
   & & 
       +\dfrac{\tilde{\lambda}_{1122}}{2! 2!} \sigma_1^2\sigma_2^2
       +\dfrac{\tilde{\lambda}_{1222}}{3!} \sigma_1^1\sigma_2^3
       +\dfrac{\tilde{\lambda}_{2222}}{4!} \sigma_2^4 .
    \nonumber\\
   & & 
\label{eq:12parameter_model2}
\end{eqnarray}
Here $\tilde{M}^2$ and $\tilde{\lambda}$ are functions
of $M^2$, $\lambda$ and $\kappa$, and are defined in
Appendix~\ref{sec:self-inter-among}. 
Comparing Eq.(\ref{eq:12parameter_model2}) with 
Eq.(\ref{eq:6parameter_model}), we find six constraints
\begin{eqnarray}
  \tilde{\lambda}_{111} &=& \dfrac{3}{v} (\tilde{M}^2)_{11}, 
\label{eq:UV-complete1}
  \\
  \tilde{\lambda}_{1111} &=& \dfrac{3}{v^2} (\tilde{M}^2)_{11}, 
  \\
  \tilde{\lambda}_{112} &=& \dfrac{1}{v} (\tilde{M}^2)_{12}, 
  \\
  \tilde{\lambda}_{1122} &=& \dfrac{1}{v} \tilde{\lambda}_{122},
  \\
  0 &=& \tilde{\lambda}_{1112}, \\
  0 &=& \tilde{\lambda}_{1222}, 
\label{eq:UV-complete6}
\end{eqnarray}
which should be satisfied to make the model UV-complete one.

\section{Effective Field Theory and Constraints on its cutoff}
\label{sec:effect-field-theory}

Varieties of effective field theory approaches have been 
proposed to describe the properties of the observed 125 
GeV Higgs particle.
In the effective field theory approaches, deviations of the 
125GeV Higgs particle are parametrized by the coefficients of 
higher dimensional operators.
These higher dimensional operators violate the perturbative
unitarity of the high energy scattering amplitudes.
They also conflict with the renormalizability of the model, 
and we need to introduce a UV cutoff in the loop 
level analysis of the effective field theory.
Perturbative unitarity and the EWPTs
are used to constrain the cutoff scale in the effective
field theory approaches.

Our approach we adopt in this paper differs from the effective
field theory approaches, since we introduce heavier
Higgs bosons other than the observed 125 GeV Higgs particle.
Moreover, the parameters of our Lagrangian are assumed to satisfy 
the unitarity sum rules, thus the scattering amplitudes are free 
from the perturbative unitarity violation even at high energies.

On the other hand, if we integrate out the heavier Higgs bosons
from our Lagrangian ({\it e.g.,} $N_0=2$ model, Eq.(\ref{eq:unitary_model})), 
we obtain an effective field theory of the 125 GeV Higgs particle:
\begin{eqnarray}
  {\cal L}_{\rm int} &=&
  \dfrac{v}{2} \kappa_V^{} h 
   \, \tr\!\! \left[ (D_\mu U)^\dagger (D^\mu U) \right]
  \nonumber\\
  & &
  +\dfrac{1}{4} \kappa_V^{hh}  hh
   \,\tr\!\! \left[ (D_\mu U)^\dagger (D^\mu U) \right],
\label{eq:effective_int}
\end{eqnarray}
with $h$ being the 125GeV Higgs particle $h=\phi_1^0$,
and
\begin{equation}
  \kappa_V^{} = \kappa_V^{\phi_1^0}, \quad
  \kappa_{V}^{hh} = \kappa_V^{\phi_1^0}\kappa_V^{\phi_1^0}  
  - \dfrac{\kappa_V^{\phi^0_2}}{M_{\phi_2^0}^2}v\lambda_{112}.
\end{equation}
Here $\lambda_{112}$ is the Higgs self-interaction coefficient
defined in Eq.(\ref{eq:12parameter_model}).\footnote{
Using the UV-completeness constraints 
Eqs.(\ref{eq:UV-complete1})--(\ref{eq:UV-complete6}), 
and the unitarity sum rule 
$(\kappa_V^{\phi_1^0})^2 + (\kappa_V^{\phi_2^0})^2 = 1$,
we are able to show $(\kappa_V^{} -1)=(\kappa_V^{hh}-1)/4$
for sufficiently large $M_{\phi_2^0}^2$.
The relation is consistent with the findings of 
Ref.\cite{Contino:2010mh} $\kappa_V^{} = 1- v^2/(2f^2) c_H$, 
$\kappa_V^{hh} = 1- 2v^2/f^2 c_H$, derived in the context of the 
Strongly Interacting Light Higgs effective theory.
}
Our approach should be understood as a systematic trial to 
construct a perturbative UV completion theory (unitary theory)
of the light Higgs effective field theory.

In this section, 
we evaluate the present constraints on the cutoff scale in the 
effective field theory using the perturbative unitarity and the
results of the EWPTs.
We then compare the cutoff constraints in the effective field theory 
method with our findings on the heavy Higgs boson mass bounds
in our approach.

\subsection{Unitarity constraints}
\label{sec:unit-constr-1}

In the effective field theory Eq.(\ref{eq:effective_int}), 
the deviation of the Higgs coupling $\kappa_V^{}$ from its SM
value affects the longitudinal gauge boson scattering amplitudes 
to violate the perturbative unitarity constraint at high energy
scale.
This is one of the reasons why we need to introduce a UV 
cutoff scale in the effective field theory framework.
We estimate the upper bound of the cutoff scale $\Lambda$ from the
$S$-wave amplitudes,
\begin{eqnarray}
  t_0^{W_L^+ W_L^- \to W_L^+ W_L^-}
   &\simeq&  \dfrac{1}{2} \tilde{{\mathscr T}}, 
  \\
  t_0^{Z_L Z_L \to W_L^+W_L^-}
   &\simeq& \dfrac{1}{\sqrt{2}} \tilde{{\mathscr T}},
  \\
  t_0^{Z_LZ_L \to Z_L Z_L}
   &\simeq& 0 , 
\end{eqnarray}
for $s \gg M_h^2$.
Here $\tilde{{\mathscr T}}$ is given by
\begin{equation}
  \tilde{{\mathscr T}} \equiv \dfrac{G_F}{8\sqrt{2} \pi} (1-\kappa_V^2) s, 
%    \qquad
%  G_F = \dfrac{1}{\sqrt{2} v^2} ,
\end{equation}
with $s$ being the square of the energy of the scattering.

The $S$-wave transition matrix among $W_L^+ W_L^-$ and $Z_L Z_L$ states
is
\begin{eqnarray}
  {\cal T} 
  &=& \left(
    \begin{array}{cc}
      t_0^{W_L^+ W_L^- \to W_L^+ W_L^-} & t_0^{W_L^+ W_L^- \to Z_L Z_L}  \\
      t_0^{Z_L Z_L \to W_L^+ W_L^-}  & t_0^{Z_L Z_L \to Z_L Z_L}
    \end{array}
  \right)
  \nonumber\\
  &=& 
      \left(
        \begin{array}{cc}
          \dfrac{1}{2}\tilde{{\mathscr T}} & \dfrac{1}{\sqrt{2}}\tilde{{\mathscr T}} \\
          \dfrac{1}{\sqrt{2}}\tilde{{\mathscr T}} & 0
        \end{array}
      \right) ,
\end{eqnarray}
and we obtain the maximum eigenvalue of the ${\cal T}$ 
matrix
\begin{equation}
  t_0^{\rm max} = \tilde{{\mathscr T}} .
\end{equation}
The perturbative unitarity requires $|t_0^{\rm max}|<1/2$,
and we thus find
\begin{equation}
  (1-\kappa_V^2) \Lambda^2 < 8\pi v^2.
\label{eq:cutoff_unitarity}
\end{equation}
Here we identified the cutoff scale $\Lambda$ as the
scattering energy scale below which the amplitudes
can be safely evaluated by using the effective theory 
framework.

Comparing Eq.(\ref{eq:cutoff_unitarity}) with 
Eq.(\ref{eq:perturbative_unitarity}), we find
the upper bound of the heavy Higgs boson mass $M_H$ as we discussed in
Sec.~\ref{sec:unit-constr} can be related with the upper bound of the
effective field theory cutoff scale $\Lambda$ :
\begin{equation}
  M_H^{\rm upper} = \sqrt{\dfrac{2}{5}} \Lambda^{\rm upper}.
\end{equation}
Noting 
\begin{equation}
  \sqrt{\dfrac{2}{5}}  \simeq 0.63, 
\end{equation}
we see that, in our model, the upper bound on the extra Higgs mass 
$M_H$ is a bit tighter than the estimation of the cutoff scale in 
the effective field theory framework.

\subsection{Electroweak Precision Tests}
\label{sec:electr-prec-tests}

We next turn to the electroweak precision constraint on the cutoff 
$\Lambda$ scale in the effective field theory approach.
Using the results of Sec.~\ref{sec:oblique}, it is straightforward
to evaluate the oblique correction parameters from the effective field
theory Lagrangian Eq.(\ref{eq:effective_int}),
\begin{eqnarray}
  S&=& \dfrac{1}{4\pi}(1-\kappa_V^2) \left[
         \frac{1}{3} \ln  \dfrac{\Lambda^2}{\mu^2} 
        -{G^{Zh}}'
       \right],
 \\
  T&=& -\dfrac{3(1-\kappa_V^2)}{16\pi^2 v^2 \alpha} (M_Z^2-M_W^2) \Biggl[
    \ln \dfrac{\Lambda^2}{\mu^2} 
   \nonumber\\
   & & \quad -\dfrac{1}{3} 
             -\dfrac{1}{3} \dfrac{G^{Zh}-G^{Wh}}{M_Z^2-M_W^2}
   \nonumber\\
   & & \quad
             +\dfrac{1}{3(M_Z^2-M_W^2)}\left[
               M_Z^2 \ln \dfrac{M_Z^2}{\mu^2} - M_W^2 \ln \dfrac{M_W^2}{\mu^2}
              \right]
  \Biggr], 
  \nonumber\\
  & &
  \\
  U&=& \dfrac{1}{4\pi}(1-\kappa_V^2) \left[
    {G^{Zh}}' - {G^{Wh}}'
  \right],
\end{eqnarray}
with $\Lambda$ being the UV cutoff scale 
as we define in Appendix.~\ref{sect-loop_integrals}.
The finite parts in the above formulas can be easily
evaluated, and we obtain
\begin{eqnarray}
  S&\simeq& \dfrac{1}{12\pi}(1-\kappa_V^2) \left[ 
      \ln \dfrac{\Lambda^2}{M_h^2}+ 1.69 \right],
\label{eq:cutoff-S}
    \\
  T&\simeq& -\dfrac{3(1-\kappa_V^2)}{16\pi^2\alpha} 
      \dfrac{M_Z^2-M_W^2}{v^2} \left[
        \ln\dfrac{\Lambda^2}{M_h^2} - 0.22
      \right],
\nonumber\\
  & &
\label{eq:cutoff-T}
    \\
  U &\simeq& \dfrac{1-\kappa_V^2}{3\pi} \times (-0.028) .
\end{eqnarray}
Again, we used $M_Z=91.2$ GeV, $M_W=80.4$ GeV %, $M_h=125$ GeV
in the estimates of the finite parts.

It should be emphasized, however, that the definition of the
cutoff parameter $\Lambda$ in the loop integrals
is not unique.
There is a non-negligible ambiguity in the size of finite
corrections in Eqs.(\ref{eq:cutoff-S}) and (\ref{eq:cutoff-T}).
Actually, Refs.\cite{Espinosa:2012im,Baak:2014ora} neglect these
finite corrections and use simpler form,
\begin{eqnarray}
  S&\simeq& \dfrac{1}{12\pi}(1-\kappa_V^2) 
      \ln \dfrac{\Lambda^2}{M_h^2},
\label{eq:cutoff-S2}
    \\
  T&\simeq& -\dfrac{3(1-\kappa_V^2)}{16\pi^2\alpha} 
      \dfrac{M_Z^2-M_W^2}{v^2} 
        \ln\dfrac{\Lambda^2}{M_h^2}.
\label{eq:cutoff-T2}
\end{eqnarray}
Note that the $T$-parameter constraint is more stringent 
than the $S$-parameter.
Comparing Eq.(\ref{eq:cutoff-T2}) with Eq.(\ref{eq:T-para2}),
we find
\begin{equation}
  M_H^{\rm upper} \simeq 1.69 \times \Lambda^{\rm upper},
\end{equation}
with $1.69\simeq e^{1.05/2}$.
We see that, in the electroweak precision constraints,
the upper bound on $M_H$ is a bit weaker than the corresponding
bound on $\Lambda$ of the effective field theory framework.

\section{Conclusions and Outlook}

In this paper we discussed how the unitarity of the 
longitudinal gauge boson scattering amplitudes is related
with the finiteness of the electroweak oblique parameters
$S$, $T$, and $U$.
Starting from general Lagrangian of the electroweak symmetry 
breaking sector with arbitrary number of neutral Higgs bosons,
we (re)derived the unitarity  sum rules among Higgs couplings,
which should be satisfied to keep the longitudinal gauge boson 
scattering amplitudes unitary at high energy.
The unitarity arguments allow us to show, without invoking
the custodial symmetry explicitly,  the tree-level $\rho$
parameter to be unity in any unitary 
EWSB model if it only possesses 
neutral Higgs bosons.   
This finding explains the reason of 
the $\rho$ parameter stability against the radiative corrections
in the septet Higgs extension model which doesn't enjoy explicit 
custodial symmetry\cite{Hisano:2013sn,Kanemura:2013mc,Alvarado:2014jva}.
Thanks to the electroweak chiral Lagrangian framework we used,
the electroweak gauge symmetry is kept manifest, which allows
us to investigate the one-loop radiative corrections to the
electroweak oblique parameters explicitly at the one-loop level.
We showed the finiteness of the oblique parameters is 
automatically guaranteed in our framework, once we impose the
unitarity sum rules among various Higgs couplings.

We also derived upper bounds on the second lightest Higgs boson
mass $M_H$ as functions of the deviation of the 125GeV Higgs
boson coupling $\Delta\kappa_V$.
We found, for $\Delta\kappa_V \lesssim -0.008$ 
($\Delta\kappa_V \lesssim -0.03$), the oblique 
parameter constraint at 95\% CL (99\% CL) gives more stringent bound on $M_H$
than the unitarity bound. 
The result of the LHC direct search of the second lightest
Higgs boson can also be combined, and 
we found a constraint on $\Delta\kappa_V$ tighter than 
the present signal strength uncertainty of the 125GeV Higgs boson measurements.
The combined results with the LHC direct search 
 give 
the strongest bound on $\kappa_V^{}$ for $M_H^{} \simeq 400$ GeV, 
while for the wide range of $M_H^{}$ region EWPTs have the best 
sensitivity. 

Finally, we compared our bounds on $M_H$ with the bounds
on the UV cutoff $\Lambda$ of the effective field theory 
approach.
Simple relationships were found between $M_H$ and $\Lambda$
bounds both in the unitarity and the oblique parameter arguments.

It should be emphasized, however, that our results
heavily rely on the assumption we made: the EWSB is perturbatively
realized only with additional neutral Higgs bosons.
We need to relax our model to include, {\it e.g.,} charged Higgs
bosons so as to make our analysis applicable to wider class of
EWSB models, 
including the triplet Higgs 
extensions\cite{Georgi:1985nv,Chanowitz:1985ug,Gunion:1989ci,Gunion:1990dt}
and the septet Higgs extensions\cite{Hisano:2013sn,Kanemura:2013mc,Alvarado:2014jva}.
It will also be interesting to utilize the Yukawa coupling 
unitarity sum rules which can be derived from the amplitudes
involving heavy fermions.
We are now preparing a complete set of the unitarity sum rules
and the oblique parameter formulas in the models including 
arbitrary number of charged Higgs bosons.  The results
will be published elsewhere.

Possibility of non-perturbative EWSB should also be investigated,
since the present experimental results still allow such a 
possibility.
For an example, as we discussed in Sec.~\ref{sec-application},
the wrong sign $\kappa_{ZZ}^h \kappa_{WW}^h$ is consistent with
the present measurements of 125GeV Higgs particle and the EWPTs.
The present measurements are sensitive only to $|\kappa_{ZZ}^h|^2$
and $|\kappa_{WW}^h|^2$, not to its relative sign.
The sign should be determined by measuring $WW\to ZZ$ cross section
at the future LHC experiments.

We finally emphasize that the 125GeV Higgs coupling measurements,
the precision oblique parameter measurements, and
the direct search of the extra Higgs bosons give complimentary
limits on the model.
Future precision measurements of these parameters at the ILC 
experiments will be able to pin down the direction of the 
new physics beyond the standard model.

\section*{Acknowledgments}
We thank Junji Hisano and Kazuhiro Tobe for useful discussions.
R.N.'s work is supported by Research Fellowships of the Japan Society 
for the Promotion of Science (JSPS) for Young Scientists No.263947.
M.T.'s work is supported in part by the JSPS Grant-in-Aid 
for Scientific Research No.23540298 and No.22224003. 
K.T.'s work is supported in part by the MEXT 
Grant-in-Aid for Scientific Research on Innovative Areas 
No. 26104704. 

\appendix

\section{Scattering amplitudes at tree level in the gaugeless limit}
\label{sec:amp_NGB}

In this appendix, we evaluate the would-be NGB two-body 
scattering amplitudes in the gaugeless limit ($g=g_Y^{}=0$) 
in the model discussed in Sec.\ref{sec:The model}. 
The equivalence theorem between the longitudinally polarized 
vector boson amplitudes and the NGB 
amplitudes then enables us to evaluate the longitudinally polarized
vector boson amplitudes in the high energy limit.

We first consider the amplitude
\begin{equation}
  w^+(p_1)\, w^-(p_2)\to w^+(p_3)\, w^-(p_4).
\end{equation}
Note that the NGBs are massless in the gaugeless limit.
We find
\begin{eqnarray}
\lefteqn{
\mathcal{A}_{w^+w^-\to w^+w^-}%(p_1,p_2,p_3,p_4)
} \nonumber\\
 &=& -\dfrac{1}{v^2} \left(4-3\dfrac{v_Z^2}{v^2}\right) u
     -\dfrac{1}{v^2} \sum_{m=1}^{N_0} (\kappa_{WW}^{\phi_m^0})^2 
                     \dfrac{t^2}{t-M_{\phi_m^0}^2}
\nonumber\\
  & & \qquad
     -\dfrac{1}{v^2} \sum_{m=1}^{N_0} (\kappa_{WW}^{\phi_m^0})^2 
                     \dfrac{s^2}{s-M_{\phi_m^0}^2},
\label{eq:amp_ww2ww}
\end{eqnarray}
with $s$, $t$ and $u$ being the usual Mandelstam variables
\begin{eqnarray*}
  s &\equiv& (p_1+p_2)^2 = (p_3+p_4)^2 , \\
  t &\equiv& (p_1-p_3)^2 = (p_2-p_4)^2 , \\
  u &\equiv& (p_1-p_4)^2 = (p_2-p_3)^2.
\end{eqnarray*}
The factor $(4-3v_Z^2/v^2)$ in the first term of 
Eq.(\ref{eq:amp_ww2ww}) agrees with the low energy theorem of 
$SU(2)\times U(1)/U(1)$ NGB scattering. It 
arises from the corresponding factor in the contact four-NGB 
vertex given in Eq.(\ref{eq:EWchirallag}).%(\ref{eq:c_vvvv}).
~The second and third terms in Eq.(\ref{eq:amp_ww2ww}) come from
the $t$- and $s$-channel exchanges of the neutral Higgs bosons, 
respectively.
We next consider the amplitude of 
\begin{equation}
  w^+(p_1)\, w^-(p_2)\to z(p_3)\, z(p_4) .
\end{equation}
It should be noted the existence of the $wwz$ vertex in
the second term of Eq.(\ref{eq:EWchirallag})%(\ref{eq:c_vvv}) 
~produces $t$- and $u$-channel 
$w$-exchange (NGB exchange) diagrams when 
$v_Z^2\ne v^2$.
The NGB pole cancels with the numerator 
at the on-shell $p_1^2 =  p_2^2 =  p_3^2 =  p_4^2 =0$
in these NGB exchange amplitudes.
Combined with the four-NGB contact interaction 
Eq.(\ref{eq:EWchirallag}),%(\ref{eq:c_vvvv}),
~these NGB exchange amplitude
reproduce the low energy theorem amplitude of $SU(2)\times U(1)/U(1)$ 
symmetry breaking.  
We now obtain
\begin{eqnarray}
\lefteqn{
\mathcal{A}_{w^+w^-\to zz}%(p_1,p_2,p_3,p_4)
} \nonumber\\
 &=& \dfrac{v_Z^2}{v^4} s - \dfrac{1}{v_Z^2}\sum_{m=1}^{N_0} 
     (\kappa_{WW}^{\phi_m^0})(\kappa_{ZZ}^{\phi_m^0})
     \dfrac{s^2}{s-M_{\phi_m^0}^2},
\label{eq:amp_ww2zz}
\end{eqnarray}
where the first term is the low energy theorem amplitude, while
the second term comes from the $s$-channel Higgs exchange diagram.

Due to the lack of the low energy theorem amplitude, 
the amplitude
\begin{equation}
  z(p_1)\, z(p_2)\to z(p_3)\, z(p_4) 
\end{equation}
behaves ${\cal O}(E^4)$ at low energy.  We find
\begin{eqnarray}
\lefteqn{
\mathcal{A}_{zz \to zz}%(p_1,p_2,p_3,p_4)
} \nonumber\\
 &=& - \dfrac{v^2}{v_Z^4}\sum_{m=1}^{N_0} 
     (\kappa_{ZZ}^{\phi_m^0})(\kappa_{ZZ}^{\phi_m^0})
     \times
 \nonumber\\
 & & \times
     \left(
       \dfrac{s^2}{s-M_{\phi_m^0}^2}
      +\dfrac{t^2}{t-M_{\phi_m^0}^2}
      +\dfrac{u^2}{u-M_{\phi_m^0}^2}
     \right).
\nonumber\\
& &
\label{eq:amp_zz2zz}
\end{eqnarray}

We next consider the amplitude
\begin{equation}
  w^-(p_1)\, w^+(p_2)\to\phi^{0}_{n_1^{}}(p_3)\, \phi^{0}_{n_2^{}}(p_4) ,
\end{equation}
which can be evaluated from the contact interaction terms
Eqs.(\ref{eq:lagphidelphi})-(\ref{eq:lagphiphi})%(\ref{eq:c_vvphiphi1})-(\ref{eq:c_vvphiphi2}) 
~and the $t$- and $u$-channel $w$ 
exchange graphs arising from Eq.(\ref{eq:lagphi}).%Eq.(\ref{eq:c_vvphi}).

We also note that there exists an $s$-channel Higgs 
exchange contribution arising from triple-Higgs couplings.  
The $s$-channel Higgs exchange graph, however, does
not grow up in high energy limit, and we neglect it
in this appendix.
We obtain
\begin{eqnarray}
\lefteqn{
\mathcal{A}_{w^-w^+\to\phi^{0}_{n_1^{}}\phi^{0}_{n_2^{}}}%(p_1,p_2,p_3,p_4)
} \nonumber\\
 &=& -\dfrac{i}{v^2} \kappa_Z^{\phi_{n_1^{}}^0\phi_{n_2^{}}^0}(t-u)
     -\dfrac{1}{v^2} \kappa_{WW}^{\phi_{n_1^{}}^0\phi_{n_2^{}}^0} s
 \nonumber\\
 & & -\dfrac{1}{v^2}
      \kappa_{WW}^{\phi_{n_1^{}}^0}\kappa_{WW}^{\phi_{n_2^{}}^0}
      \dfrac{[t-M_{\phi_{n_1^{}}^0}^2][t-M_{\phi_{n_2^{}}^0}^2]}{t} 
 \nonumber\\
 & & -\dfrac{1}{v^2}
      \kappa_{WW}^{\phi_{n_1^{}}^0}\kappa_{WW}^{\phi_{n_2^{}}^0}
      \dfrac{[u-M_{\phi_{n_1^{}}^0}^2][u-M_{\phi_{n_2^{}}^0}^2]}{u}  .
\label{eq:amp_ww2phiphi}
\end{eqnarray}
Note here that the $P$-wave final state is present 
when $\kappa_Z^{\phi^0_{n_1^{}}\phi^0_{n_2^{}}} \ne 0$.
We also note the imaginary number in the amplitude is the 
result of $CP$ violation arising from the simultaneous
existence of $\kappa_Z^{\phi_{n_1^{}}^0\phi_{n_2^{}}^0}\ne 0$
and $\kappa_{WW}^{\phi_{n_1^{}}^0}\kappa_{WW}^{\phi_{n_2^{}}^0} \ne 0$.

%~~~~~~~~~~~~~~~~~~~~~~~~~~~~~~~~~~~~~~~~~~~~~~~~~~

The amplitude
\begin{equation}
  z(p_1)\,z(p_2)\to\phi^{0}_{n_1}(p_3)\,\phi^{0}_{n_2}(p_4) 
\end{equation}
can also be evaluated in a similar manner. We find
\begin{eqnarray}
\lefteqn{
\mathcal{A}_{zz\to\phi^{0}_{n_1}\phi^{0}_{n_2}}%(p_1,p_2,p_3,p_4)
} \nonumber\\
  &=& -\dfrac{1}{v_Z^2} \kappa_{ZZ}^{\phi_{n_1}^0 \phi_{n_2}^0} s
  \nonumber\\
  & &  +\dfrac{1}{v_Z^2} \sum_m 
       \kappa_Z^{\phi_{n_1}^0 \phi_{m}^0}\kappa_Z^{\phi_{m}^0\phi_{n_2}^0}
       \dfrac{[t-M_{\phi_{n_1}^0}^2][t-M_{\phi_{n_2}^0}^2]}
             {t-M_{\phi_m^0}^2}
  \nonumber\\
  & &  +\dfrac{1}{v_Z^2} \sum_m 
       \kappa_Z^{\phi_{n_1}^0 \phi_{m}^0}\kappa_Z^{\phi_{m}^0\phi_{n_2}^0}
       \dfrac{[u-M_{\phi_{n_1}^0}^2][u-M_{\phi_{n_2}^0}^2]}
                   {u-M_{\phi_m^0}^2}
  \nonumber\\
  & &  -\dfrac{v^2}{v_Z^4} \kappa_{ZZ}^{\phi_{n_1}^0}\kappa_{ZZ}^{\phi_{n_2}^0}
       \dfrac{[t-M_{\phi_{n_1}^0}^2][t-M_{\phi_{n_2}^0}^2]}
                   {t}
  \nonumber\\
  & &  -\dfrac{v^2}{v_Z^4} \kappa_{ZZ}^{\phi_{n_1}^0}\kappa_{ZZ}^{\phi_{n_2}^0}
       \dfrac{[u-M_{\phi_{n_1}^0}^2][u-M_{\phi_{n_2}^0}^2]}
                   {u} .
\label{eq:amp_zz2phiphi}
\end{eqnarray}

We finally consider the amplitude
\begin{equation}
  {w^+(p_1)\,w^-(p_2)\to\phi^{0}_n(p_3)\,z(p_4)} .
\end{equation}
Evaluating $t$- and $u$-channel $w$ exchange graphs, contact 
interaction graphs, and the $s$-channel Higgs exchange graph,
we obtain 
\begin{eqnarray}
\lefteqn{
\mathcal{A}_{w^+w^-\to \phi^{0}_n z}%(p_1,p_2,p_3,p_4)
} \nonumber\\
 &=& -\dfrac{i}{vv_Z^{}} \left(
        \dfrac{v_Z^2}{v^2} \kappa_{WW}^{\phi^0_{n}} - \kappa_{ZZ}^{\phi^0_{n}} 
      \right)
      (t-u)
 \nonumber\\
 & &+\dfrac{1}{v v_Z^{}} \sum_{m=1}^{N_0} \kappa_Z^{\phi^0_n \phi^0_m} \kappa_{WW}^{\phi^0_m}
     \dfrac{[s-M_{\phi_n^0}^2]s}{s-M_{\phi_m^0}^2}.
\label{eq:amp_ww2phiz}
\end{eqnarray}
Again, the imaginary number in the amplitude is a consequence of 
the $CP$ violating coupling of the ``Higgs'' bosons.

In a similar manner, 
\begin{equation}
  {z(p_1)\,z(p_2)\to\phi^{0}_n(p_3)\,z(p_4)}
\end{equation}
amplitude can be evaluated from the Higgs exchange graphs.  We obtain
\begin{eqnarray}
\lefteqn{
\mathcal{A}_{zz\to \phi^{0}_n z}%(p_1,p_2,p_3,p_4)
} \nonumber\\
 &=& \dfrac{v}{v_Z^3} \sum_{m=1}^{N_0} \kappa_Z^{\phi^0_n \phi^0_m} \kappa_{ZZ}^{\phi^0_m}
     \dfrac{[s-M_{\phi_n^0}^2]s}{s-M_{\phi_m^0}^2}
 \nonumber\\
  & & +\dfrac{v}{v_Z^3} \sum_{m=1}^{N_0} \kappa_Z^{\phi^0_n \phi^0_m} \kappa_{ZZ}^{\phi^0_m}
     \dfrac{[t-M_{\phi_n^0}^2]t}{t-M_{\phi_m^0}^2}
 \nonumber\\
  & & +\dfrac{v}{v_Z^3} \sum_{m=1}^{N_0} \kappa_Z^{\phi^0_n \phi^0_m} \kappa_{ZZ}^{\phi^0_m}
     \dfrac{[u-M_{\phi_n^0}^2]u}{u-M_{\phi_m^0}^2} .
\label{eq:amp_zz2phiz}
\end{eqnarray}

\section{Evaluating $\tilde{\Pi}_{33}(0)$ and $\tilde{\Pi}_{11}(0)$}
\label{sec:eval3311}

In order to evaluate the vacuum polarization functions 
$\tilde{\Pi}_{33}(0)$ and $\tilde{\Pi}_{11}(0)$ in the electroweak 
gauged chiral Lagrangian Eq.(\ref{eq:gauge}) and Eq.(\ref{eq:EWchirallag}),
it is convenient to introduce the background field formalism.
See, {\it e.g.,} Appendix A.2 of Ref.\cite{Sekhar Chivukula:2007ic}.

We decompose the chiral field $U$ 
into background field $\bar{U}$ and dynamical fields $u^1, u^2, u^z$,
\begin{equation}
  U = \bar{U} \exp \left[ \dfrac{i(u_1 \tau_1 + u_2 \tau_2)}{v} \right]
              \exp \left[ \dfrac{i u_z \tau_3}{v_Z^{}} \right].
\end{equation}
The gauge fields ${\bf W}_\mu$ and ${\bf B}_\mu$ are also decomposed as,
\begin{equation}
  {\bf B}_\mu = \bar{\bf B}_\mu + b_\mu \dfrac{\tau_3}{2} , 
\end{equation}
and
\begin{equation}
  {\bf W}'_\mu = \bar{U}^\dagger {\bf W}_\mu \bar{U}
                -\dfrac{i}{g}\bar{U}^\dagger \partial_\mu \bar{U}
              = \bar{\bf W}_\mu + \sum_{a=1}^3 w^a_{\mu} \dfrac{\tau_a}{2} ,
\end{equation}
with
\begin{equation}
  \bar{\bf B}_\mu = \bar{B}_\mu \dfrac{\tau_3}{2}, 
  \qquad
  \bar{\bf W}_\mu = \sum_{a=1}^3 \bar{W}_\mu^a \dfrac{\tau_a}{2}.
\end{equation}
Here the background gauge fields are denoted by $\bar{\bf B}_\mu$
and $\bar{\bf W}_\mu$, while the quantum fields are $b_\mu$ and $w_\mu$.
In order to evaluate radiative corrections, we introduce
gauge fixing Lagrangian,
\begin{eqnarray}
  {\cal L}_{\rm GF} &=& 
    -\dfrac{1}{2\xi} \left[ (D_\mu w^\mu)_1 - \xi g \dfrac{v}{2} u_1 \right]^2 
  \nonumber\\
    & &
    -\dfrac{1}{2\xi} \left[ (D_\mu w^\mu)_2 - \xi g \dfrac{v}{2} u_2 \right]^2 
  \nonumber\\
    & &
    -\dfrac{1}{2\xi} \left[ (D_\mu w^\mu)_3 - \xi g \dfrac{v_Z^{}}{2} u_z \right]^2 
  \nonumber\\
    & &
    -\dfrac{1}{2\xi} \left[ \partial_\mu b^\mu 
            + \xi g_Y^{} \dfrac{v_Z^{}}{2} u_z \right]^2  ,
\label{eq:gaugefixing}
\end{eqnarray}
with
\begin{equation}
  (D_\mu w_\nu)_a \equiv \partial_\mu w_\nu^a  - g\epsilon^{abc} \bar{W}_\mu^b 
  w_\nu^c.
\end{equation}

The Lagrangian ${\cal L}_\chi$, Eq.(\ref{eq:EWchirallag}), is expanded
in terms of the fluctuating quantum field $u$.
We find the bilinear terms of $u$ can be summarized in a compact expression,
\begin{equation}
  \left. {\cal L}_{\chi} \right|_{uu}
 +\left. {\cal L}_{\rm GF} \right|_{uu}
  = \frac{1}{2} {}^t(D_\mu u) (D^\mu u) - \frac{1}{2} {}^t u \sigma u,
\label{eq:lag2_uu}
\end{equation}
with
\begin{equation}
  u \equiv \left(\begin{array}{c} u_1 \\ u_2 \\ u_z
        \end{array}
  \right) .
\end{equation}
In Eq.(\ref{eq:lag2_uu}), $D_\mu u$ is defined as
\begin{equation}
  D_\mu u \equiv \partial_\mu u + \Gamma_\mu u, 
\end{equation}
with
\begin{equation}
  \Gamma_\mu = \left(
    \begin{array}{ccc}
      \Gamma^{11}_\mu & \Gamma^{12}_\mu & \Gamma^{1z}_\mu \\
      \Gamma^{21}_\mu & \Gamma^{22}_\mu & \Gamma^{2z}_\mu \\
      \Gamma^{z1}_\mu & \Gamma^{z2}_\mu & \Gamma^{zz}_\mu \\
    \end{array}
  \right),
\end{equation}
\begin{equation}
  \Gamma^{11}_\mu = \Gamma^{22}_\mu = \Gamma^{zz}_\mu = 0, 
\end{equation}
\begin{equation}
  \Gamma^{12}_\mu = -\Gamma^{21}_\mu = 
  \dfrac{1}{2} \left(2-\dfrac{v_Z^2}{v^2}\right) g\bar{W}_{\mu}^3
       +\dfrac{1}{2} \dfrac{v_Z^2}{v^2} g_Y^{} \bar{B}_{\mu},
\end{equation}
\begin{equation}
  \Gamma^{1z}_\mu = -\Gamma^{z1}_\mu = 
  -\dfrac{v_Z^{}}{2v} g\bar{W}_{\mu}^2, 
\end{equation}
and
\begin{equation}
  \Gamma^{2z}_\mu = -\Gamma^{z2}_\mu = 
  \dfrac{v_Z^{}}{2v}  g\bar{W}_{\mu}^1 .
\end{equation}
Similarly, the matrix $\sigma$ is given by
\begin{equation}
  \sigma = \left(
    \begin{array}{ccc}
      \sigma_{11} & \sigma_{12} & \sigma_{1z} \\
      \sigma_{21} & \sigma_{22} & \sigma_{2z} \\
      \sigma_{z1} & \sigma_{z2} & \sigma_{zz}
    \end{array}
  \right),
\end{equation}
with
\begin{eqnarray}
  \sigma_{11} &=&
     \dfrac{1}{4}\left(4-3\dfrac{v_Z^2}{v^2}\right) 
     g^2 \bar{W}_{\mu}^2 \bar{W}^{2\mu}
    \nonumber\\
  & & 
    +\dfrac{1}{4} \dfrac{v_Z^4}{v^4} 
     (g\bar{W}_{\mu}^3 - g_Y^{}\bar{B}_{\mu})
     (g\bar{W}^{3\mu} - g_Y^{}\bar{B}^{\mu})
  +\xi M_W^2,
  \nonumber\\
  & &
  \\
  \sigma_{22} &=&
     \dfrac{1}{4}\left(4-3\dfrac{v_Z^2}{v^2}\right) 
     g^2 \bar{W}_{\mu}^1 \bar{W}^{1\mu}
  \nonumber\\
  & &
    +\dfrac{1}{4} \dfrac{v_Z^4}{v^4} 
     (g\bar{W}_{\mu}^3 -g_Y^{}\bar{B}_{\mu}^3)
     (g\bar{W}^{3\mu} -g_Y^{}\bar{B}^{\mu})
    +\xi M_W^2,
  \nonumber\\
  & &
  \\
  \sigma_{zz} &=&
    \dfrac{v_Z^2}{4v^2} g^2(\bar{W}_{\mu}^1 \bar{W}^{1\mu}
                       +\bar{W}_{\mu}^2 \bar{W}^{2\mu})
   +\xi M_Z^2,
  \nonumber\\
  & &
  \\
  \sigma_{12} &=& \sigma_{21}
  \nonumber\\
  &=& -\dfrac{1}{4}\left(4-3\dfrac{v_Z^2}{v^2}\right) 
       g^2 \bar{W}_{\mu}^1 \bar{W}^{2\mu},
  \\
  \sigma_{1z} &=& \sigma_{z1}
  \nonumber\\
  &=& 
    -\dfrac{1}{4} \dfrac{v_Z^3}{v^3} 
     g\bar{W}_{\mu}^1 (g\bar{W}^{3\mu} - g_Y^{}\bar{B}^{\mu}), 
  \\
  \sigma_{2z} &=& \sigma_{z2}
  \nonumber\\
  &=&
    -\dfrac{1}{4} \dfrac{v_Z^3}{v^3} 
     g\bar{W}_{\mu}^2 (g\bar{W}^{3\mu} - g_Y^{}\bar{B}^{\mu}) ,
\end{eqnarray}
with
\begin{equation}
  M_W^2 = \dfrac{g^2}{4} v^2, \qquad
  M_Z^2 = \dfrac{g^2+g_Y^2}{4} v_Z^2.
\end{equation}
In the derivation of Eq.(\ref{eq:lag2_uu}), 
we used equations of motion of the background field. 

The bilinear terms of $w^a_\mu$ and $b_\mu$ are
\begin{eqnarray}
\lefteqn{
  \left. {\cal L}_\chi \right|_{vv}
 +\left. {\cal L}_{\rm gauge} \right|_{vv}
 +\left. {\cal L}_{\rm GF} \right|_{vv} = 
} \nonumber\\
  & &   
-\dfrac{1}{2} (D_\mu w_{\nu})^a (D^\mu w^{\nu})^a 
     +\dfrac{1}{2} \left(1-\dfrac{1}{\xi}\right) 
      (D_\mu w^{\mu})^a (D_\nu w^{\nu})^a 
 \nonumber\\
 & & + \epsilon^{abc} g \bar{W}_{\mu\nu}^a w^{b\mu} w^{c\nu}
 \nonumber\\
 & & - \frac{1}{2} (\partial_\mu b_\nu) (\partial^\mu b^\nu)
     +\frac{1}{2} \left(1-\dfrac{1}{\xi}\right) 
      (\partial_\mu b^{\mu}) (\partial_\nu b^{\nu})
 \nonumber\\
  & & +\dfrac{g^2 v^2}{8} \sum_{a=1,2} w_{\nu}^a w^{a\nu}
      +\dfrac{v_Z^2}{8} (g w^3_{\nu}  - g_Y^{} b_{\nu}) 
                        (g w^{3\nu}  - g_Y^{} b^{\nu}) .
 \nonumber\\
  & &
\label{eq:lag2_vv}
\end{eqnarray}
We also find
\begin{eqnarray}
\lefteqn{
  \left. {\cal L}_\chi \right|_{uv}
% +\left. {\cal L}_{\rm gauge} \right|_{uv}
 +\left. {\cal L}_{\rm GF} \right|_{uv}
 =
} \nonumber\\
  & & - g^2 \dfrac{v}{2} \left(2-\dfrac{v_Z^2}{v^2}\right)
          \left(
             \bar{W}_{\mu}^2 w^{3\mu} u_1
           - \bar{W}_{\mu}^1 w^{3\mu} u_2
          \right)
  \nonumber\\
  & &  -gg_Y \dfrac{v_Z^2}{2v} 
          \left(
             \bar{W}_{\mu}^2 b^{\mu} u_1 
           - \bar{W}_{\mu}^1 b^{\mu} u_2
          \right)
  \nonumber\\
  & &
      - g \dfrac{v_Z^2}{2v} \left(
          (g\bar{W}_{\mu}^3-g_Y^{}\bar{B}_{\mu}) w^{1\mu} u_2 
         -(g\bar{W}_{\mu}^3-g_Y^{}\bar{B}_{\mu}) w^{2\mu} u_1
        \right) 
  \nonumber\\
  & & -g^2 \dfrac{v_Z^{}}{2} \left(
         \bar{W}_{\mu}^1 w^{2\mu} u_z 
       - \bar{W}_{\mu}^2 w^{1\mu} u_z
       \right) .
\label{eq:lag2_uv}
\end{eqnarray}

We are now ready to evaluate the vacuum polarization functions
arising from the bosonic fluctuation field ($u$, $w_\mu$ and $b_\mu$)
loops. 
We first consider the vacuum polarization functions (at zero momentum)
arising from the ${}^t u \sigma u$ term in Eq.(\ref{eq:lag2_uu})
with $u$ boson loop.
In the Feynman gauge $\xi=1$, we obtain
\begin{eqnarray}
  \Pi_{11}^{u}(0)
  &=& -\dfrac{1}{4} \left(4-3\dfrac{v_Z^2}{v^2}\right) A(M_W)
%       \int \dfrac{d^D k}{(2\pi)^D i} \dfrac{1}{M_W^2 - k^2 }
      -\dfrac{v_Z^2}{4v^2}   A(M_Z),
%       \int \dfrac{d^D k}{(2\pi)^D i} \dfrac{1}{M_Z^2 - k^2 },
  \nonumber\\
  & &
  \\
  \Pi_{33}^{u}(0)
  &=& -\dfrac{v_Z^4}{2v^4}  A(M_W),
%       \int \dfrac{d^D k}{(2\pi)^D i} \dfrac{1}{M_W^2 - k^2},
\end{eqnarray}
where the loop integral function $A$ is defined by 
Eq.(\ref{eq:def_Abare}).
In a similar manner, the $u$ boson loop contributions arising from
$\Gamma^\mu$ term in Eq.(\ref{eq:lag2_uu}) can be expressed by using
the loop integral functions $B_0$ (See Eq.(\ref{eq:def_B0bare}) for its 
definition), 
\begin{eqnarray}
  \Pi_{11}^{uu}(0) &=& \dfrac{v_Z^2}{4v^2}
  \left[ A(M_W) + A(M_Z) + B_0(M_W, M_Z; 0) \right] ,
  \nonumber\\
  & &
  \\
  \Pi_{33}^{uu}(0) &=& \dfrac{1}{4}\left(2-\dfrac{v_Z^2}{v^2}\right)^2
  \left[ 2A(M_W) + B_0(M_W, M_W; 0) \right]
  \nonumber\\
  &=& 0 .
\end{eqnarray}

We next consider the gauge boson loop diagrams arising from 
Eq.(\ref{eq:lag2_vv}).
For such a purpose, we first rearrange $w^{3\mu}$ and $b^{\mu}$
to the mass eigenfields
($z^\mu$ and $a^\mu$)
\begin{eqnarray}
  w^{3\mu} &=& \dfrac{1}{g_Z^{}}(g z^{\mu}+g_Y^{} a^{\mu}), \\
  b^{\mu} &=& \dfrac{1}{g_Z^{}}(-g_Y^{} z^{\mu}+g a^{\mu}),
\end{eqnarray}
in the Lagrangian Eq.(\ref{eq:lag2_vv}).
In the Feynman gauge $\xi=1$, we obtain
\begin{eqnarray}
  \Pi_{11}^{vv}(0)
  &=& D \left[ A(M_W) +\dfrac{g^2}{g_Z^2} A(M_Z) 
   +\dfrac{g_Y^2}{g_Z^2} A(0) \right.
  \nonumber\\
  & & \quad \left.
    + \dfrac{g^2}{g_Z^2} B_0(M_W, M_Z; 0)
                             + \dfrac{g_Y^2}{g_Z^2} B_0(M_W, 0; 0)
     \right],
  \nonumber\\
  & &
  \\
  \Pi_{33}^{vv}(0)
  &=& 2D \left[ A(M_W) + \frac{1}{2} B_0(M_W, M_W; 0) \right]
  \nonumber\\
  &=& 0. %, 
\end{eqnarray}
The effects of Faddeev-Popov ghost loop can be evaluated in a similar
manner, we obtain
\begin{eqnarray}
  \Pi_{11}^{cc}(0)
  &=& -2 \left[ A(M_W) +\dfrac{g^2}{g_Z^2} A(M_Z) 
   +\dfrac{g_Y^2}{g_Z^2} A(0) \right.
  \nonumber\\
  & & \quad \left.
    + \dfrac{g^2}{g_Z^2} B_0(M_W, M_Z; 0)
                             + \dfrac{g_Y^2}{g_Z^2} B_0(M_W, 0; 0)
     \right],
  \nonumber\\
  & &
  \\
  \Pi_{33}^{cc}(0)
  &=& -4 \left[ A(M_W) + \frac{1}{2} B_0(M_W, M_W; 0) \right]
  \nonumber\\
  &=& 0. 
\end{eqnarray}

We next consider the $u$ and gauge boson loop diagrams 
arising from Eq.(\ref{eq:lag2_uv}).
We obtain
\begin{eqnarray}
  \Pi_{11}^{uv}(0)
  &=& -\dfrac{1}{g_Z^2} \left[
       g^2 \dfrac{v}{2}\left(2-\dfrac{v_Z^2}{v^2}\right)
      -g_Y^2 \dfrac{v_Z^2}{2v} \right]^2 \times
  \nonumber\\
  & & \qquad \times
      B(M_W,M_Z;0)
  \nonumber\\
  & & -\dfrac{g^2 g_Y^2}{g_Z^2} \left[
      \dfrac{v}{2} \left(2-\dfrac{v_Z^2}{v^2}\right)+\dfrac{v_Z^2}{2v}
      \right]^2 
      B(M_W,0;0)
  \nonumber\\
  & & -\dfrac{g^2 v_Z^2}{4} 
       B(M_W,M_Z;0),
  \\
  \Pi_{33}^{uv}(0)
  &=& -\dfrac{g^2 v_Z^4}{2v^2}
       B(M_W,M_W;0),
\end{eqnarray}
with $B$ being defined by Eq.(\ref{eq:def_Bbare}).

It is now easy to evaluate $\tilde{\Pi}_{11}(0)$ and
$\tilde{\Pi}_{33}(0)$ as
\begin{eqnarray}
  \tilde{\Pi}_{11} &=& \Pi_{11}^{u} + \Pi_{11}^{uu}
    +\Pi_{11}^{vv} + \Pi_{11}^{uv},
  \\
  \tilde{\Pi}_{33} &=& \Pi_{33}^{u} + \Pi_{33}^{uu}
    +\Pi_{33}^{vv} + \Pi_{33}^{uv} .
\end{eqnarray}

\section{Loop integrals}
\label{sect-loop_integrals}

We define loop integrals in $D$ dimensions
\begin{equation}
  A(m)
  \equiv 
  \int \dfrac{d^D k}{(2\pi)^D i} \dfrac{1}{m^2-k^2},
\label{eq:def_Abare}
\end{equation}
and
\begin{eqnarray}
\lefteqn{
  B(m_1, m_2; p^2)
} \nonumber\\
 & & \equiv
  \int \dfrac{d^D k}{(2\pi)^D i} 
    \dfrac{1}{[m_1^2-(k+p)^2][m_2^2 - k^2]},
\label{eq:def_Bbare}
  \\
\lefteqn{
  g^{\mu\nu} B_0(m_1, m_2; p^2) 
} \nonumber\\
  & & \equiv
     \left. \int \dfrac{d^D k}{(2\pi)^D i} 
     \dfrac{(2k+p)^\mu (2k+p)^\nu}{[m_1^2-(k+p)^2][m_2^2 - k^2]}
     \right|_{g^{\mu\nu}},
\label{eq:def_B0bare}
\end{eqnarray}
with $I^{\mu\nu}|_{g^{\mu\nu}}$ denoting the $g^{\mu\nu}$ part of 
integral $I^{\mu\nu}(p)$, {\it i.e.,} 
\begin{displaymath}
  I^{\mu\nu}(p) = g^{\mu\nu} \left. I \right|_{g^{\mu\nu}}
                +p^\mu p^\nu \left. I \right|_{p^\mu p^\nu}.
\end{displaymath}
Note that the above definitions of the loop integrals differ
slightly from the definitions of
$\mathbf{A}$, $\mathbf{B}_0$, $\mathbf{B}_{22}$ 
used in Ref.\cite{Passarino:1978jh} : 
\begin{eqnarray}
  \mathbf{A}(m^2) &=& - (4\pi)^2 A(m) , 
  \\
  \mathbf{B}_0(p^2; m_1, m_2) &=& (4\pi)^2 B(m_1, m_2; p^2), 
  \\
  \mathbf{B}_{22}(p^2; m_1, m_2) &=& \dfrac{(4\pi)^2}{4} B_0(m_1, m_2; p^2) .
\end{eqnarray}
It is easy to see
\begin{equation}
  g^{\mu\nu} B_0(m_1, m_2; p^2) 
   =  \left. \int \dfrac{d^D k}{(2\pi)^D i} 
     \dfrac{4 k^\mu k^\nu}{[m_1^2-(k+p)^2][m_2^2 - k^2]}
     \right|_{g^{\mu\nu}}.
\end{equation}
\begin{widetext}
In the $D\to 4$ limit, these loop integrals suffer UV
divergences.
Introducing the UV cutoff momentum $\Lambda$,
they can be written as
\begin{equation}
  A(m) = \dfrac{\Lambda^2}{(4\pi)^2} 
       - \dfrac{m^2}{(4\pi)^2} \ln \dfrac{\Lambda^2}{\mu^2}
       + A_r(m),
\label{eq:defA}
\end{equation}
and
\begin{eqnarray}
 B(m_1, m_2; p^2) 
  &=& \dfrac{1}{(4\pi)^2} \ln \dfrac{\Lambda^2}{\mu^2}
     +B_r(m_1, m_2; p^2),
\label{eq:defB}
  \\
 B_0(m_1, m_2; p^2) 
  &=& -2\dfrac{\Lambda^2}{(4\pi)^2}
     +\dfrac{1}{(4\pi)^2}\left( m_1^2+m_2^2-\frac{1}{3}p^2 \right)
      \ln\dfrac{\Lambda^2}{\mu^2}
  % \nonumber\\
  % & & 
     +B_{0r}(m_1, m_2; p^2),
\label{eq:defB0}
\end{eqnarray}
with $\mu$ being a finite scale parameter.
Finite functions $A_r, B_r, B_{0r}$ can be expressed as
\begin{eqnarray}
  A_r(m) &=& -\dfrac{m^2}{(4\pi)^2} \left[
    \ln \dfrac{\mu^2}{m^2} + 1 \right],
\label{eq:A_r}
  \\
  B_r(m_1, m_2; p^2) &=&
    \dfrac{1}{(4\pi)^2} \int_0^1 dx \ln \left(
    \dfrac{\mu^2}{m_1^2 x + m_2^2(1-x) - p^2 x(1-x)}
    \right),
  \\
  B_{0r}(m_1, m_2; p^2) &=&
    \dfrac{2}{(4\pi)^2}\int_0^1 dx \left[
      m_1^2 x + m_2^2 (1-x) -p^2 x(1-x)
    \right]  
  % \times
  % \nonumber\\
  % & & \times 
  \left[
      \ln \left(
        \dfrac{\mu^2}{m_1^2 x + m_2^2(1-x) - p^2 x(1-x)}
      \right)+1 
    \right] .
  \nonumber\\
   & &
\end{eqnarray}
Performing the parameter integrals, we find
\begin{eqnarray}
  (4\pi)^2 B_r(m_1, m_2; 0)
  &=& 1 - \dfrac{1}{m_1^2-m_2^2} \left[ 
        m_1^2 \ln \dfrac{m_1^2}{\mu^2} 
      - m_2^2 \ln \dfrac{m_2^2}{\mu^2} 
      \right],
\label{eq:br}
  \\
  (4\pi)^2 B_r'(m_1, m_2; 0)
  &=& \dfrac{1}{(m_1^2-m_2^2)^2} \left[
     \dfrac{m_1^2+m_2^2}{2}
    -\dfrac{m_1^2 m_2^2}{m_1^2-m_2^2} \ln \dfrac{m_1^2}{m_2^2}
    \right],
\label{eq:brp}
  \\
  (4\pi)^2 B_{0r}(m_1, m_2; 0)
  &=& \dfrac{3}{2}(m_1^2+m_2^2)
     -\dfrac{1}{m_1^2-m_2^2} \left[
       m_1^4 \ln \dfrac{m_1^2}{\mu^2}
      -m_2^4 \ln \dfrac{m_2^2}{\mu^2}
     \right],
\label{eq:b0r}
  \\
  (4\pi)^2 B_{0r}'(m_1, m_2; 0)
  &=& -\dfrac{1}{18} \dfrac{5m_1^4 - 22 m_1^2 m_2^2 + 5m_2^4}{(m_1^2-m_2^2)^2}
      \nonumber\\
  & & +\dfrac{1}{3} \dfrac{1}{(m_1^2-m_2^2)^3} \left[
        m_1^4(m_1^2-3m_2^2) \ln \dfrac{m_1^2}{\mu^2}
       -m_2^4(m_2^2-3m_1^2) \ln \dfrac{m_2^2}{\mu^2}
      \right],
  \nonumber\\
  & &
\label{eq:b0rp}
\end{eqnarray}
with $B_r'$, $B_{0r}'$ being defined by
\begin{equation}
  B_r'(m_1, m_2; p^2) \equiv \dfrac{d}{dp^2} B_r(m_1, m_2; p^2),   
  \qquad
  B_{0r}'(m_1, m_2; p^2) \equiv \dfrac{d}{dp^2} B_{0r}(m_1, m_2; p^2) .
\end{equation}

The functions used in the expressions of
$S_f$, $T_f$ and $U_f$ are defined as
\begin{eqnarray}
F^{\phi_n \phi_m} 
  &\equiv& (4\pi)^2 \left[ 
             B_{0r}(M_{\phi_n}, M_{\phi_m}; 0)
            +A_{r}(M_{\phi_n})  + A_{r}(M_{\phi_m})
          \right]
  \nonumber\\
  &=& \dfrac{M_{\phi_n}^2+M_{\phi_m}^2}{2}
     -\dfrac{M_{\phi_n}^2 M_{\phi_m}^2}{M_{\phi_n}^2-M_{\phi_m}^2}
      \ln\dfrac{M_{\phi_n}^2}{M_{\phi_m}^2}, 
\\
{F^{\phi_n \phi_m}}' 
  &\equiv& (4\pi)^2 B_{0r}'(M_{\phi_n}, M_{\phi_m}; 0) 
  \nonumber\\
  &=& 
   -\dfrac{1}{3} \left\{
    \dfrac{4}{3}
   -\dfrac{
       M_{\phi_n}^2\ln\dfrac{M_{\phi_n}^2}{\mu^2} 
      -M_{\phi_m}^2\ln\dfrac{M_{\phi_m}^2}{\mu^2} 
    }{M_{\phi_n}^2 - M_{\phi_m}^2}
  %   \right.
  % \nonumber\\
  % & & \left.
   -\dfrac{M_{\phi_n}^2 + M_{\phi_m}^2}{\left(M_{\phi_n}^2 - M_{\phi_m}^2\right)^2}
     F^{\phi_n \phi_m} \right\},
\\
G^{V\phi}  
  &\equiv& (4\pi)^2 \left[
                B_{0r}(M_\phi, M_V;0)-4M_V^2 B_r(M_\phi, M_V; 0)
  % \right.
  % \nonumber\\ 
  % & &   \left.
               +A_r(M_\phi)+A_r(M_V)
           \right]
  \nonumber\\
  &=& F^{V\phi} 
  % \nonumber\\
  % & & 
       + 4M_V^2 \left( -1 
       +\dfrac{
          M_\phi^2 \ln \dfrac{M_\phi^2}{\mu^2}
        - M_V^2 \ln \dfrac{M_V^2}{\mu^2}}{M_\phi^2-M_V^2}\right), 
\\
{G^{V\phi}}'  
  &\equiv& (4\pi)^2 \left[
                B_{0r}'(M_\phi, M_V;0)-4M_V^2 B_r'(M_\phi, M_V; 0)
           \right]
  \nonumber\\
  &=& {F^{V\phi}}' - \dfrac{4M_V^2}{(M_V^2-M_\phi^2)^2} F^{V\phi},
\end{eqnarray}
where functions $A_r$, $B_r$, $B_{0r}$ $B'_r$ and $B'_{0r}$
are given in Eq.(\ref{eq:A_r}), 
Eq.(\ref{eq:br}), Eq.(\ref{eq:brp}), 
Eq.(\ref{eq:b0r}) and Eq.(\ref{eq:b0rp}).
\end{widetext}

For $\Delta M_{\phi_n\phi_m} \equiv |M_{\phi_n}-M_{\phi_m}| \ll
     M_{\phi_n}, M_{\phi_m}$, we find
\begin{eqnarray}
  F^{\phi_n \phi_m} 
  &=& \dfrac{2}{3} (\Delta M_{\phi_n\phi_m})^2
     -\dfrac{1}{30}\dfrac{(\Delta M_{\phi_n\phi_m})^4}
                         {\overline{M}_{\phi_n \phi_m}^2}
     + \cdots,
  \nonumber\\
  & &
  \\
  {F^{\phi_n \phi_m}}' 
  &=& \dfrac{1}{3} \ln \dfrac{\overline{M}_{\phi_n \phi_m}^2}{\mu^2}
     +\dfrac{1}{20} \dfrac{(\Delta M_{\phi_n\phi_m})^2}
                          {\overline{M}_{\phi_n \phi_m}^2}
     + \cdots,
  \nonumber\\
  & &
\end{eqnarray}
with
\begin{equation}
  \overline{M}_{\phi_n \phi_m} \equiv \dfrac{M_{\phi_n}+M_{\phi_m}}{2}.
\end{equation}

For $M_V \ll M_\phi$, we also note
\begin{eqnarray}
  G^{V\phi} &=& 
    \dfrac{1}{2} M_\phi^2 
   +\left( 3 \ln \dfrac{M_\phi^2}{\mu^2} + \ln \dfrac{M_V^2}{\mu^2} 
          -\dfrac{7}{2} \right) M_V^2 + \cdots,
  \nonumber\\
  & &
  \\
  {G^{V\phi}}' &=& 
    \dfrac{1}{3} \ln \dfrac{M_\phi^2}{\mu^2} - \dfrac{5}{18}
   -\dfrac{4}{3} \dfrac{M_V^2}{M_\phi^2} + \cdots .
\end{eqnarray}

\section{Self-interactions among Higgs bosons}
\label{sec:self-inter-among}

In this appendix, we list the formulas
of $\tilde{M}$ and $\tilde{\lambda}$ used in
Sec.~\ref{sec:uv-completion} , 
\begin{eqnarray}
  (\tilde{M}^2)_{11} &=& 
    (\kappa_V^{\phi_1^0})^2 M_{\phi_1^0}^2
   +(\kappa_V^{\phi_2^0})^2 M_{\phi_2^0}^2,
  \\
  (\tilde{M}^2)_{12} &=& 
    (\kappa_V^{\phi_1^0})(\kappa_V^{\phi_2^0})
    (M_{\phi_2^0}^2 -M_{\phi_1^0}^2 ), 
  \\
  (\tilde{M}^2)_{22} &=& 
    (\kappa_V^{\phi_2^0})^2 M_{\phi_1^0}^2
   +(\kappa_V^{\phi_1^0})^2 M_{\phi_2^0}^2,
\end{eqnarray}
\begin{eqnarray}
  \tilde{\lambda}_{111} &=& 
    \lambda_{111}(\kappa^{\phi^0_1}_V)^3
   +\lambda_{222}(\kappa^{\phi^0_2}_V)^3
   +3\lambda_{112}(\kappa^{\phi^0_1}_V)^2(\kappa^{\phi^0_2}_V)
  \nonumber\\
  & & 
   +3\lambda_{122}(\kappa^{\phi^0_1}_V)(\kappa^{\phi^0_2}_V)^2,
  \\
  \tilde{\lambda}_{112} &=& 
     -\lambda_{111}(\kappa^{\phi^0_1}_V)^2(\kappa^{\phi^0_2}_V)
     +\lambda_{222}(\kappa^{\phi^0_1}_V)(\kappa^{\phi^0_2}_V)^2
  \nonumber\\
  & & +\lambda_{112}\left[
        (\kappa^{\phi^0_1}_V)^3-2(\kappa^{\phi^0_1}_V)(\kappa^{\phi^0_2}_V)^2
      \right]
  \nonumber\\
  & & 
     -\lambda_{122}\left[
        (\kappa^{\phi^0_2}_V)^3-2(\kappa^{\phi^0_1}_V)^2(\kappa^{\phi^0_2}_V)
     \right],
  \\
  \tilde{\lambda}_{122} &=& 
     \lambda_{111}(\kappa^{\phi^0_1}_V)(\kappa^{\phi^0_2}_V)^2
    +\lambda_{222}(\kappa^{\phi^0_1}_V)^2(\kappa^{\phi^0_2}_V)
  \nonumber\\
  & & +\lambda_{112}\left[
         (\kappa^{\phi^0_2}_V)^3-2(\kappa^{\phi^0_1}_V)^2(\kappa^{\phi^0_2}_V)
       \right]
  \nonumber\\
  & & 
      +\lambda_{122}\left[
         (\kappa^{\phi^0_1}_V)^3-2(\kappa^{\phi^0_1}_V)(\kappa^{\phi^0_2}_V)^2
       \right],
  \\
  \tilde{\lambda}_{222} &=& 
      -\lambda_{111}(\kappa^{\phi^0_2}_V)^3
      +\lambda_{222}(\kappa^{\phi^0_1}_V)^3
      +3\lambda_{112}(\kappa^{\phi^0_1}_V)(\kappa^{\phi^0_2}_V)^2
  \nonumber\\
  & & 
     -3\lambda_{122}(\kappa^{\phi^0_1}_V)^2(\kappa^{\phi^0_2}_V),
\end{eqnarray}
\begin{eqnarray}
  \tilde{\lambda}_{1111} &=& 
     \lambda_{1111}(\kappa^{\phi^0_1}_V)^4
    +\lambda_{2222}(\kappa^{\phi^0_2}_V)^4
  \nonumber\\
  & &
    +6\lambda_{1122}(\kappa^{\phi^0_1}_V)^2(\kappa^{\phi^0_2}_V)^2
  \nonumber\\
  & &
    +4\lambda_{1112}(\kappa^{\phi^0_1}_V)^3(\kappa^{\phi^0_2}_V)
    +4\lambda_{1222}(\kappa^{\phi^0_1}_V)(\kappa^{\phi^0_2}_V)^3,
  \nonumber\\
  & &
  \\
  \tilde{\lambda}_{1112} &=& 
    -\lambda_{1111}(\kappa^{\phi^0_1}_V)^3(\kappa^{\phi^0_2}_V)+\lambda_{2222}(\kappa^{\phi^0_1}_V)(\kappa^{\phi^0_2}_V)^3
  \nonumber\\
  & & +3\lambda_{1122}(\kappa^{\phi^0_1}_V)(\kappa^{\phi^0_2}_V)\left[
         (\kappa^{\phi^0_1}_V)^2-(\kappa^{\phi^0_2}_V)^2
       \right]
  \nonumber\\
  & & +\lambda_{1112}(\kappa^{\phi^0_1}_V)^4
      -3\lambda_{1112}(\kappa^{\phi^0_1}_V)^2(\kappa^{\phi^0_2}_V)^2
  \nonumber\\
  & & 
      -\lambda_{1222}(\kappa^{\phi^0_2}_V)^4
      +3\lambda_{1222}(\kappa^{\phi^0_1}_V)^2(\kappa^{\phi^0_2}_V)^2,
  \nonumber\\
  & & 
  \\
  \tilde{\lambda}_{1122} &=& 
     (\lambda_{1111}+\lambda_{2222})(\kappa^{\phi^0_1}_V)^2(\kappa^{\phi^0_2}_V)^2
  \nonumber\\
  & & 
    +\lambda_{1122}\left[
        (\kappa^{\phi^0_1}_V)^4
      -4(\kappa^{\phi^0_1}_V)^2(\kappa^{\phi^0_2}_V)^2
       +(\kappa^{\phi^0_2}_V)^4
     \right]
  \nonumber\\
  & & -2(\lambda_{1112}-\lambda_{1222})
        (\kappa^{\phi^0_1}_V)(\kappa^{\phi^0_2}_V)\times
  \nonumber\\
  & & \quad \times
        \left[
           (\kappa^{\phi^0_1}_V)^2-(\kappa^{\phi^0_2}_V)^2
        \right],
  \\
  \tilde{\lambda}_{1222} &=& 
    -\lambda_{1111}(\kappa_V^{\phi_1^0}) (\kappa_V^{\phi_2^0})^3 f 
      + \lambda_{2222}(\kappa_V^{\phi_1^0})^3 (\kappa_V^{\phi_2^0}),
      \nonumber\\
    & & -3\lambda_{1122}
      \left[ (\kappa_V^{\phi_1^0})^2 - (\kappa_V^{\phi_2^0})^2 \right] 
      (\kappa_V^{\phi_1^0})(\kappa_V^{\phi_2^0})  
  \nonumber\\
  & & 
    +\lambda_{1112}\left[3(\kappa_V^{\phi_1^0})^2 -(\kappa_V^{\phi_2^0})^2 \right]
     (\kappa_V^{\phi_2^0})^2  
  \nonumber\\
  & & +\lambda_{1222}
       \left[(\kappa_V^{\phi_1^0})^2 -3(\kappa_V^{\phi_2^0})^2\right]
       (\kappa_V^{\phi_1^0})^2  ,
  \\
  \tilde{\lambda}_{2222} &=& 
       \lambda_{1111}(\kappa^{\phi^0_2}_V)^4
      +\lambda_{2222}(\kappa^{\phi^0_1}_V)^4
  \nonumber\\
  & & 
      +6\lambda_{1122}(\kappa^{\phi^0_1}_V)^2(\kappa^{\phi^0_2}_V)^2
   \nonumber\\
  & & -4\lambda_{1112}(\kappa^{\phi^0_1}_V)(\kappa^{\phi^0_2}_V)^3
      -4\lambda_{1222}(\kappa^{\phi^0_1}_V)^3(\kappa^{\phi^0_2}_V).
  \nonumber\\
  & & 
\end{eqnarray}

\end{document}